\documentclass[modern]{aastex61}
\usepackage{natbib}

\usepackage{hyperref}
\hypersetup{colorlinks=true, citecolor=blue, linkcolor=cyan, urlcolor=magenta}

\newcommand{\gm}{\phantom{\tablenotemark{a}}}
\newcommand{\CI}{[\ion{C}{1}]~($^3P_1 - {}^3P_0$)}
\newcommand{\CO}{CO~($4-3$)}
\newcommand{\smpyr}{$M_\sun~\rm yr^{-1}$}

\newcommand{\h}{$^{\rm h}$}
\newcommand{\m}{$^{\rm m}$}
\newcommand{\s}{$^{\rm s}$}

\defcitealias{2013ApJ...772...25S}{S13}
\newcommand{\Sifon}{\citetalias{2013ApJ...772...25S}}
\newcommand{\Sifonp}{\citepalias{2013ApJ...772...25S}}
\defcitealias{2015ApJ...814....9K}{K15}
\newcommand{\Kirkpatrick}{\citetalias{2015ApJ...814....9K}}

\begin{document}

\title{Herschel and ALMA Observations of Massive SZE-selected Clusters}

\author{John F. Wu}
\affiliation{Department of Physics and Astronomy, Rutgers, The State University of New Jersey, 136 Frelinghuysen Road, Piscataway, NJ 08854-8019, USA}
\email{jfwu@physics.rutgers.edu}

\author{Paula Aguirre}
\affiliation{School of Engineering, Pontificia Universidad Cat\'olica de Chile, Av. Vicu\~na Mackenna 4068, Macul, Santiago, Chile}

\author{Andrew J. Baker}
\affiliation{Department of Physics and Astronomy, Rutgers, The State University of New Jersey, 136 Frelinghuysen Road, Piscataway, NJ 08854-8019, USA}

\author{Mark J. Devlin}
\affiliation{Department of Physics and Astronomy, University of Pennsylvania, Philadelphia, PA 19104, USA}

\author{Matt Hilton}
\affiliation{Astrophysics \& Cosmology Research Unit, School of Mathematics, Statistics \& Computer Science, University of KwaZulu-Natal, Westville Campus, Durban 4041, South Africa}

\author{John P. Hughes}
\affiliation{Department of Physics and Astronomy, Rutgers, The State University of New Jersey, 136 Frelinghuysen Road, Piscataway, NJ 08854-8019, USA}
\affiliation{Center for Computational Astrophysics, Flatiron Institute, 162 Fifth Avenue, New York, NY 10010, USA}

\author{Leopoldo Infante}
\affiliation{Instituto de Astrof\'isica and Centro de Astroingenier\'ia, Facultad de F\'isica, Pontificia Universidad Cat\'olica de Chile, Vicu\~na Mackenna 4860, 7820436 Macul, Santiago, Chile}
\explain{We have added an author who was inadvertently left out.}

\author{Robert R. Lindner}
\affiliation{Department of Astronomy, The University of Wisconsin-Madison, 475 N Charter St, Madison, WI 53706}

\author{Crist\'{o}bal Sif\'{o}n}
\affiliation{Department of Astrophysical Sciences, Princeton University, Princeton, NJ 08544, USA}

	
\begin{abstract}
We present new \textit{Herschel} observations of four massive, Sunyaev-Zel'dovich Effect (SZE)-selected clusters at $0.3 \leq z \leq 1.1$, two of which have also been observed with ALMA.
We detect 19 \textit{Herschel}/PACS counterparts to spectroscopically confirmed cluster members, five of which have redshifts determined via \CO{} and \CI{} lines.
The mean [\ion{C}{1}]/CO line ratio is $0.19 \pm 0.07$ in brightness temperature units, consistent with previous results for field samples.
We do not detect significant stacked ALMA dust continuum or spectral line emission, implying upper limits on mean interstellar medium (H$_2$ + \ion{H}{1}) and molecular gas masses.
An apparent anticorrelation of $L_{\rm IR}$ with clustercentric radius is driven by the tight relation between star formation rate and stellar mass.
We find average specific star formation rate log(sSFR/yr$^{-1})=-10.36$, which is below the SFR$-M_\ast$ correlation measured for field galaxies at similar redshifts.
The fraction of infrared-bright galaxies (IRBGs; $\log (L_{\rm IR}/L_\sun) > 10.6$) per cluster and average sSFR rise significantly with redshift.
For CO detections, we find $f_{\rm gas} \sim 0.2$, comparable to those of field galaxies, and gas depletion timescales of about $2$~Gyr.
We use radio observations to distinguish active galactic nuclei (AGNs) from star-forming galaxies.
At least four of our 19 \textit{Herschel} cluster members have $q_{\rm IR} < 1.8$, implying an AGN fraction $f_{\rm AGN} \gtrsim 0.2$ for our PACS-selected sample.
\end{abstract}

\setcounter{footnote}{0}

\section{Introduction}

Galaxy clusters are the most massive virialized structures in the universe.
Due to interactions with the hot intracluster medium (ICM) and with other galaxy members, cluster galaxies are likely to evolve differently from field galaxies.
As a result, nearby galaxy clusters are almost entirely devoid of the gas-rich, star-forming galaxies that we often see in the field; instead, they are full of quiescent early-type galaxies, particularly in their dense cores \citep[see, e.g.,][]{1980ApJ...236..351D}.

The impacts of cluster environment on star formation properties have been studied extensively in the local universe \citep[see, e.g.,][]{1993AJ....106..473C, 2002MNRAS.334..673L, 2004MNRAS.353..713K, 	
2004ApJ...613..851K, 2006PASP..118..517B}.
One such example is the Virgo Cluster, in which \ion{H}{1}-rich galaxies are still making their first passes through the cluster \citep[][]{2004AJ....127.3361K, 2004ApJ...613..866K, 2009AJ....138.1741C}.
Ongoing ram-pressure stripping by the hot ICM truncates infalling galaxy cold gas disks \citep{1972ApJ...176....1G}, and stripping of hot gas in their surrounding halos depletes gas reservoirs that would otherwise cool and replenish their disks \citep[starvation/strangulation;][]{1980ApJ...237..692L}.
These mechanisms catalyze the evolution of cluster members, leaving them with old stellar populations after their cold gas components are exhausted.
Other effects, such as collisional interactions between galaxies \citep{1996Natur.379..613M, 2004cgpc.symp..277M}, tidal interactions \citep{1999ApJ...510L..15B}, viscous or turbulent stripping \citep{1982MNRAS.198.1007N}, or thermal evaporation \citep{1977Natur.266..501C}, may play a part in their evolution as well.
Most $z \sim 0$ virialized clusters reflect these quenching processes and are full of ``red and dead'' elliptical galaxies.

Observations of clusters at increasing redshifts reveal that their populations are more likely to include blue, star-forming galaxies \citep[Butcher-Oemler effect;][]{1978ApJ...219...18B}.
Star formation rate (SFR) trends with redshift have also been measured at infrared wavelengths in intermediate-redshift clusters \citep[see, e.g.,][]{2009ApJ...704..126H, 2010ApJ...720...87F}.
At $z \gtrsim 1.4$, cluster galaxies appear to be forming new stars at exceptionally high rates \citep[$\sim 100~$\smpyr{};][]{2013ApJ...779..138B}, typical of obscured field galaxies at $z \sim 2$ \citep[i.e.,][]{2013A&A...553A.132M}.
Furthermore, the well-known local morphology-density relation reverses at high redshifts; in fact, galaxy SFRs in high-$z$ cluster cores exceed field galaxy SFRs at the same epochs \citep[][but see also, e.g., \citealt{2014MNRAS.437..458Z, 2015A&A...579A.132P}]{2010ApJ...719L.126T, 2010ApJ...718..133H, 2010MNRAS.402.1980H, 2013ApJ...779..138B, 2014MNRAS.437..437A}.
The fraction of active galactic nuclei (AGNs) in clusters also appears to evolve significantly with redshift \citep[analogous to the Butcher-Oemler effect;][]{2009A&A...508..645G, 2013ApJ...768....1M}.
\cite{2009A&A...508..645G} find that the prevalence of X-ray-selected AGNs increases by at least a factor of 3 from clusters at $0.5 < z < 1$ to clusters at $1 < z < 1.5$, and \cite{2009ApJ...696..891H} and \cite{2016ApJ...825...72A} find similar results with AGNs selected at other wavelengths.

\cite{2013ApJ...779..138B} suggest that clusters undergo an epoch of galaxy merger-driven starbursts and AGN activity at $z \gtrsim 1.4$.
In this scenario, abundant cool gas in high-$z$ cluster galaxies, carried by in-falling galaxies and filaments, induces star formation activity.
As stellar mass assembly ramps up in galaxies replete with dense gas, tidal forces and galaxy mergers also promote gas accretion onto the super-massive black holes at the centers of galaxies, causing AGN feedback.
AGNs provide efficient heating mechanisms to quench star formation and mature post-starburst galaxies to quiescent early-types \citep[e.g.,][]{2008ApJS..175..356H,2008MNRAS.391..481S}.
As cluster galaxies grow in stellar mass and cold gas reservoirs are depleted, they also become quiescent, and SFRs plummet.

How does galaxy density or cluster halo mass shape the star formation histories of cluster galaxies, and how do these histories compare to those of field galaxies? 
\cite{2011ApJ...743...34C} find that, in nearby clusters, the specific SFR ($\textrm{sSFR}~\equiv ~\textrm{SFR}/M_\ast$) and fraction of star-forming galaxies increase with projected distance from the cluster center --- i.e., SFR anticorrelates with density even when stellar mass is taken into account. 
However, they find that cluster halo masses do not correlate with integrated sSFR, indicating that ram pressure stripping and galaxy harassment --- both of which scale with cluster halo mass --- are \textit{not} important mechanisms for galaxy evolution in low-$z$ clusters.
Such conclusions may not be applicable to star-forming populations in more massive clusters. 
For example, \cite{2006PASP..118..517B} find that ram pressure stripping and other ICM processes are \textit{most} relevant to the evolution of late-type cluster galaxies in local \textit{massive} clusters.
Although tidal forces and galaxy-galaxy interactions induce star-formation and AGN activity and eventual gas consumption in higher-$z$ clusters with lower velocity dispersions, ICM processes such as ram pressure stripping and thermal evaporation are thought to be the main quenching mechanisms for galaxies in massive, evolved clusters.

In addition to environmental variables such as density and cluster halo mass, how do redshift and cluster dynamical state (which correspond to larger scales) impact AGN prevalence and SFR?
\cite{1999AJ....118..625D} study two $z \sim 0.25$ clusters with different dynamical states in order to understand their star-forming and AGN populations as traced by radio observations.
The cluster with significant substructure, indicative of an ongoing merger, contains a larger fraction of star-forming galaxies, as traced by blue optical color or low radio luminosity, than the other virialized, passive cluster.
The authors propose that the different dynamical states in the two clusters are responsible for the differences in star-forming populations.
\cite{2005A&A...431..433C} find an anomalously high fraction of mid-infrared sources in a $z \sim 0.5$ cluster, which they attribute to the cluster's recent merger.
By juxtaposing the populations of obscured star-forming sources of this and another cluster at similar redshift, \cite{2006ApJ...649..661G} conclude that a combination of ICM effects and dynamical state are responsible for triggering recent star formation.
Similar examples are also seen in the local universe, such as the still-merging Virgo cluster and the evolved Coma cluster. 
Abundant signs of recent galaxy evolution can be found in the former \citep{2009AJ....138.1741C}, but few are seen in the latter (e.g., \citealt{2004ApJ...601..197P}).

Detailed observations and analyses are necessary for understanding the growth of cluster galaxies during the transition phase at intermediate ($0.3 \leq z \leq 1.4$) redshifts in the mass regime where environmental effects are most pronounced.
The existence of cold gas reservoirs, which may be traced by molecular lines or dust emission, is vital for continued star formation in clusters.
One might ask how gas and star formation properties of cluster galaxies are impacted by an extreme cluster environment (i.e., how galaxies might evolve in the cores versus near the outskirts of clusters).
Other properties, such as cluster dynamical state (whether a cluster is merging, dynamically disturbed, or virialized) and total cluster mass, are worth investigating in order to explore their impacts on star formation.
Questions of how cluster galaxies evolve at intermediate redshifts can be comprehensively answered by studying both their star formation \textit{and} their gas/dust properties.

We present a study of galaxies in four $0.3 \lesssim z \lesssim 1.1$ massive clusters selected via the Sunyaev-Zel'dovich effect \citep[SZE;][]{1972CoASP...4..173S}.
The clusters are selected from the LABOCA/ACT Survey of Clusters at All Redshifts \citep[LASCAR;][]{2015ApJ...803...79L} sample.
Our sample of clusters ranges in mass from $(5 - 13) \times 10^{14}~M_\odot$, nearly an order of magnitude greater than the masses of clusters used in previous studies \citep[e.g., the IRAC Shallow Cluster Survey;][]{2008ApJ...684..905E, 2013ApJ...779..138B, 2014MNRAS.437..437A}.
By targeting the most massive clusters, we can see how the most extreme environments affect their galaxies -- and because many clusters in the LASCAR sample are not dynamically relaxed, we can additionally study the effects of cluster mergers on galaxy properties.
We focus on infrared observations that trace obscured star formation, as well as (in the case of our two $z \sim 1$ clusters) the dust and cold gas content that are detectable at millimeter wavelengths.

In Section \ref{sec:observations}, we introduce new \textit{Herschel} and Atacama Large Millimeter/submillimeter Array (ALMA) Band 6 observations, and present observations at other wavelengths used in the analysis.
We describe new detections in Section \ref{sec:results} and the results of a stacking analysis in Section \ref{sec:stacking}. 
In Section \ref{sec:discussion}, we discuss the implications of these findings and compare with other results in the literature.
Detailed descriptions of individual galaxy detections are presented in the Appendix.

We assume a flat, concordance $\Lambda$CDM cosmology with $H_0 = 70 {\rm\ km\ s^{-1}\ Mpc^{-1}}$, $\Omega_\Lambda = 0.7$, and $\Omega_M  = 0.3$.
All magnitudes are reported in the AB system.

\section{Observations} \label{sec:observations}

\begin{figure*}
\plotone{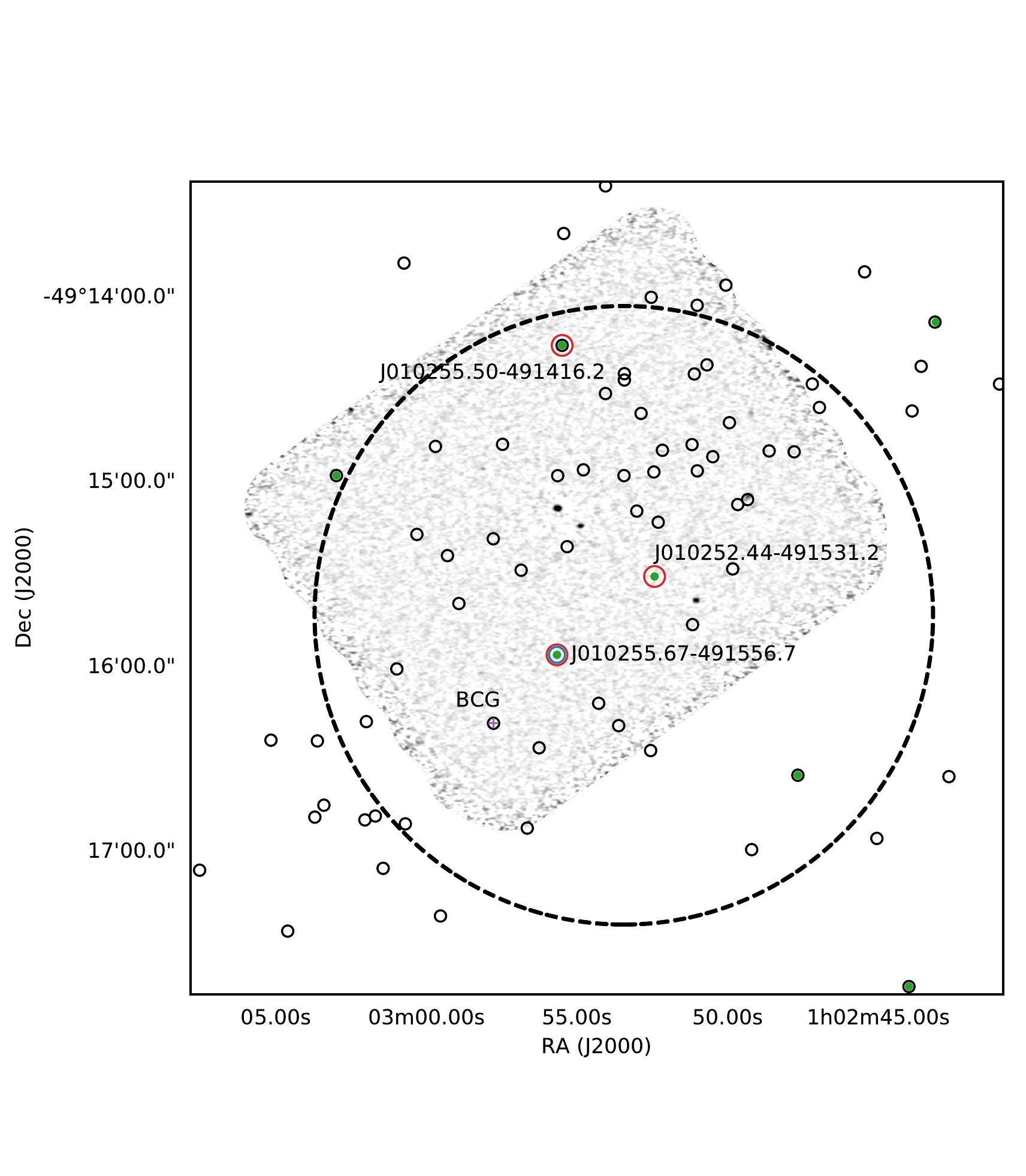}
\caption{ALMA Band 6 continuum mosaic imaging of J0102 shown in grayscale. Confirmed cluster members are marked by circles: black circles are galaxies found via optical spectroscopy (\Sifon{}); smaller green filled circles are galaxies with \textit{Herschel}/PACS counterparts; larger red (blue) circles are galaxies with ALMA \CO{} (\CI{}) line detections (labeled by name). A purple cross marks the brightest cluster galaxy (also labeled). The large dashed circle encloses $0.5\times R_{\rm 200c}$ and is centered on the midpoint between the two peaks in the mass distribution \citep[determined by weak lensing;][]{2014ApJ...785...20J}. 
\vspace{1em}
}
\label{fig:alma overview}
\end{figure*}

\begin{figure*}
\plotone{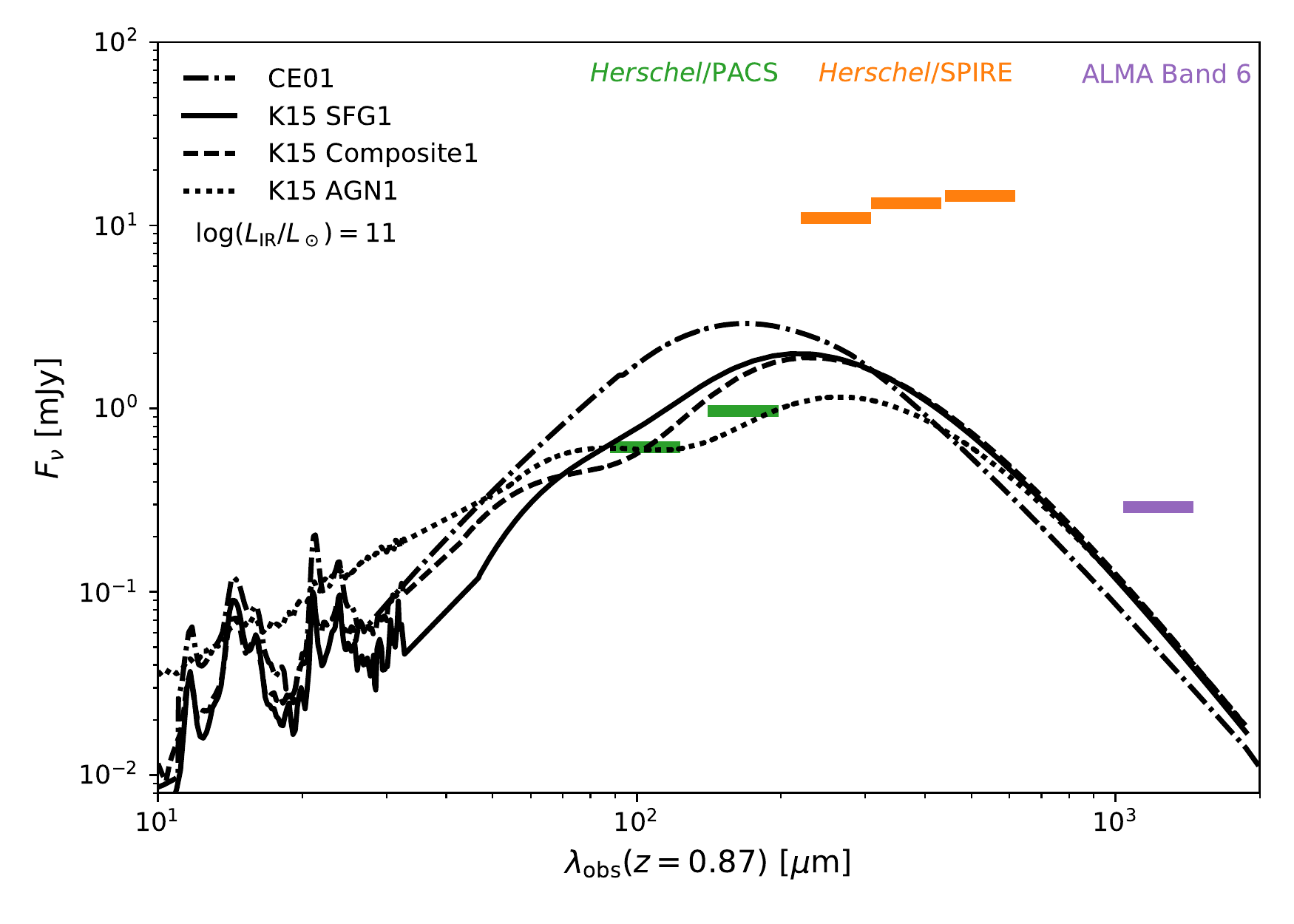}
\caption{\citet[][CE01]{2001ApJ...556..562C} and three example \citet[][K15]{2015ApJ...814....9K} template SEDs with $L_{\rm IR} = 10^{11}~L_\sun$, redshifted to $z=0.87$. We show the sensitivities ($3~\sigma$ upper limits) of our FIR/millimeter observations (point-source subtracted maps) in colored horizontal lines.}
\label{fig:continuum sensitivities}
\end{figure*}

We use multiple datasets to study the star formation, dust, and cold gas properties of galaxies in our cluster sample. 
In the following subsections, we describe the relevant datasets and all steps needed to clean data and generate data products for our analysis.
In Figure~\ref{fig:alma overview}, we show an example ALMA continuum image of J0102 with \textit{Herschel} and ALMA spectral line detections of spectroscopically confirmed cluster members. 
We show in Figure \ref{fig:continuum sensitivities} fiducial SEDs from the \cite{2001ApJ...556..562C} and \cite{2015ApJ...814....9K} template libraries after redshifting to $z=0.87$ in order to compare to our J0102 observations.
\textit{Herschel} and ALMA continuum sensitivities ($3~\sigma_{\rm rms}$ values) of J0102 are also shown for comparison.
Details about new ALMA Band 6 observations are listed in Table~\ref{tab:ALMA observations}. 

\subsection{The sample of massive clusters}

The Atacama Cosmology Telescope \citep[ACT;][]{2011ApJS..194...41S} observed at 148 GHz a 455 $\rm deg^2$ patch of the southern sky in a region spanning right ascension $\rm 00^h12^m$ to $\rm 07^h08^m$ and declination $-56\arcdeg~11\arcmin$ to $-49\arcdeg00\arcmin$, identifying 23 SZE decrements in the cosmic microwave background (CMB) as cluster candidates \citep{2011ApJ...737...61M}.
Optical and X-ray follow-up observations confirmed these SZE detections to be rich clusters \citep{2010ApJ...723.1523M}, and spectroscopy for 16 clusters provided precise redshifts and dynamical mass estimates \citep[][hereafter \Sifon{}]{2013ApJ...772...25S}. 
\cite{2016MNRAS.461..248S} found that the median dynamical mass of the 16 clusters was $M_{\rm 200c} \approx 8.2 \times 10^{14} M_\sun$,\footnote{$M_{\rm 200c} \equiv 200 (4\pi/3) \rho_c r_{\rm 200c}^3$, where $\rho_c$ is the critical density of the universe at the redshift of the cluster, and $r_{\rm 200c}$ is the radius enclosing a density of 200 times $\rho_c$.} consistent with expectations that the ACT SZE survey would detect the most massive clusters in its volume.

\floattable
\begin{deluxetable}{c c c c}
	\tablecaption{Our sample of clusters\label{tab:clusters}}
	\tablecolumns{5}
	\tablehead{Cluster name & $z$\tablenotemark{a} & $N_{\rm gal}$\tablenotemark{a}\gm & $M_{\rm 200c}$\tablenotemark{b}\\
		& & &  $[10^{14}~M_\sun]$} 
	
	\startdata
	ACT-CL J0235-5121 & $0.2777$ & $82$ & \phn$8.0 \pm 2.9$ \\
	ACT-CL J0438-5419 & $0.4214$ & $65$ & $12.9 \pm 3.2$ \\
	ACT-CL J0102-4915 & $0.8701$ & $89$ & $11.3 \pm 2.9$ \\
	ACT-CL J0546-5345 & $1.0663$ & $49$ & \phn$5.5 \pm 2.3$ \\
	\enddata
	
	\tablenotetext{a}{Spectroscopic redshifts and numbers of spectroscopically confirmed members ($N_{\rm gal}$) from \cite{2013ApJ...772...25S}.}
	\tablenotetext{b}{Dynamical mass estimates ($M_{\rm 200c}$) from \cite{2016MNRAS.461..248S}.}
	\vspace{-1em}
\end{deluxetable}

The LASCAR sample was selected from the highest signal-to-noise ratio (SNR) of these SZE decrements that had not yet been observed at submillimeter wavelengths as of 2011. 
Nine out of ten LASCAR clusters were undiscovered before the ACT or SPT surveys \citep{2010ApJ...723.1523M, 2011ApJ...737...61M, 2010ApJ...722.1180V}.  
These clusters have redshifts $z \approx 0.3-1.1$ and masses $M_{\rm 200c} \sim (5-13) \times 10^{14} M_\sun$ \citep{2016MNRAS.461..248S}.

We focus our analysis on a subsample of four LASCAR clusters: two are at low redshift ($z\sim 0.3$) and two are at high redshift ($z\sim 1$).
ACT-CL~J0102-4915 (J0102 for short; also known as ``El Gordo''; \citealt{2012ApJ...748....7M}) appears to be caught in the midst of a spectacular merger, and ACT-CL~J0546-5345 (J0546) is the highest-redshift cluster in the LASCAR sample.
The two clusters at lower redshift are of comparable mass.
All four are considered ``disturbed'' on the basis of their galaxy dynamics (see Section 4.2 of \Sifon{}).
These clusters are listed in Table \ref{tab:clusters}.

\subsection{Optical and near-infrared}
\subsubsection{Ground-based optical}

\Sifon{} selected galaxies visually by color and brightness \citep[from optical \textit{gri} imaging by][]{2010ApJ...723.1523M} as targets for optical spectroscopy.
Their precise ($\Delta z/z \lesssim 0.005$) spectroscopic observations led to the identification of a few dozen galaxies per cluster via emission-line and/or absorption-line features; these objects form our sample of galaxies. 
Most galaxies in this catalog are red-sequence, absorption-line systems.
It so happens that all \Sifon{} emission-line galaxies in the two higher-redshift clusters (J0102 and J0546) also show \ion{Ca}{2}~(K,H) 3950~\AA{} absorption features. 
In Section~\ref{sec:selection}, we discuss possible biases and resulting effects from the optical selection.

\cite{2016MNRAS.461..248S} produced revised catalogs based on new cluster membership criteria inspired by simulation results and redshifts corrected to the heliocentric frame.
We use updated cluster properties in our analysis (e.g., dynamical mass and $R_{\rm 200c}$), but base our analysis on the original, larger spectroscopic catalog of cluster members in order to study a larger sample of star-forming galaxies and AGNs.
For spectral line stacking (Section~\ref{sec:spectral stacking}), we repeat our analysis with both catalogs and find that the results do not change.

\subsubsection{Hubble Space Telescope} \label{sec:hst}
We make use of \textit{HST}/ACS F625W, F775W, F850LP images of J0102 \citep[Program ID: 12755, PI: Hughes;][]{2013ApJ...770L..15Z} and \textit{HST}/ACS F606W and F814W imaging of J0546 \cite[Program ID: 12477, PI: High;][]{2015ApJS..216...27B}.
\textit{HST} images in the paper use all three bands (\textit{rgb}) for J0102 and F814W for J0546.
There is a small offset between the \Sifon{} and \textit{HST} astrometry, so we manually re-registered the former to align with the latter, and with \textit{Spitzer} images for galaxies outside \textit{HST} coverage.
The average corrections are about $\Delta\theta \sim 0\farcs{}7$, and maximum offsets were $\Delta\theta \sim1\farcs{}6$.
The difference between the imprecise positions used for long-slit spectroscopy (which were rounded to the nearest $0\fs{}1 \approx 1\farcs{}0$ in RA; \Sifon{}) and the high-resolution \textit{Hubble} and \textit{Spitzer} observations accounts for the astrometric shift.
We then compare the new positions to interferometric observations (ATCA mapping, see \textsection\ref{sec:atca}) to verify revised \Sifon{} catalog positions by matching with radio-loud AGN and other radio sources.

\subsubsection{Spitzer Space Telescope}\label{sec:spitzer}
For our analysis, we also include \textit{Spitzer}/IRAC 3.6 and 4.5 \micron{} observations to complement optical photometry and spectroscopy \citep{2013MNRAS.435.3469H}.
\textit{Spitzer} catalogs are produced using the \texttt{SExtractor} software \citep{1996A&AS..117..393B}.
Because rest-frame near-infrared (NIR) emission traces total stellar mass, the \textit{Spitzer} observations are useful for cross-matching detections at long wavelengths, and particularly helpful for crude estimation of photometric redshifts.
\textit{Spitzer} detections of optical-wavelength dropouts are excluded as high-redshift galaxies, i.e., contaminants in our study of star-forming cluster members (see Appendix~\ref{sec:DSFGs} for results on high-redshift submillimeter galaxies).

\cite{2013MNRAS.435.3469H} have estimated stellar mass ($M_{\ast}$) from 3.6~\micron{} photometry by employing a \cite{2003MNRAS.344.1000B} $\tau = 0.1$~Gyr burst model beginning at a formation time $z_f = 3$, and assuming a \cite{1955ApJ...121..161S} IMF (which we convert to a \citealt{2003PASP..115..763C} IMF by multiplying the final masses by $0.61$; see Figure~4 of \citealt{2014ARA&A..52..415M}). 
Uncertainties are estimated by computing $M_\ast$ for a range of formation redshifts from $z_f=2-5$ and measuring the resulting scatter (additional systematic errors from choice of IMF and star formation history are neglected). 
The same process is used for new detections found via ALMA spectral line emission.
We choose not to employ optical and NIR spectral energy distribution (SED) fitting because our optical wavelength observations do not sufficiently cover all cluster galaxies, potentially leading to a bias in our stellar mass estimates as a function of clustercentric distance.

\subsection{ATCA} \label{sec:atca}
We use Australia Telescope Compact Array (ATCA) continuum observations centered at 2.1~GHz as presented by \cite{2015ApJ...803...79L}.
Radio flux densities and uncertainties are measured using the \texttt{IMFIT} task in the Common Astronomy Software Applications package \citep[CASA;][]{2007ASPC..376..127M}.
Re-registered galaxy positions agree with bright radio sources to sub-pixel (1\arcsec{}) precision.

\subsection{Herschel PACS and SPIRE} \label{sec:herschel data}
\textit{Herschel Space Observatory} \citep{2010A&A...518L...1P} observations using the Photoconductor Array Camera and Spectrometer \citep[PACS;][]{2010A&A...518L...2P} and Spectral and Photometric Imaging Receiver \citep[SPIRE;][]{2010A&A...518L...3G} instruments were obtained for our clusters through Program ID \texttt{OT2\_abaker01\_2}( PI: Baker).
PACS 100 and 160~\micron{} images were produced using version~10.3.0 of HIPE \citep{2010ASPC..434..139O} and have diffraction-limited beamwidths of $7\farcs{}2$ and $11\farcs5$, respectively.
SPIRE 250, 350, and 500 \micron{} observations of our clusters were introduced in \cite{2015ApJ...803...79L}.

Point sources with SNR~$> 4.0$ are extracted and catalogued from each of the PACS maps using a matched-filter algorithm \cite[e.g.,][]{2003MNRAS.344..887S, 2015ApJ...803...79L}.
In Section~\ref{sec:FIR continuum results}, we show that the number of false positives is expected to be very low.
100~\micron{} and 160~\micron{} point source-subtracted maps are generated by subtracting model images, constructed from the point source catalogs, from the observed sky maps. 
The point source-subtracted maps contain residual positive signal due to incomplete subtraction, and we consider its effects on our stacking analysis in Section~\ref{sec:PACS stacking}.
Rms values in point source-subtracted maps range from $13-15~\rm\mu Jy~pixel^{-1}$ (100~\micron{}) and $8-12 ~\rm\mu Jy~pixel^{-1}$ (160~\micron{}).

\subsection{ALMA Band 6}

We have obtained new ALMA Cycle 2 observations (Program 2013.1.01358.S; PI: Baker) of the two highest redshift clusters in our sample, J0102 and J0546.
Band 6 observations were taken between 2015 January 3 and 2015 April 6.
Table~\ref{tab:ALMA observations} summarizes the observing frequencies and dates for these observations.
We show a continuum image of J0102, including the positions of cluster members detected via ALMA spectral lines, in Figure~\ref{fig:alma overview}.

\floattable
\begin{deluxetable*}{cr crcrcr ccc}
\tablewidth{0pt}
\tablecolumns{11}
\tablecaption{ALMA Band 6 Observations \label{tab:ALMA observations}}
\tablehead{
    Cluster && 
    \multicolumn{5}{c}{Observing frequency [GHz]} && 
    \multicolumn{3}{c}{Observations} 
    \\
    \cline{3-7} \cline{9-11} \vspace{-1.5em}
    \\
    (Phase center) && 
    Continuum &&
    \CO{} & &
    \CI{} && 
    Date & 
    Start time [UT] & 
    Duration [min]}

    \startdata
    ACT-CL J0102-4915                               && 254.6 && $244.7-248.3$ && $260.7-264.3$ && 2015 Jan \gm 3 & 21:54 & 51\\
    (01\h02\m55\s, -49\arcdeg15\arcmin14\arcsec)    &&       &&               &&               && 2015 Jan \gm 4 & 00:33 & 68\\
                                                    &&       &&               &&               && 2015 Jan 13    & 23:27 & 68\\
     \\
    ACT-CL J0546-5345                               && 230.7 && $221.3-224.9$ && $236.3-240.0$ && 2015 Jan 15    & 04:46 & 68\\
    (05\h46\m38\s, -53\arcdeg45\arcmin28\arcsec)    &&       &&               &&               && 2015 Jan 15    & 06:07 & 78\\
                                                    &&       &&               &&               && 2015 Apr \gm 4 & 23:25 & 79\\
                                                    &&       &&               &&               && 2015 Apr \gm 6 & 00:14 & 68\\
    \enddata
\end{deluxetable*}

\subsubsection{Observing strategy}
For J0102, 35 available antennas provided baseline lengths ranging from 42~m to 1082~m.
For J0546, around 30 antennas were consistently unflagged, providing baseline lengths ranging from 36~m to 859~m.
We requested two adjacent spectral windows (1.875~GHz usable bandwidth at 15.625~MHz resolution) in each of the lower and upper sidebands for each of the two clusters.
The spectral window pairs spanned 3.75~GHz and were centered at 246.6~GHz and 262.6~GHz for J0102, allowing the detection of redshifted \CO{} ($\nu_{\rm rest}=461.04$~GHz) and \CI{} ($\nu_{\rm rest}=492.16$~GHz) emission, respectively.
Similarly, spectral window pairs were centered on 223.2~GHz and 238.2~GHz in order to detect the same lines in J0546 cluster galaxies.
For J0102, we chose to image a $2.5\arcmin \times 2.0\arcmin$ rectangular mosaic angled $55\arcdeg$ east of north, covering 41 spectroscopically confirmed cluster members, one of which is the brightest cluster galaxy (BCG). 
The 150 mosaic pointings are arranged in a hexagonal grid to maximize sensitivity.
For J0546, we chose to image a $3.0\arcmin \times 1.7\arcmin$ rectangular mosaic, angled $35\arcdeg$ east of north, with 142 pointings in a hexagonal grid that covers 40 cluster galaxies (including the BCG).
Each pointing received $\sim 40$ seconds of integration time.
The mosaic centers and orientations were chosen to maximize the numbers of spectroscopically confirmed members they contained.

\subsubsection{Calibration and imaging}
We manually reduced the J0102 data using CASA version 4.3.1 under the guidance of ALMA data analysts/NRAO staff in Charlottesville, VA.
The J0546 data were automatically reduced by the ALMA pipeline, which used CASA version 4.2.2, although we tweaked some default settings for both clusters.\footnote{Standard calibration settings automatically flag eight channels at both ends of each 128-channel spectral window; because we separated spectral windows by 1.875 GHz, this flagging left 125 MHz-wide gaps at the centers of our combined spectral windows. 
To remedy the problem, we flagged four rather than eight edge channels and included the noisier edge-channel data.
In those channels, sensitivities are a factor of $\sim 1.2$ - $1.5$ worse than those in the rest of the cube.}
To calibrate the J0102 data, the quasars J0334-4008 and J2357-5311 were used for bandpass calibration, and the latter was used for gain calibration at $\sim 6-10$ minute intervals. 
Uranus was observed for flux calibration \citep{1993Icar..105..537G, 2013ApJS..204...19P}.
For the J0546 observations, the quasars J0538-4405, J0519-454, J0854+2006, and J1107-4449 were used for bandpass calibrations, J0549-5246 was used for gain calibration, and Ganymede and Callisto were used for flux calibration.
System temperatures for both datasets were $\sim 80$~K, except in the vicinity of narrow atmospheric features at 237.2~GHz, 239.1~GHz, 248.2~GHz, and 263.7~GHz, where they increased to $160-200$~K.

Continuum images and data cubes centered on the CO and [\ion{C}{1}] lines were produced using the standard deconvolution task in CASA, \texttt{clean}.
Both clusters were continuum imaged with natural weighting, which yielded a synthesized beam of $\sim 1.58\arcsec \times 0.98\arcsec$ ($1.85\arcsec \times 1.08\arcsec $) and a $1~\sigma$ continuum sensitivity of $0.11\ \rm mJy~ beam^{-1}$ ($ 0.09\ \rm mJy~ beam^{-1}$) for J0102 (J0546).
The center frequency of the J0102 (J0546) continuum map is $254.585$~GHz ($230.683$~GHz).
Continuum-subtracted spectral line cubes were also imaged with natural weighting and with 50 MHz channel widths. 
The synthesized beam size is $1.5 \arcsec \times 1.0\arcsec$ ($1.9\arcsec \times 1.1 \arcsec$) for J0102 (J0546), and the line cube has $1.5\ \rm mJy~ beam^{-1}$ ($1.0\ \rm mJy~ beam^{-1}$) $1~\sigma$ sensitivity per channel, apart from the narrow frequency intervals noted in the preceding paragraph.

\subsubsection{Astrometry}
Our interferometric source centroids are in agreement with optical and NIR imaging to within one \textit{Spitzer}/IRAC 3.6~\micron{} pixel ($0\farcs{}6$) for the limited number of objects detected in both images.
Most long-wavelength counterparts of optical/NIR sources are also found in the ATCA maps (pixel size~$=1\arcsec$).
If we only examine comparisons between high-resolution observations, i.e., ALMA \CO{} sources (pixel size~$=0\farcs{}15$) with \textit{HST} (pixel size~$=0\farcs{}04$) counterparts, we find offsets of $0\farcs{}34$ (six sources). 
When we compare ALMA continuum sources and their \textit{HST} counterparts, we measure a mean offset of $0\farcs{}26$ (six sources, three of which do not have CO counterparts). 
\cite{2017MNRAS.466..861D} report offsets of $\sim 0\farcs{}3 - 0\farcs{}6$ between deep \textit{HST} and ALMA imaging of the Hubble Ultra Deep Field.
We conclude that the astrometry in \textit{HST} observations is in agreement with that of our interferometric images.

\section{Results} \label{sec:results}

We find six ALMA \CO{} line detections corresponding to cluster galaxies, of which four are accompanied by \CI{} detections.
Three ALMA Band 6 continuum sources are matched with cluster members in J0546; no counterparts are found in J0102.
We also catalog 19 \textit{Herschel}/PACS counterparts to cluster members.
In Table~\ref{tab:measurements}, we summarize measurements at infrared and radio wavelengths for cluster galaxies with \textit{Herschel}/PACS detections.
\textit{Herschel} and ALMA continuum observations are fit to \citet[][hereafter K15]{2015ApJ...814....9K} infrared SED templates in order to estimate IR luminosity and identify AGN.
In Table~\ref{tab:properties}, we report astrophysical properties derived from NIR, FIR, submillimeter, and radio observations for cluster members with \textit{Herschel} detections.

In the following subsections, we present our strategies for finding long-wavelength sources and our methods of converting measurements (flux and flux density) into physical quantities (mass and luminosity).
Additionally, Appendices~\ref{sec:ALMA line detections}, \ref{sec:ALMA dust detections}, and \ref{sec:Herschel detections} present detailed descriptions of ALMA line and continuum and \textit{Herschel} sources, including information on radio-wavelength and morphological properties when available.
The six ALMA line sources are displayed in Figures~\ref{fig:J0102 CO detections} and \ref{fig:J0546 CO detections}.

\floattable
\begin{deluxetable*}{l rrr l rr rrr r}
	\tablecaption{Measurements of detected cluster members \label{tab:measurements}}
	\rotate
	\tablecolumns{11}
	\tablehead{
		\colhead{Object} &
		\colhead{RA} &
		\colhead{Dec} &
		\colhead{$z$} &
		\colhead{Spectral} &
		\colhead{$S_{100~\micron{}}$} &
		\colhead{$S_{160~\micron{}}$} &
		\colhead{$S_{{\rm CO}\,(4-3)} \Delta v$} &
		\colhead{$S_{{\rm [CI]}\,{}^3P_1-{}^3P_0} \Delta v$} &
		\colhead{$S_{\rm ALMA}$} &
		\colhead{$S_{\rm 2.1~GHz}$}
		\\
		\colhead{} &
		\colhead{[$\arcdeg$]} &
		\colhead{[$\arcdeg$]} &
		\colhead{} &
		\colhead{features\tablenotemark{a}} &
		\colhead{[mJy]} &
		\colhead{[mJy]} &
		\colhead{[Jy~km~s$^{-1}$]} &
		\colhead{[Jy~km~s$^{-1}$]} &
		\colhead{[mJy]} &
		\colhead{[$\mu$Jy]}
	}
	\startdata
	J010243.99-491744.4 & 15.68328 & -49.29565 & 0.87535 & AGN & $2.58 \pm 0.41$\gm & $3.77 \pm 0.45$\gm & \nodata\gm & \nodata\gm & \nodata\gm & $256 \pm 13$\gm \\
	J010247.68-491635.7 & 15.69867 & -49.27659 & 0.88977 & [\ion{O}{2}]; \ion{Ca}{2} & $1.22 \pm 0.19$\gm & $1.54 \pm 0.26$\gm & \nodata\gm & \nodata\gm & \nodata\gm & $119 \pm 21$\gm \\
	J010252.44-491531.2 & 15.71849 & -49.25867 & 0.86100 & \nodata & $0.87 \pm 0.16$\gm & $1.80 \pm 0.24$\gm & $0.97 \pm 0.26$\gm & \nodata\gm & $<0.36$\gm & $<27$\gm \\
	J010255.50-491416.2 & 15.73126 & -49.23782 & 0.87117 & [\ion{O}{2}]; \ion{Ca}{2} & $<0.78$\gm & $1.38 \pm 0.34$\tablenotemark{b} & $1.16 \pm 0.31$\gm & $<0.81$\gm & $<0.36$\gm & $<24$\gm \\
	J010255.67-491556.7 & 15.73198 & -49.26574 & 0.86780 & \nodata & $0.74 \pm 0.16$\gm & $1.17 \pm 0.23$\gm & $0.95 \pm 0.25$\gm & $0.58 \pm 0.11$\tablenotemark{c} & $<0.33$\gm & $<24$\gm \\
	J010302.99-491458.4 & 15.76246 & -49.24956 & 0.87433 & [\ion{O}{2}]; \ion{Ca}{2} & $0.64 \pm 0.18$\gm & $1.37 \pm 0.27$\gm & $<0.72$\gm & $<0.78$\gm & $<0.39$\gm & $<30$\gm \\
	J023540.10-512255.5 & 38.91708 & -51.38208 & 0.27350 & \ion{Ca}{2} & $1.04 \pm 0.21$\gm & $2.15 \pm 0.50$\tablenotemark{b} & \nodata\gm & \nodata\gm & \nodata\gm & $<33$\gm \\
	J023542.40-512101.0 & 38.92667 & -51.35028 & 0.27277 & \ion{Ca}{2} & $0.80 \pm 0.18$\gm & $1.82 \pm 0.34$\tablenotemark{b} & \nodata\gm & \nodata\gm & \nodata\gm & $<36$\gm \\
	J023547.60-512029.4 & 38.94833 & -51.34150 & 0.28608 & \ion{Ca}{2} & $1.04 \pm 0.18$\gm & $1.46 \pm 0.29$\tablenotemark{b} & \nodata\gm & \nodata\gm & \nodata\gm & $<33$\gm \\
	J023549.20-511920.6 & 38.95500 & -51.32239 & 0.27379 & \ion{Ca}{2} & $0.95 \pm 0.20$\gm & $1.89 \pm 0.41$\tablenotemark{b} & \nodata\gm & \nodata\gm & \nodata\gm & $<33$\gm \\
	J023557.20-511820.8 & 38.98833 & -51.30578 & 0.28160 & \ion{Ca}{2} & $2.21 \pm 0.50$\tablenotemark{b} & $2.89 \pm 0.71$\gm & \nodata\gm & \nodata\gm & \nodata\gm & $<36$\gm \\
	J043810.40-542008.1 & 69.54333 & -54.33558 & 0.41830 & [\ion{O}{2}]; \ion{Ca}{2} & $4.66 \pm 0.22$\gm & $6.76 \pm 0.52$\gm & \nodata\gm & \nodata\gm & \nodata\gm & $130 \pm 27$\gm \\
	J043824.40-541716.0 & 69.60167 & -54.28778 & 0.41748 & \ion{Ca}{2} & $1.24 \pm 0.26$\gm & $2.00 \pm 0.49$\tablenotemark{b} & \nodata\gm & \nodata\gm & \nodata\gm & $<39$\gm \\
	J054627.43-534433.6 & 86.61429 & -53.74267 & 1.05660 & \nodata & $4.24 \pm 0.17$\gm & $7.03 \pm 0.30$\gm & $1.37 \pm 0.35$\tablenotemark{c} & $0.72 \pm 0.24$\tablenotemark{d} & $1.21 \pm 0.18$\gm & $101 \pm 13$\gm \\
	J054635.39-534541.1 & 86.64745 & -53.76143 & 1.07116 & [\ion{O}{2}]; \ion{Ca}{2} & $0.76 \pm 0.16$\gm & $1.10 \pm 0.22$\gm & $<0.42$\gm & $<0.42$\gm & $<0.30$\gm & $<24$\gm \\
	J054636.61-534405.9 & 86.65252 & -53.73499 & 1.05774 & \ion{Ca}{2} & $0.90 \pm 0.16$\gm & $1.47 \pm 0.28$\gm & $<0.45$\gm & $<0.63$\gm & $<0.30$\gm & $<27$\gm \\
	J054638.87-534613.6 & 86.66196 & -53.77045 & 1.08130 & \nodata & $1.13 \pm 0.17$\gm & $1.70 \pm 0.27$\gm & $0.84 \pm 0.17$\tablenotemark{c} & $0.49 \pm 0.13$\tablenotemark{d} & $1.08 \pm 0.15$\gm & $51 \pm 11$\gm \\
	J054642.12-534543.9 & 86.67550 & -53.76220 & 1.05240 & [\ion{O}{2}]; \ion{Ca}{2} & $<0.69$\gm & $2.15 \pm 0.27$\gm & $<0.39$\gm & $<0.48$\gm & $<0.42$\gm & $<30$\gm \\
	J054644.15-534608.7 & 86.68397 & -53.76908 & 1.06590 & \nodata & $1.47 \pm 0.17$\gm & $2.18 \pm 0.28$\gm & $1.47 \pm 0.24$\tablenotemark{c} & $0.67 \pm 0.14$\gm & $0.81 \pm 0.11$\gm & $<24$\gm \\
	\enddata
	\tablecomments{Thresholds for detection are $3~\sigma$ for ALMA line sources and $4~\sigma$ for \textit{Herschel}, ALMA, and ATCA continuum sources. 
		For all non-detections, we report $3~\sigma$ upper limits.}
	\tablenotetext{a}{Optical spectral features reported by \Sifon{}.}
	\tablenotetext{b}{Measured from peak brightness because no detection was found by our matched-filter source extractor.}
	\tablenotetext{c}{Identified using SoFiA \citep{2015MNRAS.448.1922S}.}
	\tablenotetext{d}{Measured from peak brightness because \texttt{IMFIT} did not converge on a solution.}
\end{deluxetable*}

\floattable
\begin{deluxetable*}{lrlrrrrrrrrrr}
	\tablecaption{Derived properties of detected cluster members \label{tab:properties}}
	\rotate
	\tablecolumns{13}
	\tablehead{
		\colhead{Object} &
		\colhead{$r$} &
		\colhead{K15 IR} &
		\colhead{$\log L_{\rm IR}$} &
		\colhead{$\log L_{\rm IR}^{\rm SF}$} &
		\colhead{$\log M_\star$} &
		\colhead{sSFR} &
		\colhead{$\log M_{\rm mol}$} &
		\colhead{$f_{\rm gas}$} &
		\colhead{$\log \tau_{\rm dep}$} &
		\colhead{$\log M_{\rm ISM}$} &
		\colhead{$\log L_{\rm 1.4~GHz}$} &
		\colhead{$q_{\rm IR}$}
		\\
		\colhead{} &
		\colhead{[$R_{\rm 200c}$]} &
		\colhead{template} &
		\colhead{[$L_\sun$]} &
		\colhead{[$L_\sun$]} &
		\colhead{[$M_\sun$]} &
		\colhead{[yr$^{-1}$]} &
		\colhead{[$M_\sun$]} &
		\colhead{} &
		\colhead{[yr]} &
		\colhead{[$M_\sun$]} &
		\colhead{[W~Hz$^{-1}$]} &
		\colhead{}
	}
	\startdata
	J010243.99-491744.4 & 0.76 & \textsc{agn2} & $11.64 \pm 0.13$ & $11.42 \pm 0.07$ & $12.17 \pm 0.09$ & $-10.53 \pm 0.11$ & \nodata & \nodata & \nodata & \nodata & $24.34 \pm 0.02$ & $1.3 \pm 0.3$ \\
	J010247.68-491635.7 & 0.38 & \textsc{agn1} & $11.32 \pm 0.17$ & $10.96 \pm 0.12$ & $11.03 \pm 0.09$ & $-9.85 \pm 0.14$ & \nodata & \nodata & \nodata & \nodata & $24.03 \pm 0.08$ & $1.3 \pm 0.4$ \\
	J010252.44-491531.2 & 0.08 & \textsc{composite2} & $11.02 \pm 0.19$ & $10.92 \pm 0.06$ & $10.85 \pm 0.09$ & $-9.71 \pm 0.10$ & $10.4 \pm 0.3$ & $0.36$ & $9.3$ & $< 10.3$ & $< 23.35$ & $> 1.7$ \\
	J010255.50-491416.2 & 0.45 & \textsc{composite1} & $10.85 \pm 0.15$ & $10.82 \pm 0.14$ & $11.46 \pm 0.09$ & $-10.42 \pm 0.16$ & $10.5 \pm 0.3$ & $0.11$ & $9.5$ & $< 10.3$ & $< 23.31$ & $> 1.6$ \\
	J010255.67-491556.7 & 0.13 & \textsc{composite3} & $10.91 \pm 0.22$ & $10.72 \pm 0.10$ & $11.27 \pm 0.09$ & $-10.34 \pm 0.13$ & $10.4 \pm 0.3$ & $0.14$ & $9.5$ & $< 10.3$ & $< 23.30$ & $> 1.6$ \\
	J010302.99-491458.4 & 0.52 & \textsc{composite2} & $10.92 \pm 0.19$ & $10.82 \pm 0.09$ & $11.02 \pm 0.09$ & $-9.98 \pm 0.12$ & $< 10.3$ & $< 0.19$ & $< 9.3$ & $< 10.4$ & $< 23.41$ & $> 1.5$ \\
	J023540.10-512255.5 & 0.29 & \textsc{agn1} & $10.00 \pm 0.18$ & $9.64 \pm 0.16$ & $11.49 \pm 0.03$ & $-11.63 \pm 0.16$ & \nodata & \nodata & \nodata & \nodata & $< 22.11$ & $> 1.9$ \\
	J023542.40-512101.0 & 0.07 & \textsc{agn1} & $9.92 \pm 0.18$ & $9.56 \pm 0.15$ & $11.43 \pm 0.03$ & $-11.65 \pm 0.15$ & \nodata & \nodata & \nodata & \nodata & $< 22.15$ & $> 1.8$ \\
	J023547.60-512029.4 & 0.10 & \textsc{composite1} & $9.76 \pm 0.10$ & $9.72 \pm 0.06$ & $10.87 \pm 0.03$ & $-10.94 \pm 0.07$ & \nodata & \nodata & \nodata & \nodata & $< 22.16$ & $> 1.6$ \\
	J023549.20-511920.6 & 0.27 & \textsc{agn1} & $9.96 \pm 0.18$ & $9.61 \pm 0.16$ & $9.77 \pm 0.03$ & $-9.94 \pm 0.16$ & \nodata & \nodata & \nodata & \nodata & $< 22.11$ & $> 1.9$ \\
	J023557.20-511820.8 & 0.48 & \textsc{sfg1} & $10.00 \pm 0.15$ & $9.98 \pm 0.08$ & $11.15 \pm 0.03$ & $-10.95 \pm 0.08$ & \nodata & \nodata & \nodata & \nodata & $< 22.18$ & $> 1.8$ \\
	J043810.40-542008.1 & 0.23 & \textsc{sfg1} & $10.78 \pm 0.12$ & $10.76 \pm 0.02$ & $11.50 \pm 0.04$ & $-10.53 \pm 0.04$ & \nodata & \nodata & \nodata & \nodata & $23.17 \pm 0.09$ & $1.6 \pm 0.3$ \\
	J043824.40-541716.0 & 0.40 & \textsc{agn1} & $10.51 \pm 0.18$ & $10.15 \pm 0.16$ & $11.01 \pm 0.04$ & $-10.64 \pm 0.17$ & \nodata & \nodata & \nodata & \nodata & $< 22.65$ & $> 1.9$ \\
	J054627.43-534433.6 & 0.78 & \textsc{agn1} & $12.11 \pm 0.16$ & $11.75 \pm 0.03$ & $11.30 \pm 0.09$ & $-9.33 \pm 0.09$ & $10.7 \pm 0.3$ & $0.27$ & $8.8$ & $11.0 \pm 0.3$ & $24.17 \pm 0.06$ & $1.9 \pm 0.4$ \\
	J054635.39-534541.1 & 0.16 & \textsc{agn3} & $11.27 \pm 0.25$ & $10.85 \pm 0.18$ & $11.13 \pm 0.09$ & $-10.07 \pm 0.20$ & $< 10.2$ & $< 0.13$ & $< 9.2$ & $< 10.4$ & $< 23.57$ & $> 1.7$ \\
	J054636.61-534405.9 & 0.62 & \textsc{composite3} & $11.26 \pm 0.22$ & $11.06 \pm 0.09$ & $11.26 \pm 0.09$ & $-9.98 \pm 0.13$ & $< 10.3$ & $< 0.10$ & $< 9.0$ & $< 10.4$ & $< 23.60$ & $> 1.7$ \\
	J054638.87-534613.6 & 0.31 & \textsc{agn1} & $11.58 \pm 0.17$ & $11.22 \pm 0.10$ & $11.45 \pm 0.09$ & $-10.01 \pm 0.14$ & $10.5 \pm 0.3$ & $0.12$ & $9.1$ & $11.0 \pm 0.3$ & $23.91 \pm 0.10$ & $1.7 \pm 0.4$ \\
	J054642.12-534543.9 & 0.30 & \textsc{composite1} & $11.33 \pm 0.10$ & $11.30 \pm 0.07$ & $10.74 \pm 0.09$ & $-9.23 \pm 0.11$ & $< 10.2$ & $< 0.28$ & $< 8.7$ & $< 10.6$ & $< 23.64$ & $> 1.7$ \\
	J054644.15-534608.7 & 0.49 & \textsc{agn1} & $11.67 \pm 0.17$ & $11.31 \pm 0.08$ & $11.24 \pm 0.09$ & $-9.72 \pm 0.12$ & $10.8 \pm 0.3$ & $0.34$ & $9.2$ & $10.9 \pm 0.3$ & $< 23.56$ & $> 2.1$ \\
	\enddata
	\tablecomments{For all non-detections, we report $3~\sigma$ upper limits.}
\end{deluxetable*}

\subsection{Dust continuum sources}

The Rayleigh-Jeans tail of the modified blackbody emission can reliably trace cluster galaxy dust mass while also serving as a proxy for total (\ion{H}{1} + molecular) gas mass, assuming a constant dust-to-gas ratio \citep{2016ApJ...820...83S}.
We use a dust-to-gas scaling law (equation~8.26) from \cite{2013seg..book..491S} to convert the ALMA continuum flux density into a total gas mass, $M_{\rm ISM}$:
\begin{equation}\label{eq:ISM mass}
\frac{M_{\rm ISM}}{M_\odot} = 1.12 \times 10^{10} \left (\frac{S_\nu}{\rm mJy}\right ) \left (\frac{1}{1+z}\right )^{3+\beta}  \left (\frac{350~\micron{}}{\nu_{\rm obs}}\right )^{2 + \beta} \left (\frac{d_L}{\rm Gpc}\right )^{2},
\end{equation}
where $S_{\nu}$ is the ALMA continuum flux density, $\beta = 1.8$ parameterizes the dust absorption coefficient $\kappa \propto \nu^{-\beta}$, $\nu_{\rm obs}$ is the observed wavelength (i.e., the continuum map central frequency), and $d_L$ is the luminosity distance.
Equation~\ref{eq:ISM mass} is approximately equal to equation~(A14) in \cite{2016ApJ...820...83S} for galaxies at $z=1$ with dust temperature $T_{\rm d} = 20~$K (see their Figure~A1), assuming a Galactic $X_{\rm CO} = 2\times10^{20} \rm~cm^{-2}~(K~km~s^{-1})^{-1}$ conversion factor \citep{2013ARAA..51..207B}.\footnote{We neglect uncertainties in the CO-to-gas mass conversion factor -- although we note that adopting the \cite{1997ApJ...478..144S} value, $\alpha_{\rm CO} = 0.8~M_\sun~({\rm K~km~s^{-1}~pc^{2}})^{-1}$, results in five-fold smaller gas mass estimates. The same is true for computation of gas mass in Section~\ref{sec:ALMA line results}.}

We search for ALMA Band 6 continuum sources in the J0102 and J0546 fields by identifying $>4~\sigma$ pixels in the ALMA maps and checking for \textit{HST} or \textit{Spitzer} counterparts.
23 ALMA sources are found this way, of which three are cluster members and three are later determined to be contaminants near cluster galaxies on the sky.
These detections are described in Appendices~\ref{sec:ALMA line detections} and \ref{sec:ALMA dust detections}.
The remaining sources are not detected at optical wavelengths.
We use the CASA \texttt{IMFIT} task to measure flux densities from mosaicked ALMA continuum maps.
In Table~\ref{tab:measurements}, we list the dust flux densities or $3~\sigma$ upper limits for all cluster members that have PACS counterparts.
Additional ALMA continuum detections that are not associated with known cluster galaxies are described in Appendix~\ref{sec:DSFGs}.
These (sub)millimeter sources are described in Table~\ref{tab:alma continuum detections}, and some examples are shown in Figure~\ref{fig:DSFGs}.

\subsection{Millimeter line detections} \label{sec:ALMA line results}

We use the Source Finding Application package \citep[SoFiA;][]{2015MNRAS.448.1922S} to search for spectral line sources in the ALMA data cubes.
We use a SNR threshold of 5.0 and smooth over 3 channels and 3 spatial pixels with Gaussian kernels.
Detections without optical counterparts are disregarded.
Additionally, we roughly estimate the number of false detections by extracting ``sources'' from versions of each data cube that have been multiplied by $-1$.
The numbers of SoFiA detections found in true data cubes and in the negative versions (before assessment of counterparts) are shown in Table~\ref{tab:SoFiA fidelity}.
Given these results, we expect greater fidelity for the SoFiA CO line detections relative to the [\ion{C}{1}] sources, which are less likely to be real.
We indeed find that CO line detections are more likely to have short-wavelength counterparts.
None of the negative CO line cube detections align with any optical or NIR sources, and only one [\ion{C}{1}] negative line detection is within $0\farcs{}5$ of a faint \textit{HST} source in J0546.

\floattable
\begin{deluxetable}{lr cr cr}
\tablecaption{Numbers of SoFiA detections \label{tab:SoFiA fidelity}}
\tablehead{
	\colhead{Cluster + line} && 
	\colhead{cube} && 
	\colhead{$-1\times$~cube}
	\\
	\colhead{} &&
	\colhead{(true)} &&
	\colhead{(false)}
	}
\startdata
J0102 [\ion{C}{1}] && 36 && 35 \\
J0102 CO            && 17 && 11 \\
J0546 [\ion{C}{1}] && 12 && 12 \\
J0546 CO            &&  8 &&  4 \\
\enddata
\end{deluxetable}

For J0102, two CO and one [\ion{C}{1}] lines detected by SoFiA have optical counterparts; however, further assessment suggest that the two CO lines are spurious (Appendix~\ref{sec:ALMA line detections}).
We also manually search for ALMA spectral line counterparts of \Sifon{} cluster members and find that one [\ion{O}{2}] emitter has CO emission at its systemic velocity.
A third source is found by examining the ALMA spectra of bright optical sources with \textit{Herschel} detections.
In J0546, three SoFiA-detected CO lines have optical counterparts, one of which (J054638.87-534613.6) is also the lone optical and NIR counterpart of a [\ion{C}{1}] line detection.
No additional ALMA sources are found as counterparts to \Sifon{} cluster members or to other bright optical and \textit{Herschel} sources.
For the [\ion{C}{1}] SoFiA line detection in J0102, we search for and find CO emission at the same redshift and position. 
We similarly find positive [\ion{C}{1}] line emission for four out of the five CO-selected sources; for the final CO source, redshifted [\ion{C}{1}] emission would lie beyond frequency range of the spectral cube, so no [\ion{C}{1}] data are available.
All detections, including spurious sources, are described in detail in Appendix~\ref{sec:ALMA line detections}.

Once detected sources have been validated, they are examined to determine the spectral extent of their line emission, and moment zero (integrated flux) maps are produced using channels with spatially coherent emission.
We use \texttt{IMFIT} to measure flux and estimate uncertainties.
If \texttt{IMFIT} fails to converge on a solution, we use the source's peak flux and compute uncertainties using the standard deviation of spatially-adjacent pixels (off-galaxy) in the moment zero map.
The upper limit of a non-detection is computed by multiplying three times the rms in a moment zero map generated assuming a fiducial line width of $\Delta v = 300~\rm km~s^{-1}$ \cite[e.g., similar to CO line widths found in Virgo spirals, low-$z$ ULIRGs, and high-$z$ quasar hosts;][]{2003A&A...398..959H, 1997ApJ...478..144S, 2006AJ....131.2763C}.
We report CO and [\ion{C}{1}] flux measurements in Table~\ref{tab:measurements}.

We derive molecular gas masses, $M_{\rm mol}$, whenever \CO{} is detected using a Galactic $X_{\rm CO}$ factor and a \CO{}/CO $(1-0)$ brightness temperature ratio $r_{4,1} = 0.4$ \citep[see, e.g.,][]{2013ARA&A..51..105C, 2014MNRAS.437.1434T} to convert CO line flux into molecular gas mass \citep{2013ARAA..51..207B}:
\begin{equation} \label{eq:molecular gas mass}
\frac{M_{\rm mol}}{M_\sun} = 1.05 \times 10^{4} \left (\frac{I_{\rm CO~(1-0)}}{\rm Jy~km~s^{-1}}\right ) \left (\frac{d_L}{\rm Mpc}\right )^{2} \left (\frac{1}{1 + z}\right ).
\end{equation}
An additional 0.3~dex uncertainty as recommended by \cite{2013ARAA..51..207B} is included in all of this paper's calculations that depend on $M_{\rm mol}$.

Along with CO rotational transitions, the [\ion{C}{1}] line can also be used as a tracer of dense gas.
\cite{2000ApJ...537..644G} analyze local galaxies to derive a line luminosity ratio of $L'($\CI{}$)/L'(\rm CO(1-0)) = 0.2 \pm 0.2$.
\cite{2005A&A...429L..25W} find ratios between $0.15-0.32$ at $z\sim 2.5$, and \cite{2011ApJ...730...18W} find a ratio of $0.29 \pm 0.12$ in their sample of $z > 2$ SMGs and quasar host galaxies.
Using a fiducial ratio of  $L'($\CI{}$)/L'(\rm CO(1-0)) = 0.25$ and $r_{4,1} = 0.4$, we expect that $I_{\rm [CI]} = 0.71 \times I_{\rm CO(4-3)}$.
We can thus estimate a molecular gas mass by measuring the neutral atomic carbon line flux and using Equation~\ref{eq:molecular gas mass} as a conversion.
However, because the CO emission is generally stronger than [\ion{C}{1}], and traces a denser phase of the gas (molecular versus atomic), we use CO measurements to characterize the molecular gas reservoirs that fuel star formation.

The gas fraction, $f_{\rm gas} \equiv M_{\rm mol} / M_\star$, quantifies a galaxy's available gas for forming new stars relative to its existing stellar mass.
The gas depletion time, $\tau_{\rm dep} \equiv M_{\rm mol} / \textrm{SFR}$ (proportional to $L_{\rm CO}' / L_{\rm IR}$), is another metric of how much gas can fuel star formation at its current rate.
In Table~\ref{tab:properties}, we report molecular gas masses, gas fractions, and gas depletion times computed from CO line fluxes for J0102 and J0546 cluster members.

\subsection{Far-infrared continuum sources} \label{sec:FIR continuum results}

We catalog \textit{Herschel}/PACS SNR~$> 4.0$ sources using a matched-filter algorithm as described in Section~\ref{sec:herschel data}.
We cross-match the 100~\micron{} and 160~\micron{} catalogs with the \Sifon{} confirmed cluster members and new ALMA line sources using search radii of $3\farcs{}6$, which allows the inclusion of sources that may be been assigned incorrect centroid positions. 
For example, \cite{2011ApJ...737...83L} used a FWHM$~=11\arcsec$ matched filter to recover artificially injected sources and found $1-3\arcsec$ offsets between injected and recovered source centroids.
We remove contaminants such as strongly lensed PACS sources or neighboring sources of emission, which in some cases are blended with the true counterparts.
PACS sources that are closer in projection to other optical or NIR neighbors than to cluster members are considered contaminants.
For cluster members without matches, we search for $4~\sigma$ peaks relative to the local noise.
After rejecting contaminants, we find 14 PACS 100 and/or 160~\micron{} counterparts to \Sifon{} cluster members, and PACS counterparts to all five new ALMA CO line sources.
The majority of matched PACS counterparts are offset from optical centroids by less than $2\arcsec$, and none are offset by more than $3\arcsec$.
All 19 \textit{Herschel}/PACS detections and $3~\sigma$ upper limits are shown in Table~\ref{tab:measurements} (not including two contaminants, which we describe in the Appendix).

We also extract sources from a negative version of the PACS images in order to check the fidelity of our catalogs. 
If negative ``detections'' are purely due to noise, and the noise profile is centered around zero, then we can use the negative catalogs to quantify the likelihood of false positives.
However, we have found that the noise profile is not centered around zero, and in fact contains excess positive signal.
Whether or not the signal is astrophysical in nature, it will bias our fidelity estimate of the PACS catalogs; we therefore add uniform signal to the negative PACS image such that the mean value is equal to that in the original (positive) image.
We extract negative sources using the same matched-filter technique.
Only a single 4~$\sigma$ detection is found in the negative J0438 100~\micron{} map, and it is not near any known cluster members.
Based on the results of this test, the probability of contaminants appearing in our catalog is small (i.e., the number of contaminants is likely less than one).
The numbers of both positive and negative sources found using the matched-filter algorithm on the PACS maps are shown in Table~\ref{tab:matched filter fidelity}.

\floattable
\begin{deluxetable}{lr cccc}
	\tablecaption{Numbers of PACS detections \label{tab:matched filter fidelity}}
	\tablehead{
		\colhead{Cluster} && 
		\colhead{100~\micron{} map} &
		\colhead{$-1\times$~100~\micron{} map} &
		\colhead{160~\micron{} map} &
		\colhead{$-1 \times$~160~\micron{} map}
		\\
		\colhead{} &&
		\colhead{(true)} &
		\colhead{(false)} &
		\colhead{(true)} &
		\colhead{(false)}
	}
	\startdata
	J0102 && 62 & 0 & 46 & 0 \\
	J0235 && 88 & 0 & 44 & 0 \\
	J0438 && 49 & 1 & 24 & 0 \\
	J0546 && 65 & 0 & 54 & 0 \\
	\enddata
\end{deluxetable}

\cite{2010A&A...518L..29E} have shown that the total infrared luminosity in dusty galaxies can be derived from a single PACS band at intermediate redshifts ($z \sim 0.5-1$) using the \cite{2001ApJ...556..562C} library of template SEDs.
However, it has also been reported that the \cite{2001ApJ...556..562C} SED templates, which are primarily assembled from observations of local (U)LIRGs, underestimate the cold dust content for $z \sim 1-2$ star-forming galaxies \cite[see, e.g.,][]{2009ApJ...692..556R, 2012ApJ...759..139K}. 
Because the \cite{2001ApJ...556..562C} templates are constructed from a disproportionately large fraction of local major mergers, they generally suffer from a dearth of cold dust emission at $\lambda_{\rm rest} > 200~\micron{}$.
After fitting 160~\micron{} PACS flux densities to these redshifted template SEDs, we indeed find that the \cite{2001ApJ...556..562C} SED templates are inconsistent with observations at long wavelengths based on comparison of ALMA continuum measurements against SED predictions.\footnote{Note that we ported the \cite{2001ApJ...556..562C} IDL routine to Python, which uses a different interpolation routine for numerical integration. Computed $L_{\rm IR}$ values differ randomly by about $0.03\%$ ($1~\sigma$ scatter). The code can be found at \url{http://github.com/jwuphysics/chary_elbaz_python}.}

We also use the \Kirkpatrick{} comprehensive library of infrared SED templates, which are empirically constructed from observations of $0.3 < z < 2.8$ (U)LIRGs, a population that includes ordinary star-forming galaxies, AGNs, and merger-induced starbursts.
This library  contains star-forming galaxies (\textsc{sfg}), AGNs (\textsc{agn}) and combinations of the two (\textsc{composite}); each of these categories are further divided by luminosity and redshift.
To compare monochromatic fits to \cite{2001ApJ...556..562C} SEDs, we also fit 160~\micron{} flux densities using only the \textsc{sfg} subset of templates in the \Kirkpatrick{} library.
In both cases, 160~\micron{} flux density is a better observational constraint for fitting because it lies near the SED peak, and is detected for all 19 cluster members.
In order to find the best fitting template, $\chi^2$ is minimized while the overall SED normalization is allowed to vary as a free parameter.
We find that $L_{\rm IR}$ derived using \cite{2001ApJ...556..562C} templates is systematically 0.3~dex lower compared to $L_{\rm IR}$ computed from \Kirkpatrick{} star-forming galaxy templates.
Thus, we proceed with fitting SEDs using the comprehensive \Kirkpatrick{} library, using $\chi^2$ to discriminate among templates.
The best-fit template for each detection is reported in Table~\ref{tab:properties}.

Once the best-fit template is known, we use \texttt{emcee}, a Markov chain Monte Carlo (MCMC) code \citep{2013PASP..125..306F}, to sample the posterior probability distribution of the normalization free parameter. 
For each galaxy, we verify that the samples follow a normal distribution.
Uncertainties in $L_{\rm IR}$ are estimated by taking a standard deviation of the distribution. 

Finally, we follow \cite{2012ARA&A..50..531K} in deriving SFR from IR luminosity assuming a Chabrier initial mass function \citep{2003PASP..115..763C}:
\begin{equation} \label{eq:SFR}
\log\left (\frac{\rm SFR}{M_\sun~ {\rm yr}^{-1}}\right ) = \log\left (\frac{L_{\rm TIR}(3-1100~\micron{})}{\rm erg~s^{-1}}\right ) - 43.41
\end{equation}
In this paper, we approximate $L_{\rm TIR}(3-1100~\micron{}) \approx 1.1\times L_{\rm IR}(8-1000~\micron{})$ \cite[see, e.g., section 2.5 of][who find a $4-15\%$ difference between the two IR luminosity conventions]{2016MNRAS.457.2703R}.
In practice, we use the conversion SFR~$= 1.65~M_\sun {\rm~yr}^{-1} \times (L_{\rm IR} / 10^{10}~L_\sun)$.
This approach, although limited, is justified because we have no mid-IR observational constraints.

\subsection{Radio-bright sources}

The hot and warm dust components of AGN are bright in the PACS wavebands and can contaminate any study of star-forming cluster galaxies.
Out of all the \Sifon{} galaxies in our sample, only one (J010243.99-491744.4) was found to match an optical AGN spectral template.
However, it seems likely that our sample of clusters contains more AGNs than the one singled out by optical-wavelength observations.
We therefore search through the catalogs of $4~\sigma$ ATCA point sources to identify radio-loud AGN \citep{2015ApJ...803...79L}.
For counterparts to cluster members, we use \texttt{IMFIT} to measure 2.1~GHz flux density.
Contamination rates are expected to be very low based on the fact that we do not find any $4~\sigma$ negative detections near cluster galaxies.
ATCA flux densities or upper limits are shown in Table~\ref{tab:measurements}.

We also calculate the (rest-frame) 1.4~GHz luminosity by measuring $S_{\rm 2.1~GHz,\,obs}$ the observed 2.1~GHz flux density, and extrapolating to a rest frequency of 1.4~GHz by assuming $S_\nu \propto \nu^{-\alpha}$ with spectral index $\alpha = 0.8$:
\begin{equation}
L_{1.4\,\rm GHz} = (4 \pi d_L^2) \times S_{2.1\,\rm GHz,\,obs} \left(\frac{1.4 \,\rm GHz}{2.1\,\rm GHz}\right)^{-\alpha} (1+z)^{\alpha},
\end{equation}
where $d_L$ is the luminosity distance.
We estimate the AGN fraction in our sample based on the FIR-radio correlation, parameterized by $q_{\rm IR}$, in Section~\ref{sec:q parameter}.
Both $L_{\rm 1.4\,GHz}$ and $q_{\rm IR}$ are reported in Table~\ref{tab:properties}.

\section{Stacking analysis} \label{sec:stacking}

\subsection{\textit{Herschel}/PACS continuum stacking} \label{sec:PACS stacking}

\begin{figure*}
\plotone{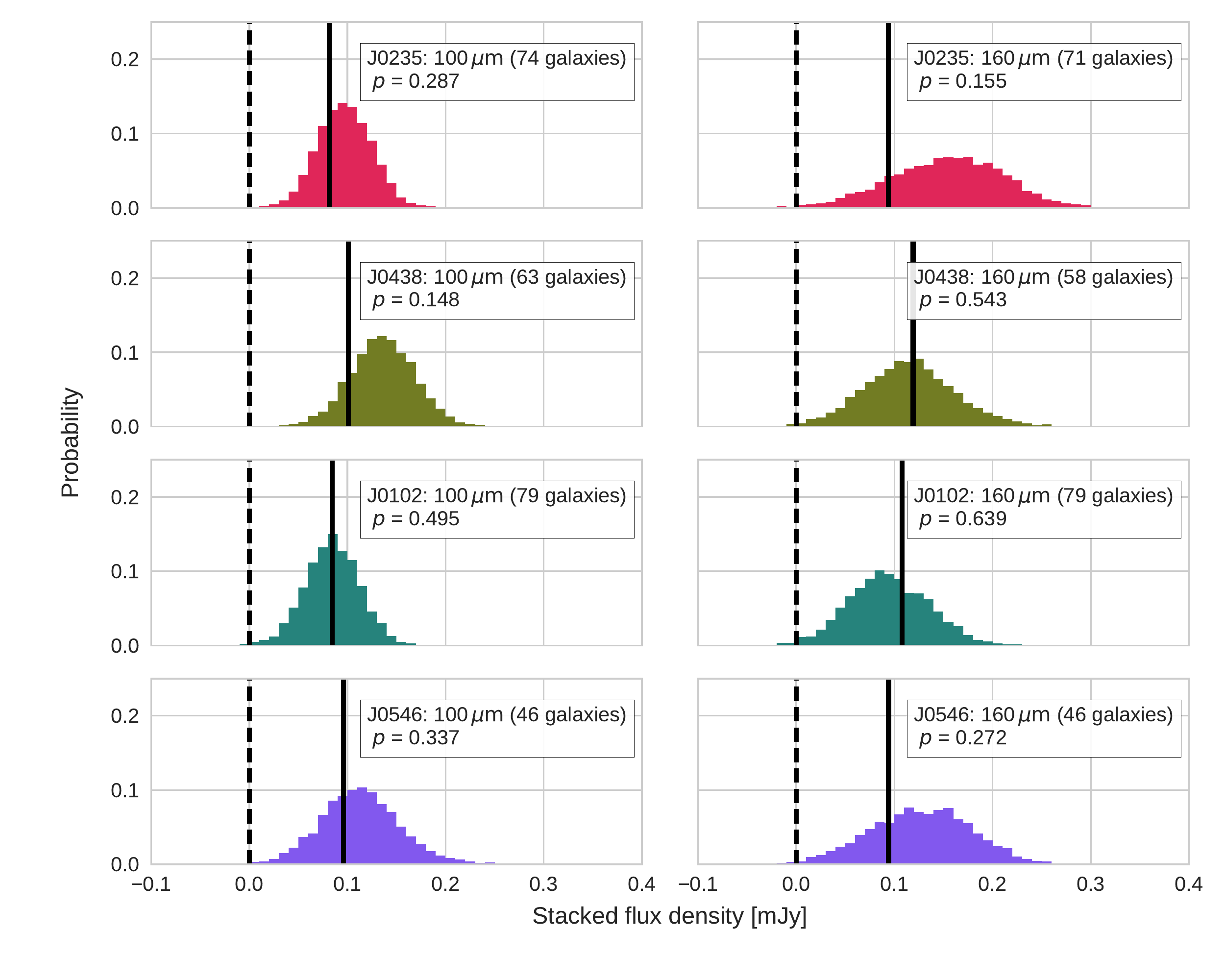}
\caption{Histograms of bootstrapped stacked \textit{Herschel}/PACS 100~\micron{} and 160~\micron{} flux densities for 5000 iterations. Outlying fluxes in excess of 4~MADs have been removed. Flux densities are taken from the ``residual'' point source-subtracted PACS maps. The black dashed vertical line signifies zero flux, and the solid black line indicates the mean flux of blank-sky pixels. We also report the number of galaxies per stack and $p$-values determined by computing the percentile of the blank-sky mean within the bootstrapped distribution of stacked flux densities.}
\label{fig:pacs stack}
\end{figure*}

By stacking a collection of galaxies, we trade off information about individual properties for more sensitive \textit{average} measurements of otherwise undetected galaxies.
We stack, or co-add, flux densities of non-detections in order to reduce noise, which should decrease as $N_{\rm gal}^{-1/2}$ barring correlations between pixels.
For 271 out of 285 cluster galaxies in our optically selected sample, no PACS emission was individually detected.
We use 100~\micron{} and 160~\micron{} PACS ``residual'' point source-subtracted maps to extract fluxes from pixels centered at the locations of these undetected cluster members.
Outliers have been removed by using a cut of 4~MADs, where the median absolute deviation (MAD) of a dataset, $\vec x$, is defined 
\begin{equation}
{\textrm{MAD} \equiv \textrm{median}( | \vec x - \textrm{median}(\vec x)| )}.
\end{equation}
We use bootstrapping (with replacement) to calculate stacking uncertainties or confidence intervals.
Histograms of PACS 100~\micron{} and 160~\micron{} stacks (with 5000 resampling iterations) are shown in Figure~\ref{fig:pacs stack} in color.

Separate stacks of pixels are also selected at random from the ``blank-sky'' PACS images, which are the point source-subtracted maps with (1) all \textit{Spitzer} detections masked, in order to remove FIR contaminants, and (2) all regions beyond the projected radius of the furthest cluster galaxy masked, to exclude noisy regions in the maps.
We then measure the mean blank-sky flux density after removing outliers using the 4~MAD cut.
In Figure~\ref{fig:pacs stack}, bootstrapped distributions of the stacked galaxy flux density (colored histograms) are compared to the mean blank-sky flux densities (vertical black lines).

We see that, although the stacked galaxy signal is inconsistent with \textit{zero} at the $2-3~\sigma$ level, it is not significantly greater than the \textit{blank-sky} mean flux density.
In each panel of Figure~\ref{fig:pacs stack}, we also report the $p$-value of the stack relative to the mean blank-sky value; the $p$-value is simply the (fractional) percentile rank of the blank-sky flux compared to the bootstrapped distribution of stacked flux.
In every case, $p > 0.1$, signifying that the stacked fluxes do not statistically exceed the blank-sky values in the point source-subtracted PACS maps.
Excluding the BCG from each stack does not significantly change our results.
Additionally, stacking only [\ion{O}{2}] emitters undetected by \textit{Herschel} yields non-detections at both 100~\micron{} and 160~\micron{} for each cluster.
Table~\ref{tab:PACS stacking limits} shows the PACS $3~\sigma$ upper limits for non-detection along with $L_{\rm IR}$ computed using a monochromatic fit to the \textsc{sfg1} star-forming galaxy \Kirkpatrick{} template.
We find that the $4~\sigma$ flux limits correspond to luminosities below $\log (L_{\rm IR} / L_\sun) < 10.6$, which we adopt as the luminosity completeness limit.

\floattable
\begin{deluxetable}{l r r r r}
\tablecaption{PACS stacking $3~\sigma$ upper limits \label{tab:PACS stacking limits}}
\tablewidth{0pt}
\tablecolumns{3}
\tablehead{
	\colhead{Cluster} &
	\colhead{$S_{\rm 100~\micron{}}$} &
	\colhead{$\log L_{\rm IR}^{\rm 100~\mu m}$} &
	\colhead{$S_{\rm 160~\micron{}}$} &
	\colhead{$\log L_{\rm IR}^{\rm 160~\mu m}$} 
	\\
	&
	\colhead{[mJy]} &
	\colhead{[$L_\sun$]} & 
	\colhead{[mJy]} &
	\colhead{[$L_\sun$]}
	}
	
	\startdata
	J0235 & $<0.08$ &  $<8.6$ & $<0.18$ & $<8.8$ \\
	J0438 & $<0.10$ &  $<9.1$ & $<0.15$ & $<9.1$ \\
	J0102 & $<0.09$ &  $<10.0$ & $<0.12$ & $<9.9$ \\
	J0546 & $<0.12$ & $<10.5$ & $<0.16$ & $<10.2$ \\
	\enddata
\end{deluxetable}

We also divide galaxies into bins of clustercentric radius, where the radii are normalized by $R_{\rm 200c}$ and centered on the BCG (or, for J0102, the midpoint of the two merging mass distributions).
For each radial bin, we compute the bootstrapped distribution of stacked galaxy flux densities and compare to the blank-sky flux densities at each radius.
Although we are stacking on smaller samples of galaxies than before, the goal is to examine star formation trends over a range in galaxy density (which decreases with distance from the cluster center).
We still do not find any significant differences between the clustercentric stacks and the positive excess blank-sky flux densities.

\subsection{ALMA continuum stacking}

\begin{figure*}
\plottwo{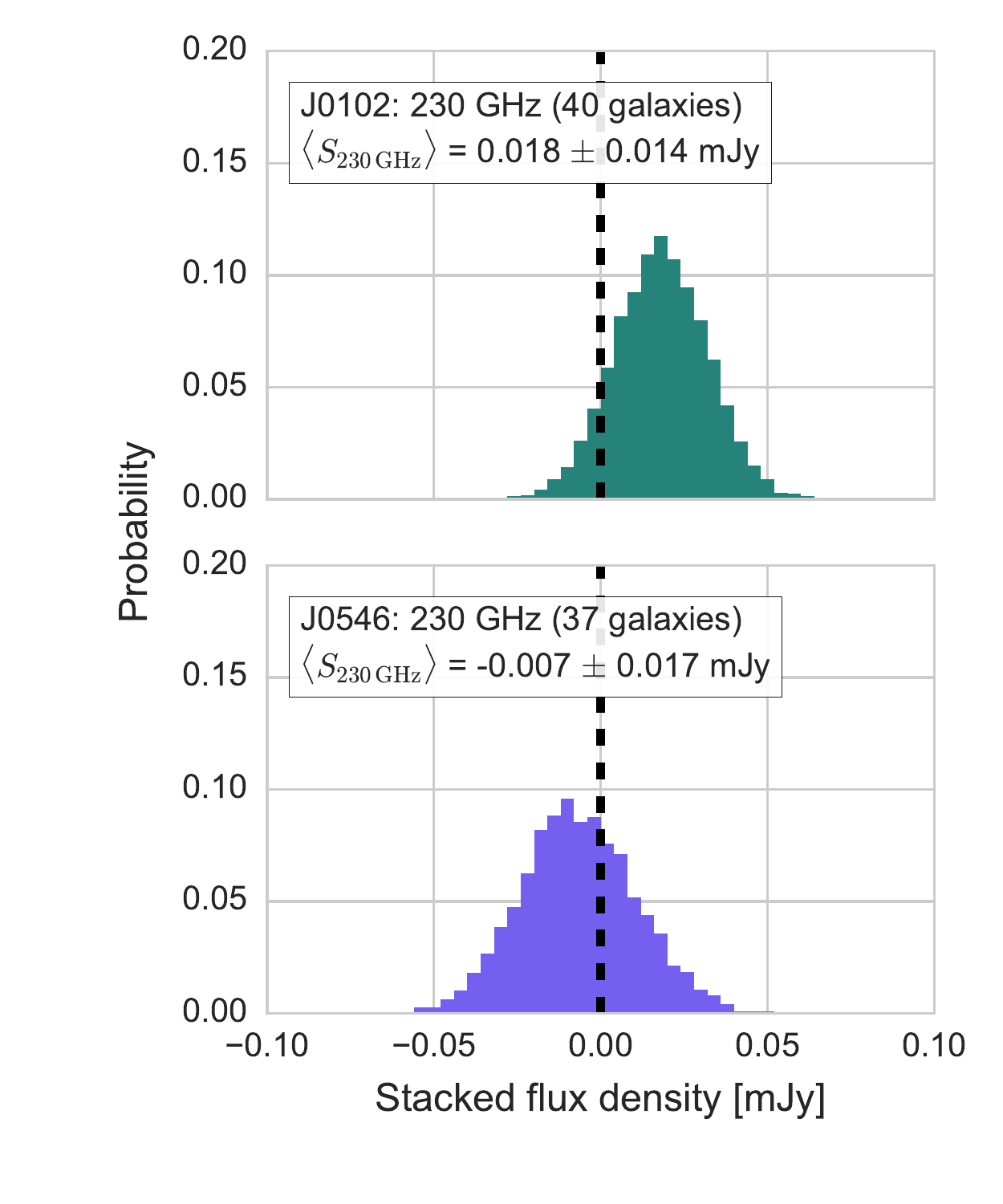}{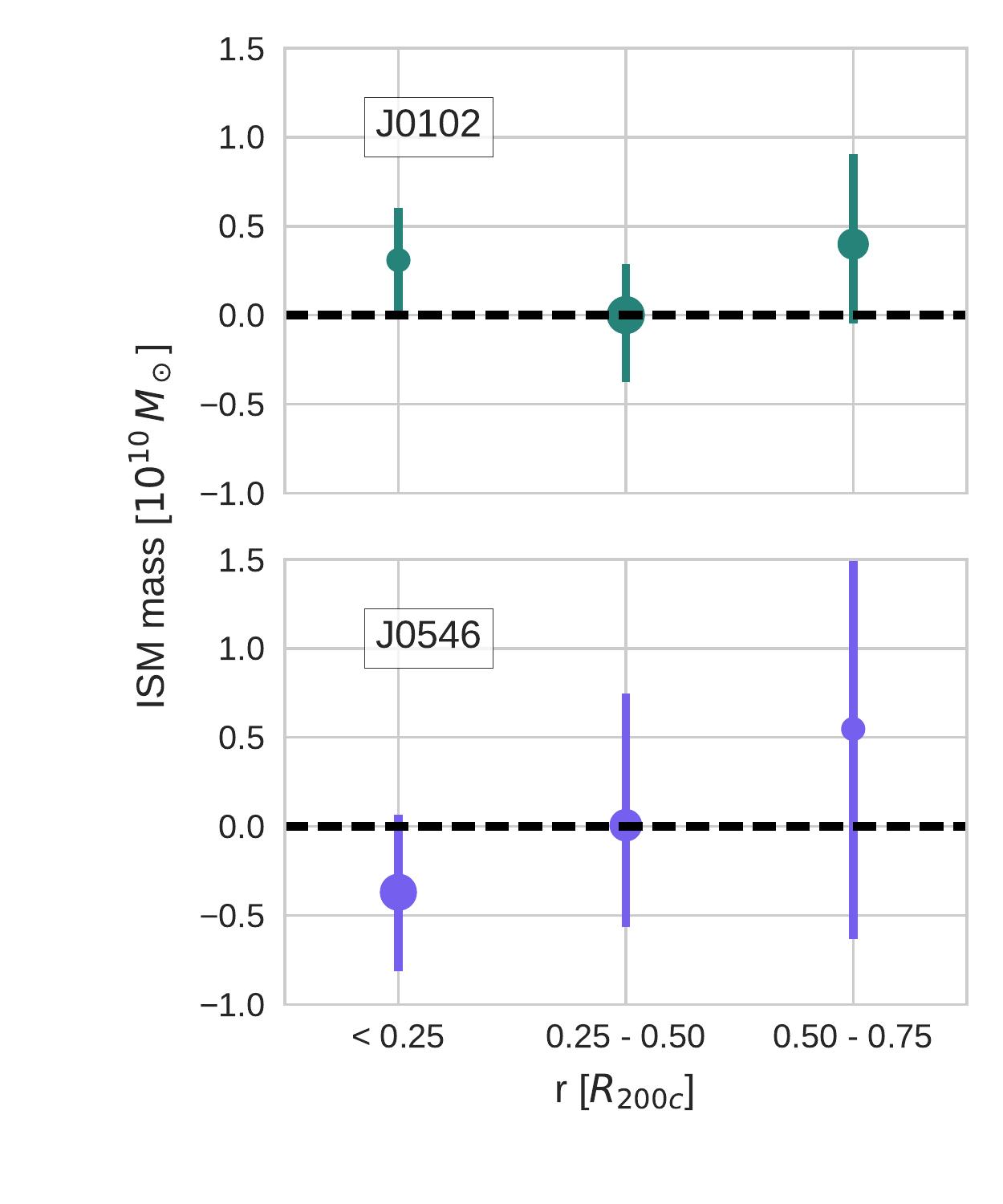}
\caption{(\textit{Left}) Bootstrapped histograms of stacked ALMA dust continuum flux densities for galaxies in our two high-$z$ clusters. 
(\textit{Right}) Continuum fluxes converted to $M_{\rm ISM}$ values and stacked in bins of clustercentric radius. 95\% confidence intervals (computed by bootstrap resampling for 5000 iterations) are shown for each cluster. A dashed line marks zero. In both cases, 4-MAD outliers were removed.}
\label{fig:alma continuum stack}
\end{figure*}

We stack the ALMA continuum flux densities of J0102 and J0546 cluster galaxies, and display them in the left panel of Figure~\ref{fig:alma continuum stack}.
Bootstrapping results show that the stacked flux is consistent with zero.
Blank-sky flux densities are not displayed in the figure because the mean values are very near zero. 
ISM masses derived from continuum fluxes binned by clustercentric radius as in Section 4.1 (right panel of Figure~\ref{fig:alma continuum stack}) are also consistent with zero at the $3~\sigma$ level.

Our $3\sigma$ limits on ISM mass are $\log (M_{\rm ISM}/M_\sun) < 10.0$ for J0102 and $\log (M_{\rm ISM}/M_\sun) < 9.9$ for J0546.
Intriguingly, the J0102 stacks in two radial bins are positive at the $\sim 2~\sigma$ level: we measure $\log (M_{\rm ISM}/M_\sun) = 9.5$ ($p = 0.022$) for galaxies at $r < 0.25~R_{\rm 200c}$ and $\log(M_{\rm ISM}/M_\sun) = 9.6$ ($p=0.036$) for galaxies at $r > 0.5~R_{\rm 200c}$.
We calculate $p$-values using bootstrapping.

\subsection{Radio continuum stacking}

We stack the observed 2.1~GHz flux densities in our ATCA maps, but bootstrapping reveals that the mean fluxes are consistent with zero.
We also consider stacks in bins of clustercentric distance, but again find no evidence for non-zero stacks at any radius.

\subsection{ALMA spectral stacking} \label{sec:spectral stacking}

Spectral line stacking uses positional and velocity information for confirmed (cluster) galaxies to measure their averaged spectral line properties \citep[see, e.g.,][]{2001A&A...372..768C}.
From the ALMA data cubes, we extract individual spectra positionally centered on confirmed cluster members (not including ALMA detections) and spectrally centered on their \CI{} and \CO{} lines.
We shift the spectra to a common velocity frame and co-add them in order to produce stacked spectra for each cluster and line.
Outliers are removed at each channel using a 4~MAD cut. 
Because we include in the analysis galaxies that are near the frequency edges of the data cubes, with spectra that are incomplete over the velocity range of the stack, different velocity channels in the stacked spectrum have different sensitivities (scaling as $\sigma_{\rm channel~rms} = N_{\rm channel}^{-1/2} \sigma_{\rm cube~rms} $).
The number of channels going into each stacked spectrum lies in the range $23-31$ ($27-29$) in the J0102 (J0546) [\ion{C}{1}] data cube, and in the range $28-34$ ($28-29$) in the J0102 (J0546) CO cube.

\begin{figure*}
\plotone{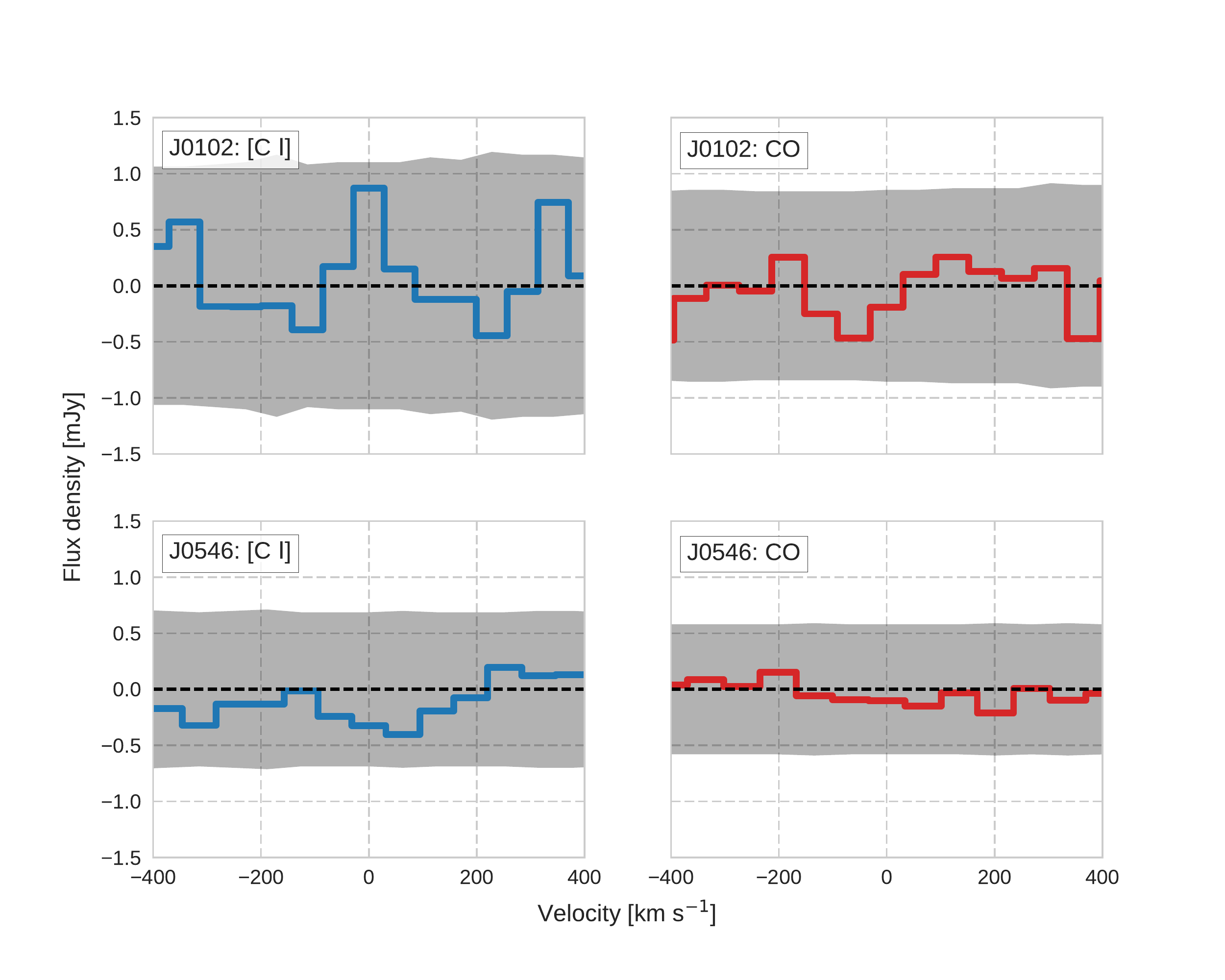}
\caption{[\ion{C}{1}] (blue) and CO (red) stacked spectra from the J0102 and J0546 ALMA data cubes.
Grey regions show $3~\sigma$ limits (centered around zero). 
All outlying flux densities in excess of 4 MADs were removed from calculation of stacked fluxes and uncertainties.}
\label{fig:alma spectral stack}
\end{figure*}

The stacked spectra and $3~\sigma$ limits are shown in Figure~\ref{fig:alma spectral stack}.
Stacked \CI{} and \CO{} line emission are consistent with zero flux in both clusters.
We stack galaxies in the \cite{2016MNRAS.461..248S} catalogs, which are presented in the heliocentric frame, and estimate $\pm 3~\sigma$ confidence intervals based on the sensitivities at the corresponding positions and frequencies. (Stacks based on the earlier \Sifon{} catalog also yield non-detections.)
It is possible that we excluded legitimate line detections by excising 4~MAD outliers, but visual inspection reveals that all outlier fluxes were due to large primary-beam corrections (near the mosaicked image edges) or random noise.

CO non-detections offer more stringent constraints on the total gas mass, so we start from the CO channel rms $\sigma_{\rm channel~rms}$ to calculate upper limits on $M_{\rm mol}$.
We assume a conservative fiducial line width, $\Delta v = 300~{\rm km~s^{-1}}$, and use $r_{4,1} = 0.4$ as before.
We find $\log (M_{\rm mol}/M_\sun) < 9.9$ for average galaxies in both J0102 and J0546.
We also stack the CO and [\ion{C}{1}] non-detections together, assuming a constant flux ratio of $I_{{\rm [CI]}~(^3 P_1 - ^3P_0)} / I_{\rm CO~(4-3)} = 0.71$ (see Section~\ref{sec:ALMA line results}), to find upper limits: $\log (M_{\rm mol}/M_\sun) < 9.8$ in both J0102 and J0546.

\section{Discussion} \label{sec:discussion}

\subsection{Atomic carbon abundance} \label{sec:carbon abundance}

\begin{figure}
\plotone{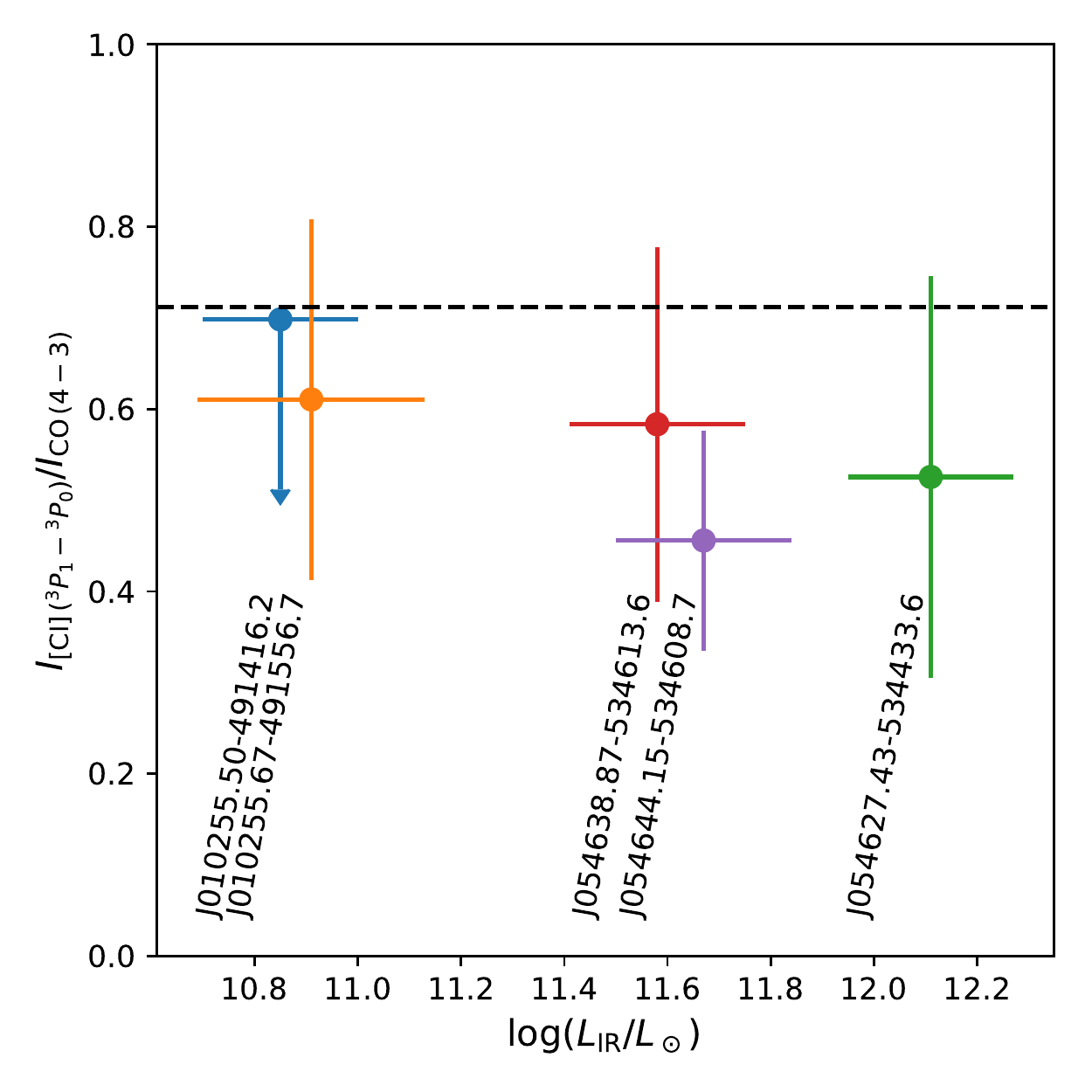}
\caption{The ratio of [\ion{C}{1}] and CO in integrated flux units plotted vs. IR luminosity ($1~\sigma$ error bars shown). 
The fiducial literature-based flux ratio of 0.71 adopted in this paper is shown as a dashed line.}
\label{fig:CI abundance}
\end{figure}

Among our six CO detections, we are able to measure four [\ion{C}{1}] line fluxes. We compare integrated line fluxes in Figure~\ref{fig:CI abundance}. 
We find a mean (median) flux density ratio of $I_{\rm [CI]}/I_{\rm CO~(4-3)} = 0.54 \pm 0.20$ ($0.55$), and an upper limit of $0.70$ in one galaxy, which roughly agree with the fiducial [\ion{C}{1}]-to-\CO{} ratio of $0.71$ used in Section~\ref{sec:results}.
In temperature units, the mean line luminosity ratio is $L'_{[{\rm CI}]~({}^3P_1 - {}^3P_0)}/L'_{\rm CO(4-3)} = 0.48 \pm 0.17$. 
Converted to a \CI{}/CO~$(1-0)$ ratio (again using $r_{4,1}=0.4$), we find a mean $L'_{[{\rm CI}]~({}^3P_1 - {}^3P_0)}/L'_{\rm CO(1-0)} = 0.19 \pm 0.07$. 
Our results agree with the line ratios found in samples of nearby \citep[i.e., $0.2 \pm 0.2$;][]{2000ApJ...537..644G} and high-$z$ \citep[i.e., $0.29 \pm 0.12$;][]{2011ApJ...730...18W} field galaxies, indicating a consistent carbon abundance and similar excitation state in our star-forming cluster members and AGNs.

\subsection{Gas mass, infrared luminosity, and gas depletion timescale}
\explain{In this section we now talk about gas masses, IR luminosities, and gas depletion timescale rather than only gas mass/ISM mass/IR luminosity}

\begin{figure}
\plotone{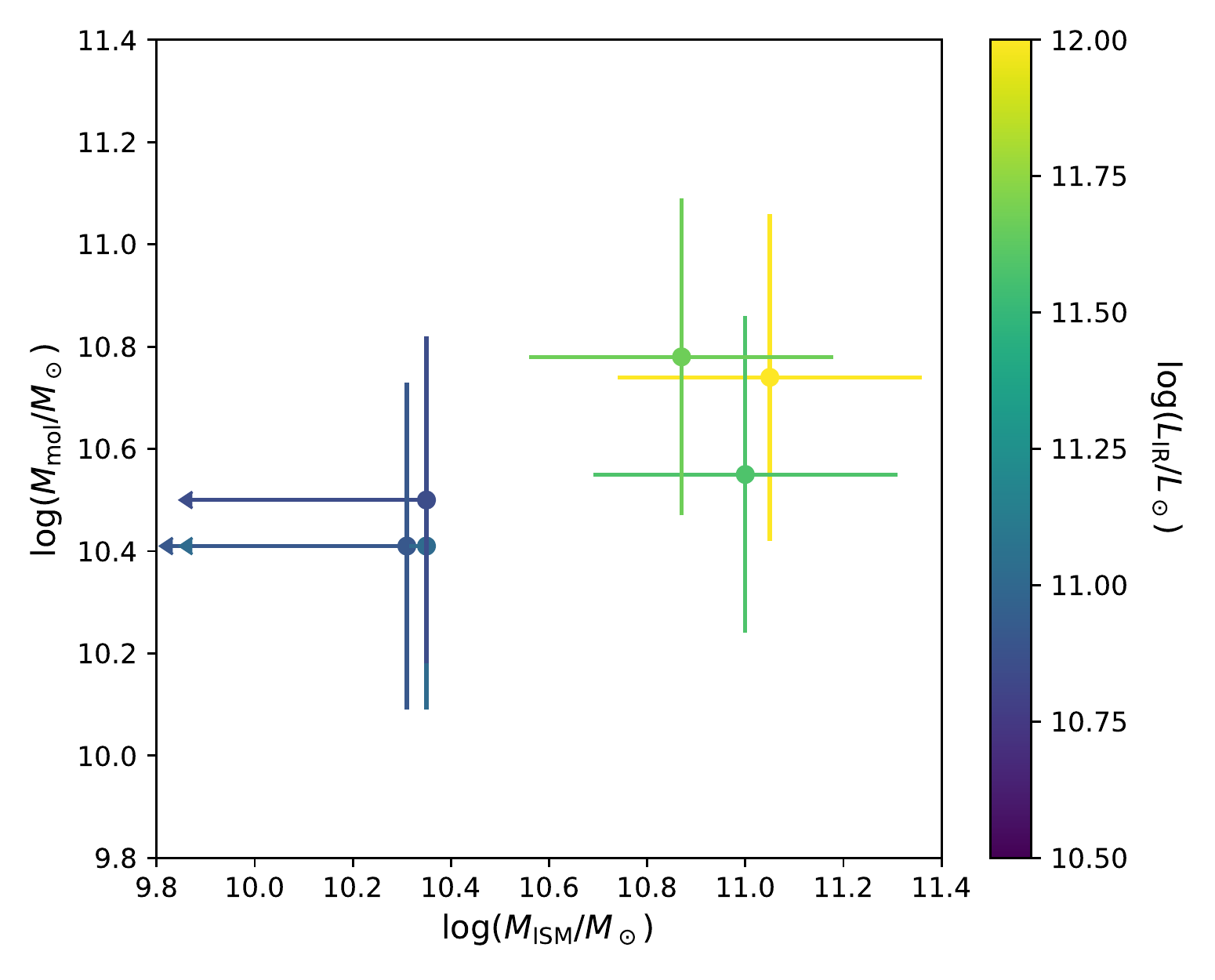}
\caption{
	ISM mass (vertical) vs. molecular mass (horizontal) vs. IR luminosity (color). 
Error bars denote $1~\sigma$, and leftward-pointing arrows indicate $3~\sigma$ upper limits on $M_{\rm ISM}$.}
\label{fig:gas, dust, and IR luminosity}
\end{figure}

We plot all ALMA detections in Figure~\ref{fig:gas, dust, and IR luminosity} along with their IR luminosities.
For the three galaxies that have both ALMA CO and continuum detections, all of which belong to J0546, we have used conversion factors to estimate molecular gas mass and ISM (dust/molecular + atomic gas) mass, respectively \citep{2013ARAA..51..207B, 2013seg..book..491S}.
The mean $M_{\rm mol}/M_{\rm ISM}$ ratio in J0546 is $0.55^{+0.58}_{-0.34}$.
For the three CO sources in J0102 , none of which have matching dust continuum detections, we find a nominal $M_{\rm mol}/M_{\rm ISM} > 1.33$ based on $3~\sigma$ ISM mass upper limits.
For comparison, the ratio of molecular gas mass to total ISM mass predicted by simulations is $\sim 2/3$ for massive halos ($M_{\rm vir} \geq 10^{12}~M_\odot$) at $z = 1$ \citep[i.e.,][]{2015MNRAS.449..477P}.

Disagreement between the mass ratios in the two clusters is possibly a result of small sample size, as \cite{2016ApJ...820...83S} find that this ratio is constant over a wide range in mass and redshift. 
However, it is also plausible that physical differences between the ISM of galaxies in J0102 and J0546 may impact the gas-to-dust ratio. 
Notably, J0102 is actively undergoing a merger, which may affect its galaxy members by increasing their CO excitation and thus appearing to elevate $M_{\rm mol}$ (as calculated from a mid-$J$ CO line flux) relative to $M_{\rm ISM}$ at an unphysical level.
A higher ratio of $r_{4,1} \sim 0.8$, as is seen in local (U)LIRGs \citep[e.g., M~82;][]{2000A&A...358..433M,2005A&A...438..533W,2012MNRAS.426.2601P} or in QSOs \citep{2013ARA&A..51..105C}, is able to resolve the conflict.
Analyses of these ULIRGs have found a \textit{lower} $X_{\rm CO}$ conversion factor than what we have used \citep{1998ApJ...507..615D}, implying less massive gas reservoirs in ULIRG-like objects.
Depending on the conversion factor used, gas masses may be lowered by a factor of $2 - $a few \citep[see, e.g.,][]{2013ARAA..51..207B, 2017arXiv170909679V}.
The ratio of $M_{\rm mol}/M_{\rm ISM}$ will not be further affected by choice of $X_{\rm CO}$, since both mass estimates depend on it.

CO-detected galaxies in J0102 have star formation rates nearly an order of magnitude lower than those in J0546. 
As a result, we find longer $\tau_{\rm dep}$ for J0102 galaxies in which molecular gas is measured.
The average depletion time for our sample is $\tau_{\rm dep} \sim 2~$Gyr.
However, for one CO non-detection in J0546, we find $\tau_{\rm dep}$ as low as $0.5~$Gyr.
Our CO sources are characterized by longer depletion timescales than normal star-forming field galaxies at $z \sim 1$ \citep[e.g.,][]{2013ARA&A..51..105C,2015ApJ...800...20G}, for which $\tau_{\rm dep} \sim 0.1 - 1$~Gyr.
We compare our sources to two CO detections of luminous infrared galaxies in the outskirts of a rich $z = 0.4$ cluster \citep{2009MNRAS.395L..62G, 2011ApJ...730L..19G}.
They find gas depletion timescales of $\sim 300 - 900$~Myr (calculated using a Galactic CO-to-H$_2$ conversion factor), which are on par with only the most efficiently star-forming members of J0546.
Similar gas depletion timescales are found in three gas-rich galaxies within the virial radius of two $z \sim 0.5$ clusters \cite[with $M_{\rm 200c} \sim 4 \times 10^{14}~M_\sun$;][]{2013A&A...557A.103J}, and in a $z \sim 1$ massive ($\gtrsim 10^{14}~M_\sun$) $z \sim 1$ cluster \citep{2012ApJ...752...91W}.
At higher redshifts ($z \sim 1.6$), however, \cite{2017ApJ...842L..21N} and \cite{2017arXiv170906963R} have found longer $\tau_{\rm dep} \sim 0.7-3.0$~Gyr in cluster galaxies, which agree well with the timescales in our two massive clusters.

\subsection{Substructure: planes of infrared-bright galaxies?}
The $4~ \sigma$ detection threshold in our PACS maps corresponds to galaxy luminosities of $\log (L_{\rm IR}/L_\sun) \approx 9.5$ for our two lower-$z$ clusters and to $\log(L_{\rm IR}/L_\sun) \approx 10.6$ for our two higher-$z$ clusters.
For convenience, we define an infrared-bright galaxy (IRBG) as one with IR luminosity in excess of $\log(L_{\rm IR}/L_\sun) = 10.6$ in order to facilitate a complete census of PACS sources across all clusters.
In both J0102 and J0546, IRBGs appear to lie along planes comprising five and six galaxies, with those in J0102 appearing to lie along a plane perpendicular to its merging axis.
Although the planes look significant by eye, they turn out to be fairly common occurrences by chance selection of galaxies.
For each cluster, we estimate significance by (1) summing distances (i.e., computed residuals) from the IR-bright galaxies to a best-fit plane, and (2) selecting similar subsamples of random cluster members and calculating residuals to their respective best-fit planes.
Our planes of IRBGs were less significant than randomly-chosen cluster members the majority of the time ($p=0.88$ for J0102 and $p = 0.63$ for J0546).
Therefore, chance alignments in our higher-$z$ clusters are likely to form planes of IRBGs (or in other words, a handful of galaxies in a cluster with $50 - 90$ members can easily trace a statistically insignificant ``plane'' configuration).

\subsection{Biases from optical selection of cluster members}\label{sec:selection}

The original targets for the optical spectroscopy we have used to identify cluster members were visually selected, on the basis of their \textit{gri} colors and brightnesses, to favor redder, brighter galaxies \Sifonp{}.
Thus, it is no surprise that cluster members with [\ion{O}{2}] emission lines comprise only a small fraction of the original sample ($ 17 / 285 \approx 6\%$).
For comparison, the Gemini Observations of Galaxies in Rich Early Environments survey \citep[GOGREEN;][]{2017MNRAS.470.4168B} have also observed J0546; their fraction of [\ion{O}{2}] emitters found so far is 7/28 (M. Balogh, private communication), which is about twice the fraction in our sample (6/49).
Since detection of [\ion{O}{2}] serves as a viable indicator of \textit{Herschel} emission,\footnote{$6 / 14$ \textit{Herschel} sources in the \Sifon{} sample are [\ion{O}{2}] emitters, and $6 / 17$ [\ion{O}{2}] emitters have secure \textit{Herschel} counterparts.} we have grounds for concern that our optical selection induces a general bias in our search for dust-obscured galaxies, and thus, in our stacking analysis.
As an extreme example, in J0235, our lowest-$z$ cluster, we find a low fraction of [\ion{O}{2}] emitters (1/82 members) relative to the other clusters, with all \textit{Herschel} detections in that cluster having absorption-line counterparts.
It is important to note that we needed to perform an astrometric shift of $\sim 0\farcs{}7$ in order to align \Sifon{} positions with \textit{HST} imaging when available (i.e., for our higher-$z$ clusters, Section~\ref{sec:hst}). 
If the spectroscopic slits were not originally centered on the nuclear regions of cluster galaxies, then the observations may have systematically missed  spectral line emission from star formation, resulting in our low [\ion{O}{2}] fractions in J0235 and J0546 (and perhaps in our other two clusters as well).

An additional population of optically faint or obscured sources, some of which are detectable via dust emission, may lurk below the [\ion{O}{2}] detection threshold.
The [\ion{O}{2}] completeness limit in \Sifon{} is not provided, but observations to similar depth ($g < 22.8$) at $z \sim 0.3$ yield detections down to $\log (L_{\rm [OII]}/{\rm erg~s^{-1}}) \sim 41$ \citep[see, e.g., the SCUSS+SDSS sample used in][]{2015A&A...575A..40C}. 
We would thus expect to detect star-forming galaxies with \textit{unobscured} SFR $\sim 1$~\smpyr{} \citep[e.g.,][]{1998ARA&A..36..189K, 2004AJ....127.2002K}.
We can compare this with our J0235 detection threshold for \textit{Herschel}/PACS-derived \textit{obscured} SFR, which is approximately $0.6$~\smpyr{}.
If we make the broad assumption that unattenuated and attenuated SFR components are roughly equal, our \textit{Herschel}-selected sample should probe slightly lower SFRs than the [\ion{O}{2}]-selected sample.
This asymmetry would disappear if star-forming cluster members were dust-poor, but would be more pronounced for heavily obscured galaxies.
If such observational systematics have left emission-line galaxies underrepresented, it may not be surprising that \textit{Herschel} counterparts have been found for cluster members lacking [\ion{O}{2}] emission.
This appears to be the case for \ion{Ca}{2} absorbers with significant \textit{Herschel}/PACS emission in J0235; fits to \Kirkpatrick{} suggest that their average SFR is a modest $1.4$~\smpyr{}, which is only just above the expected [\ion{O}{2}] detection threshold.

Part of the selection bias can be remedied with our new ALMA observations, which reveal dusty and gas-rich sources.
Our analysis is insensitive to the most heavily obscured sources because we require all cluster members to be detected at optical wavelengths.
However, six ALMA sources selected via CO emission are found to be cluster members (and have optical counterparts), and only one of them was previously identified in \Sifon{} as an [\ion{O}{2}] emitter.\footnote{We find that J010255.50-491416.2, the galaxy detected in both CO and [\ion{O}{2}] line emission, has the lowest IR luminosity of all higher-$z$ cluster members detected by \textit{Herschel}.
J010255.50-491416.2 appears to be a face-on galaxy, for which the line-of-sight dust column is expected to be minimized, making it more likely that bluer light (and therefore, the [\ion{O}{2}] doublet emission) would be detected.
For more discussion and \textit{HST} imaging of this object, see Section~\ref{sec:ALMA line detections}.}
The small samples of CO and [\ion{O}{2}] detections mean that we can not securely conclude whether or not the ALMA observations correct for the selection effects on our optical sample.
Nonetheless, their relatively disjoint populations indicate that long-wavelength spectroscopy is important for facilitating a more complete census of star-forming cluster members.

The members detected in CO and [\ion{C}{1}] have rich molecular gas reservoirs with masses $M_{\rm mol} \sim (2-6) \times 10^{10}~M_\sun$.
Other measurements in the literature \citep[e.g.,][]{2011ApJ...730L..19G,2012ApJ...752...91W,2013A&A...557A.103J}, as well as our own stacking results of [\ion{O}{2}]-detected galaxies, show that these gas-rich sources are rare in massive clusters at intermediate redshifts.
Future (sub)millimeter surveys will likely identify star-forming galaxies based on optical (emission line) priors, as we have seen.
Otherwise, FIR or millimeter spectroscopy is needed to discover cluster galaxies that are optically faint due to heavy dust obscuration \citep[see, e.g.,][]{2017arXiv170906963R}.

\subsection{Redshift evolution of infrared-bright galaxies} \label{sec:IRBGs}

For our massive clusters, we find that the fraction of galaxies that are IRBGs increases with redshift: 0/82 in J0235, 1/65 in J0438, 6/91 in J0102, and 6/52 in J0546. We consider IRBGs (rather than all \textit{Herschel} counterparts) because they are a luminosity-limited population.
For a fairer comparison, we restrict our analysis to galaxies within $r < 0.5~R_{\rm 200c}$ so that the higher-$z$ clusters can be compared to lower-$z$ clusters (for which the same FOV corresponds to a smaller physical area).
The number of IRBGs then rises from 0/82 in J0235, to 1/63 in J0438, to 4/45 in J0102, and to 4/29 in J0546.
To assess the significance of this apparent redshift evolution, we calculate 68\% binomial confidence intervals for the IRBG fraction in each cluster: $f_{\rm IRBG} = 0^{+0.012}, 0.016_{-0.009}^{+0.025}, 0.089_{-0.034}^{+0.052},$ and $0.138_{-0.052}^{+0.076}$. 
Using these uncertainties, we minimize $\chi^2$ to find the maximum a posteriori power-law model fitting $f_{\rm IRBG}$ to redshift.
A multiplicative coefficient and power-law spectral index are allowed to vary as free parameters and we force the model always to output a positive IRBG fraction.
We sample the marginalized distribution of the power-law index parameter using \texttt{emcee}.
Despite the small number of IRBGs in our galaxy sample, we find that $p < 0.005$ for a zero or negative slope, indicating a definite redshift evolution in the IRBG population of our massive clusters.
This evolution is consistent with a picture in which the prevalence of obscured, strongly star-forming and accreting systems increases with redshift (e.g., in the field; \citealt{2005ApJ...632..169L, 2010ApJ...718.1171R}; and in cluster environments; \citealt{2009ApJ...704..126H}).

\subsection{The FIR-radio correlation and AGN fraction} \label{sec:q parameter}

\begin{figure}
\plotone{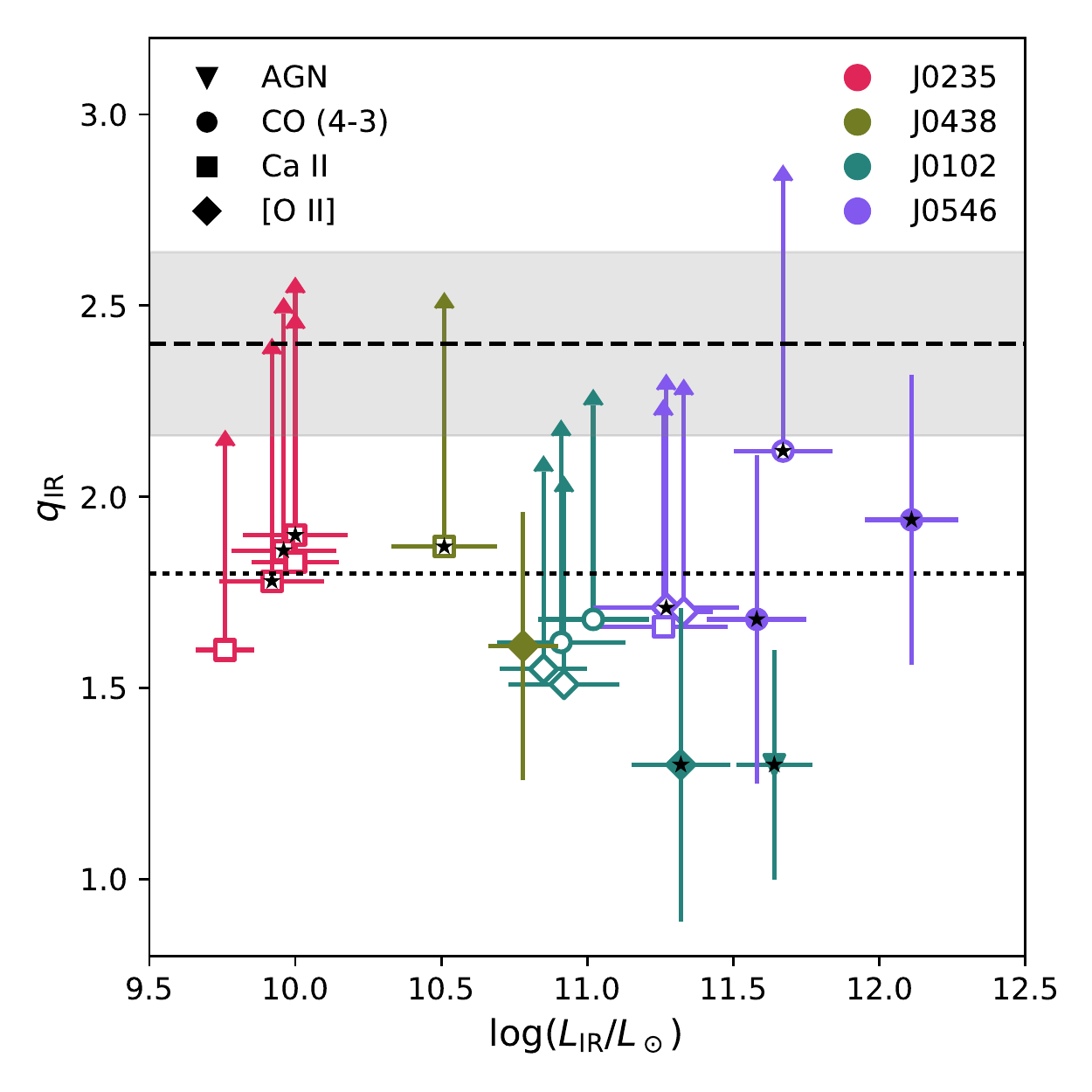}
\caption{
FIR-radio correlation $q_{\rm IR}$ parameter is plotted for 19 \textit{Herschel} detections as a function of IR luminosity. 
The horizontal dashed line and shaded region show the median $q_{\rm IR}$ value and $\pm 1~\sigma$ uncertainties for 250~\micron{} selected star-forming galaxies \citep{2010A&A...518L..31I}, and the horizontal dotted line shows $q_{\rm IR} = 1.8$ used to separate star-forming galaxies from AGNs. 
Error bars are $\pm 1~\sigma$; note that the error in $q_{\rm IR}$ is correlated with errors in $L_{\rm IR}$ and $L_{\rm 1.4~GHz}$. 
Empty markers with upward-pointing arrows are $3~\sigma$ lower limits on $q_{\rm IR}$ due to radio non-detections. 
Marker colors and symbols correspond to particular galaxy clusters and radio/optical spectral classifications, respectively.
AGN-dominated systems as determined from best-fitting \Kirkpatrick{} IR spectral templates are marked with black stars.}
\label{fig:q_IR}
\end{figure}

The far-infrared and radio emission of star-forming galaxies are tightly correlated over many orders of magnitude in luminosity and flux density \citep{1985A&A...147L...6D, 1985ApJ...298L...7H, 1991ApJ...376...95C}.
Following \cite{1985ApJ...298L...7H}, we define the logarithmic ratio between the FIR flux and non-thermal radio ($\nu_{\rm rest} = 1.4$~GHz) flux density as 
\begin{equation} \label{eq:q_IR}
q_{\rm IR} = \log\left ( \frac{S_{\rm IR}  / \left (3.75 \times 10^{12}~\rm W~ m^{-2}\right ) }{  S_{\nu,1.4~\rm GHz} / \left (\rm W~ m^{-2}~ Hz^{-1} \right )}\right ),
\end{equation}
where the FIR flux is $S_{\rm IR} = L_{\rm IR} / 4\pi d_L^2$ (and $d_L$ the luminosity distance), and $S_{\nu,\rm 1.4~GHz}$ is the $k$-corrected ATCA flux density.

The radio detection limit in our high-$z$ clusters is $\log(L_{1.4~\rm GHz}/{\rm W~Hz^{-1}}) \approx 23.5$. 
At these radio luminosities, AGNs exceed star-forming galaxies by at least an order of magnitude in the field \citep{2002AJ....124..675C, 2007MNRAS.375..931M, 2010A&A...518L..31I}.
However, our sample of \textit{Herschel} and/or ALMA-detected cluster galaxies are selected by infrared rather than radio fluxes.
We find that six cluster members have measurable radio emission and 13 do not (although in some cases, we see $\sim 3~\sigma$ peaks in the radio surface brightness; see Appendix~\ref{sec:new detections} for details on individual galaxies).

We calculate $q_{\rm IR}$ values, or assign $3~\sigma$ lower limits when no radio counterpart is detected, and plot them against $L_{\rm IR}$ in Figure \ref{fig:q_IR}.
In line with expectation, the optically confirmed AGN in the \Sifon{} catalog has the highest radio luminosity and the lowest $q_{\rm IR} = 1.3 \pm 0.3$ within our sample.
For the other radio detections, we find a mean $q_{\rm IR} = 1.54 \pm 0.22$.
Previous studies have established $q \approx 2.3$ for star-forming galaxies \citep{2001ApJ...554..803Y, 2002AJ....124..675C, 2003ApJ...586..794B}.
\cite{2010A&A...518L..31I} find a median $q_{\rm IR} = 2.40 \pm 0.24$ when studying a sample of \textit{Herschel} 250~\micron{}-selected galaxies (we show this as a dashed black line in Figure \ref{fig:q_IR}).
Additionally, a separator of $q_{\rm IR} \approx 1.8$ is sometimes used to differentiate between star-forming galaxies and AGNs \citep[see, e.g.,][]{2002AJ....124..675C}.

Four out of the 19 \textit{Herschel} detections in our study have $q_{\rm IR} < 1.8$ and appear to be AGNs, in addition to all four being IRBGs.
This AGN fraction $f_{\rm AGN} \gtrsim 0.20 - 0.30$ is large compared to values found in other studies of IR-selected AGN in $z\sim 1$ clusters \citep[e.g., $f_{\rm AGN} \sim 0.1$ in][]{2013ApJ...768....1M, 2016ApJ...825...72A}.
One cluster member, J054638.87-534613.6, is a new CO detection and we cannot examine its optical spectrum to check for emission lines.
Its IR SED is best fit by an AGN template, although this classification may be incorrectly driven by its bright dust continuum emission.
A star-forming galaxy template with less warm dust would imply a lower $L_{\rm IR}$ and thereby drive down $q_{\rm IR}$, which would reinforce its classification as a radio AGN.
The other three radio-loud AGNs have both [\ion{O}{2}] and radio emission.
Note, however, that the uncertainties on $q_{\rm IR}$ are large, mainly reflecting scatter in the IR luminosity fit. 
The two AGNs in J0102 appear to be secure classifications, but the other two are less certain since they have $q_{\rm IR}$ within $1~\sigma$ of the dividing line.

\subsection{Star formation rate as a function of environment?}

\begin{figure}
\plotone{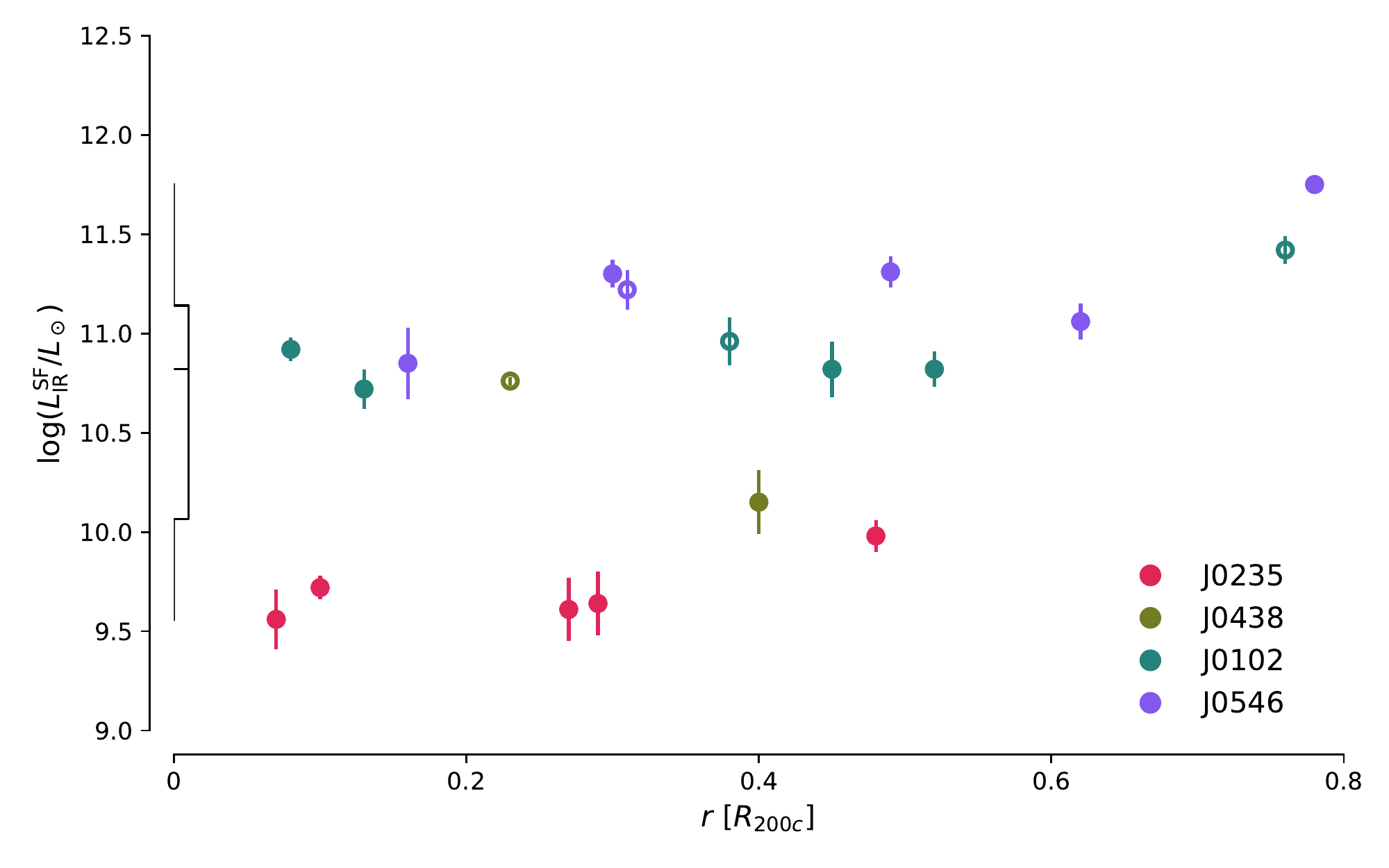}
\caption{IR luminosity due to star formation is plotted against clustercentric radius for 19 \textit{Herschel} detections.
Marker color and shape indicate cluster membership. 
Clustercentric radius is computed as the projected distance from the BCG for J0235, J0438, and J0546, and from the midpoint between weak lensing peaks for J0102 \citep{2014ApJ...785...20J}.
Unambiguous radio AGNs ($q_{\rm IR} < 1.8$) are shown with open markers.
On the $y$-axis, a box plot shows the interquartile range in $L_{\rm IR}^{\rm SF}$.
}
\label{fig:L_IR with environment}
\end{figure}

\begin{figure}
\plotone{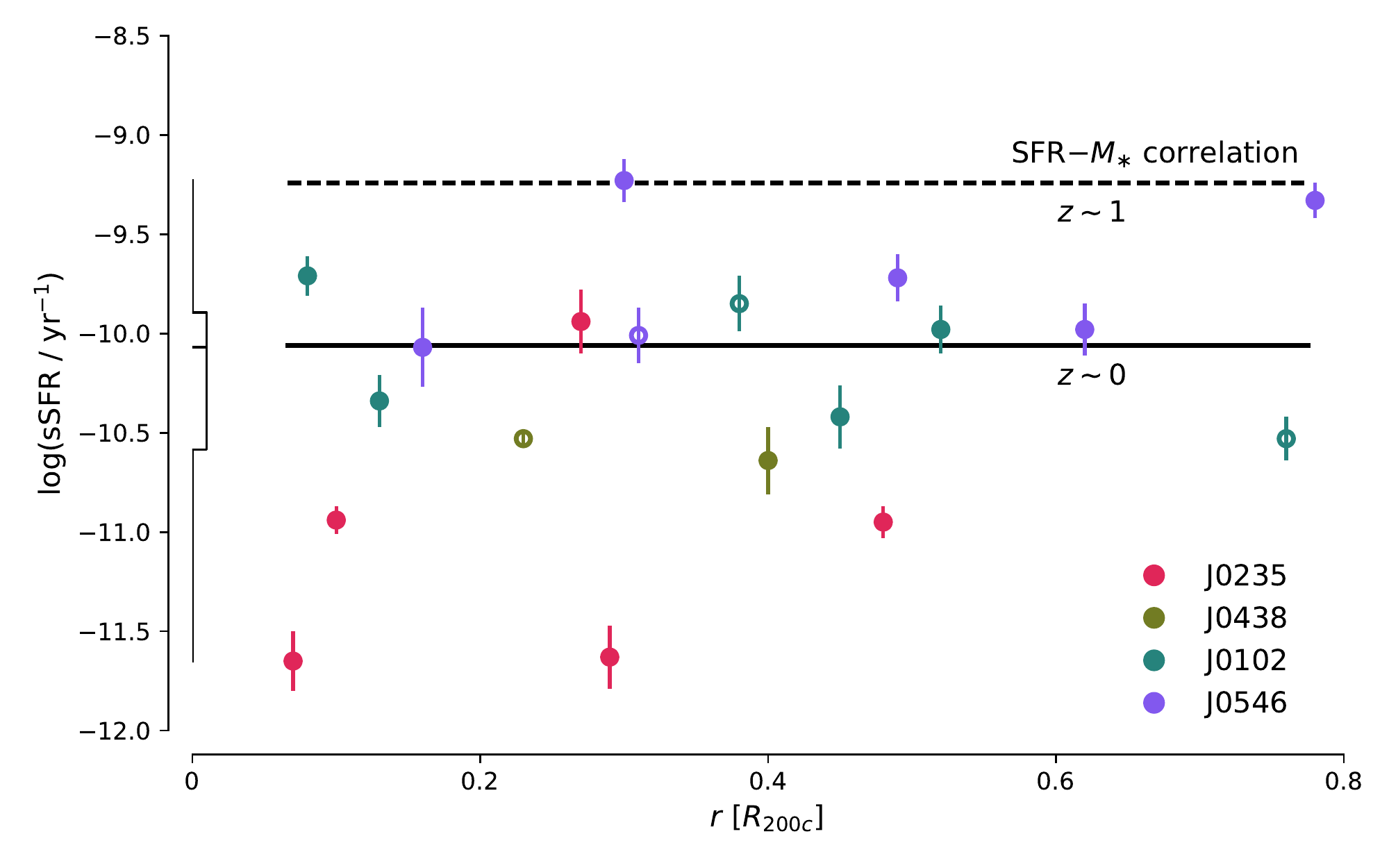}
\caption{Specific SFR (sSFR) is plotted against clustercentric radius for 19 \textit{Herschel} detections.
Colors and markers are the same as in Figure~\ref{fig:L_IR with environment}, where radio AGNs are shown with open markers.
The dashed (solid) horizontal black line illustrates the measured SFR$-M_\ast$ correlation for $\log(M_\ast/M_\sun) = 11.0$ at $z\sim 1$ ($z \sim 0$) as reported by \cite{2007A&A...468...33E}. 
On the $y$-axis, a box plot shows the interquartile range in sSFR.
}
\label{fig:sSFR with environment}
\end{figure}

In Figure~\ref{fig:L_IR with environment}, we plot the luminosity contribution from dust-obscured star formation, $L_{\rm IR}^{\rm SF}$, against clustercentric radius for all cluster galaxies with \textit{Herschel} detections.
We also include sources that are labeled as AGNs ($q_{\rm IR}<1.8$; shown as open circles); all infrared luminosities have been corrected for the contributions from AGN implied by the best-fit \Kirkpatrick{} templates.
For J0102, we compute clustercentric radius using the midpoint between its mass peaks determined via weak lensing measurements \citep{2014ApJ...785...20J}.
For the other clusters, we use the BCG as the cluster center.
There appears to be a modest trend of increasing $L_{\rm IR}^{\rm SF}$ with radius, from which we might be tempted to infer a relationship between SFR and environmental galaxy density.
However, the decrease in SFR with clustercentric distance could also be a reflection of (or driven by) the SFR$-M_\ast$ correlation \citep[see, e.g.,][]{2007ApJ...660L..43N, 2007A&A...468...33E, 2007ApJ...670..156D}.
Stellar mass tends to increase with density, although the effects of mass and of environment should be separable \citep[see, e.g.,][]{2010ApJ...721..193P}. 
Other authors have argued that the correlation between SFR and environment is preserved even when stellar mass is controlled for \citep{2013MNRAS.434..423K, 2015MNRAS.450.2749G}: the cluster environment affects the \textit{fraction} of star-forming galaxies, but has little effect on an individual galaxy's \textit{star formation rate}, which is instead correlated with that galaxy's halo and stellar properties.

In order to separate the effects of cluster environment and galaxy mass, we compute specific SFR (${\rm sSFR \equiv SFR}/M_{\ast}$).
Stellar mass is derived as described in Section~\ref{sec:spitzer}; the average (median) $\log(M_\ast / M_\sun) = 11.0$ (11.2).
Both $M_\ast$ and sSFR are included in Table~\ref{tab:properties}.
For our sample, we compute a median $\log({\rm sSFR / yr}^{-1}) = -10.07$, and average $\log({\rm sSFR / yr}^{-1}) = -10.36 \pm 0.57$.
We compare our sample to those found at $z\sim 1$ in the GOODS field and at $z\sim 0$ in the Sloan Digital Sky Survey \citep[SDSS;][]{2007A&A...468...33E}: SFR$_{z\sim 1}=7.2^{+7.2}_{-3.6} \times (M_\ast / 10^{10}~M_\sun)^{0.9}$~\smpyr{}, and SFR$_{z \sim 0}=8.7^{+7.4}_{-3.4} \times (M_\ast / 10^{11}~M_\sun)^{0.77}$~\smpyr{}.
The SFR$-M_\ast$ correlations at $z\sim 1$ and $z\sim 0$ for $\log(M_{\ast}/M_\sun) = 11.0$ are respectively shown as dashed and solid horizontal lines in Figure~\ref{fig:sSFR with environment}, in which we show the sSFR plotted against clustercentric radius.
The relationship between star formation and radius seen in Figure~\ref{fig:L_IR with environment} has completely vanished once we normalize the SFR by mass.
(These results do not change for J0102 even if we consider the BCG to be its cluster center.)
sSFR appears to increase with cluster redshift, similar to the results seen in Section~\ref{sec:IRBGs}.
We fit a straight-line model to the cluster-averaged log sSFR against redshift, varying the slope and intercept as free parameters. 
We determine the best-fit slope to be positive at $p < 0.001$ significance using \texttt{emcee} to sample its posterior distribution.
Therefore, the redshift evolution in sSFR is still evident, and can been seen in Figure~\ref{fig:sSFR with environment}.
It appears that the SFR-radius trend seen before is driven by a radial gradient in stellar mass.
We do not see a difference in the effects of environment on sSFR between the cluster center and $0.8\times$ the virial radius.
However, our results are consistent with a scenario in which the cluster environment depresses sSFRs below the SFR$-M_{\ast}$ correlation measured for field galaxies at similar redshift.

We have also examined the radial distributions of the gas depletion time and gas fraction in the two higher-$z$ clusters.
Remarkably, the galaxy with highest gas fraction, J010252.44-491531.2, is located near  (projected distance~$< 0.1~R_{\rm 200c}$) the J0102 cluster merging center.
However, with only six CO detections, we do not find any measurable correlations between cold gas properties and environment.
The mean gas fraction is $f_{\rm gas} = 0.22$.

Most luminous star-forming galaxies live in the outskirts of intermediate redshift clusters \citep{2009ApJ...704..126H}.
CO has likewise been detected at large clustercentric distances, $\sim 1-3 \times$ the virial radius, in other intermediate-redshift clusters (\citealt{2011ApJ...730L..19G, 2012ApJ...752...91W}; but see also \citealt{2013A&A...557A.103J}).
If we assume that gas fraction decreases as we move toward the cluster center, then it would not be surprising if the average gas fraction $f_{\rm gas} \sim 0.2-0.3$ found in the field at $z \sim 1$ \citep[see, e.g.,][]{2013ApJ...768...74T,2015ApJ...800...20G} is also characteristic of cluster galaxies at large clustercentric distances.
Even if cold gas reservoirs are not preferentially less massive near cluster cores, the same stellar mass gradient that dominated our SFR-environment trend might be expected to reduce gas fractions at small radii.
However, we have found six CO detections with $f_{\rm gas} \sim 0.1-0.3$ located within $R_{\rm 200c}$ of their cluster centers.
It is noteworthy that these gas-rich, star-forming cluster members, with gas fractions comparable to those of $z \sim 1$ field galaxies, reside in the cores of the most massive galaxy clusters at the same redshifts.

\section{Conclusions}

We have presented new \textit{Herschel} PACS 100/160~\micron{} and ALMA Band 6 observations of four massive, SZE-selected galaxy clusters.
Based on our analysis, we conclude the following about the star formation, cold gas, and dust properties of our galaxy sample:

\begin{enumerate}
\item We detect 19 \textit{Herschel}/PACS counterparts to galaxies in our sample of four massive SZE-selected clusters. 
Five are newly confirmed as cluster members by \CO{} and/or \CI{} detections in ALMA data cubes.

\item We detect dust emission in three cluster galaxies, which we use to calculate ISM (H$_2$ + \ion{H}{1}) masses.
All detections are from galaxies in the J0546 cluster.
ALMA-traced gas and dust masses correlate as expected with \textit{Herschel}-derived IR luminosities.
While ISM and molecular gas masses largely agree, the mass ratios differ between clusters: $M_{\rm mol}/M_{\rm ISM} > 1.33$ for J0102 and $M_{\rm mol}/M_{\rm ISM} = 0.55^{+0.58}_{-0.34}$ for J0546.
This apparent discrepancy may be a result of enhanced CO excitation in the violent merger in J0102, or may simply reflect the small size of the detected sample.

\item We find a mean [\ion{C}{1}]-to-\CO{} ratio of $0.54 \pm 0.20$ in flux units, corresponding to a line ratio of $L'_{[{\rm CI}]~({}^3P_1 - {}^3P_0)}/L'_{\rm CO(1-0)} = 0.19 \pm 0.07$.
Our results agree with previous literature measurements for both local and high-$z$ field galaxies.
We also find that cluster-averaged sSFR increases significantly with redshift.

\item We find strong evidence for an increase in the prevalence of infrared-bright galaxies (IRBGs; $\log (L_{\rm IR}/L_\sun) > 10.6$) with redshift, for the inner regions of our cluster sample.

\item We place upper limits on dust and dense gas mass via stacking ALMA continuum and line non-detections in the higher-$z$ clusters.
Our $3~\sigma$ limits on ISM mass are $\log (M_{\rm ISM}/M_\sun) < 10.0$ for J0102, and $\log(M_{\rm ISM}/M_\sun) < 9.9$ for J0546.
By stacking CO and [\ion{C}{1}] lines together, we constrain the mean molecular gas mass for galaxies in our sample: $\log (M_{\rm mol}/M_\sun) < 9.8$ for both J0102 and J0546.

\item By using radio interferometric observations, we estimate the FIR-radio correlation parameter, $q_{\rm IR}$, for our 19 IR-detected cluster members.
Five have both \textit{Herschel}/PACS and ATCA counterparts, including one optical AGN, and we find that the mean $\langle q_{\rm IR}\rangle = 1.54 \pm 0.24$.
Four galaxies are clearly below $q_{\rm IR} = 1.8$ -- a threshold commonly used to separate AGNs from star-forming galaxies -- all of which are also IRBGs.
This lower limit on the AGN fraction, $f_{\rm AGN} \gtrsim 0.2$, is approximately twice the fraction seen for FIR-selected galaxies in less-massive, $z \sim 1$ clusters.

\item We find modest radial trends in SFR, with decreased SFR at small distances from the cluster center. 
However, no significant correlation is found when we examine sSFR binned by radial distance.
The mean (median) log(sSFR/yr$^{-1}$) of our sample is $-10.36 \pm 0.57$ ($-10.07$), consistent with the SFR$-M_\ast$ correlation seen at low redshift.

\item  For CO detections, we find an average gas fraction $f_{\rm gas} \approx 0.2$ in the cores of massive $z\sim 1$ clusters, consistent with those of field galaxies at the same redshift.
However, the average gas depletion timescale, $\tau_{\rm dep} \approx 2$~Gyr, is long relative to those of $z\sim 1$ field galaxies.

\end{enumerate}

\acknowledgments
We thank the anonymous referee for detailed and useful comments that have greatly improved this paper.
J.F.W. and A.J.B. thank Amanda Kepley for valuable support and guidance during the ALMA data reduction process.
P.A. acknowledges support from CONICYT through grant FONDECYT Iniciaci\'on 11130590.
J.P.H.\ acknowledges the hospitality of the Flatiron Institute which is supported by the Simons Foundation.
This paper makes use of the following ALMA data: ADS/JAO.ALMA\#2013.1.01358.S. ALMA is a partnership of ESO (representing its member states), NSF (USA) and NINS (Japan), together with NRC (Canada), MOST and ASIAA (Taiwan), and KASI (Republic of Korea), in cooperation with the Republic of Chile. The Joint ALMA Observatory is operated by ESO, AUI/NRAO and NAOJ.
The National Radio Astronomy Observatory is a facility of the National Science Foundation operated under cooperative agreement by Associated Universities, Inc.
This work has been supported by (i) an award issued by JPL/Caltech in association with \textit{Herschel}, which is a European Space Agency Cornerstone Mission with significant participation by NASA, (ii) the National Science Foundation through grant AST-0955810 and award GSSP SOSPA2-018 from the National Radio Astronomy Observatory, which is operated under cooperative agreement by Associated Universities, Inc., and (iii) the National Aeronautics and Space Administration (NASA) through Chandra Award numbered GO2-13156X issued to Rutgers University by the Chandra X-ray Observatory Center, which is operated by the Smithsonian Astrophysical Observatory for and on behalf of NASA under contract NAS8-03060.

\software{APLpy \citep{2012ascl.soft08017R}, AstroPy \citep{2013A&A...558A..33A}, emcee \citep{2013PASP..125..306F}, Jupyter \citep{10.3233/978-1-61499-649-1-87}, matplotlib \citep{matplotlib}, pandas \citep{pandas}, SciPy/NumPy \citep{NumPy&SciPy}.}

\newpage

\appendix

\section{New ALMA and \textit{Herschel} detections} \label{sec:new detections}

We describe new galaxies in our high-$z$ clusters that have ALMA and/or \textit{Herschel} source detections in Sections~\ref{sec:ALMA line detections}, \ref{sec:ALMA dust detections}, and \ref{sec:Herschel detections}.
All new line detections are shown in Figures~\ref{fig:J0102 CO detections} (J0102) and \ref{fig:J0546 CO detections} (J0546).
In some cases, cluster galaxies are at small angular separations from background or foreground galaxies that are bright at long wavelengths, and it becomes challenging to deblend the sources of emission.
Cluster galaxy names preceded by an asterisk (*) are those that have PACS or ALMA emission dominated by such contaminants, or are otherwise spurious, and are excluded from the stacking analysis of Section~\ref{sec:stacking}.

\subsection{ALMA line detections} \label{sec:ALMA line detections}

\begin{figure*}
	\plotone{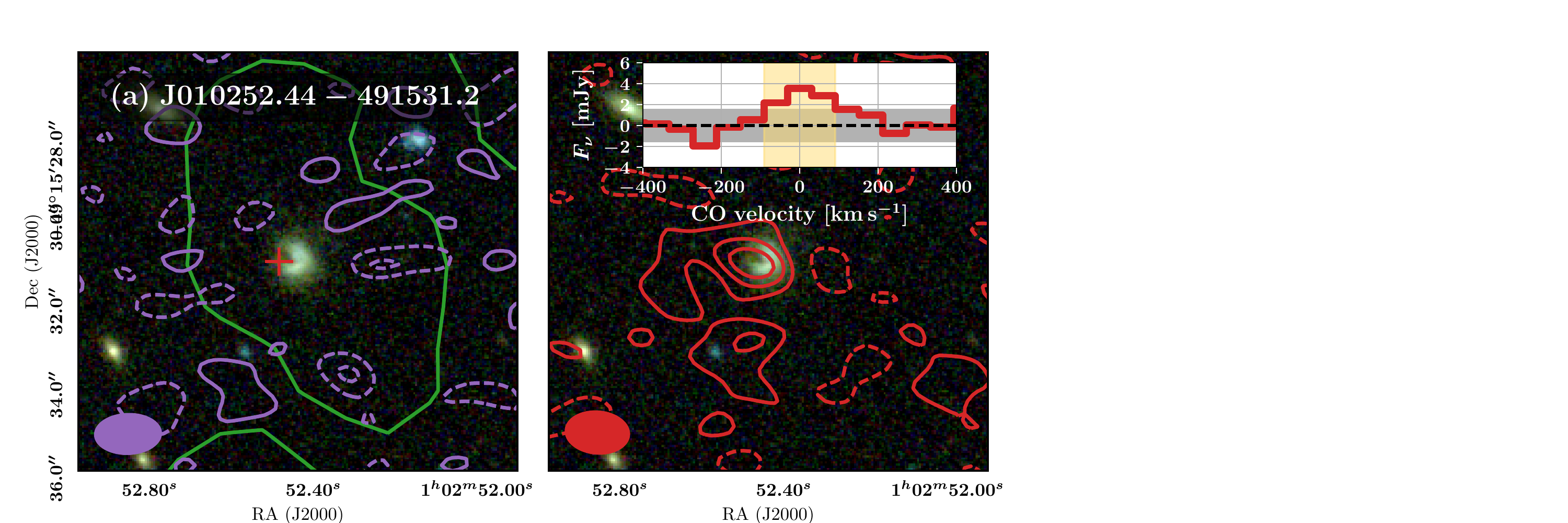}
	\plotone{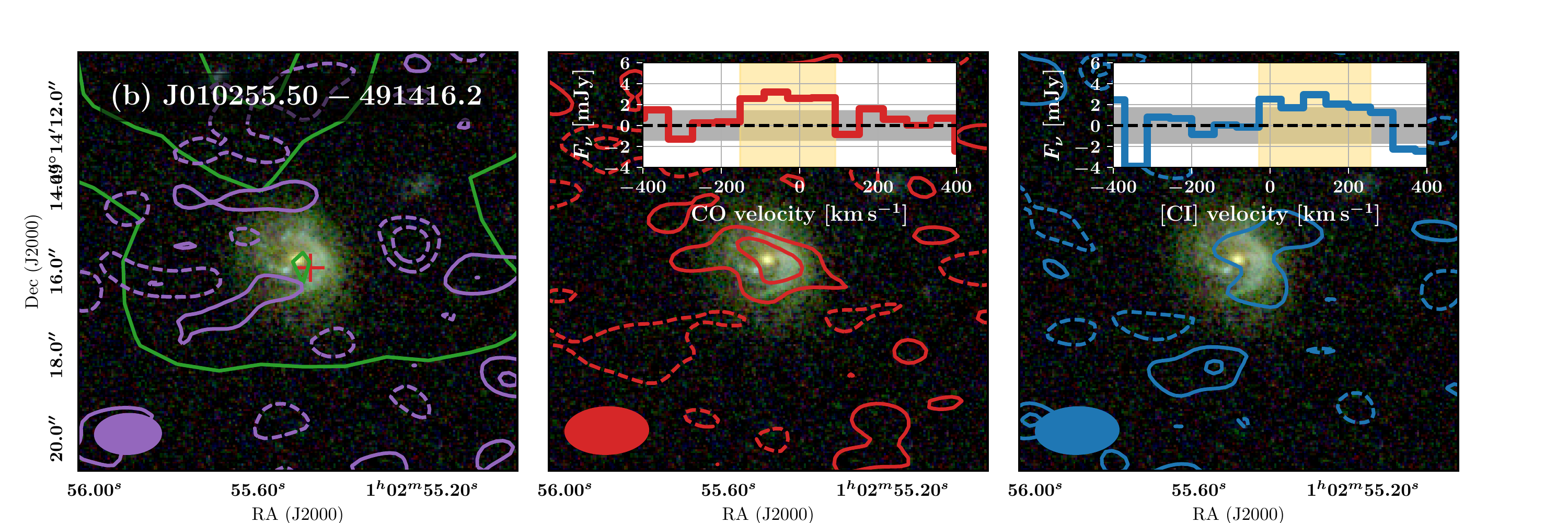}
	\plotone{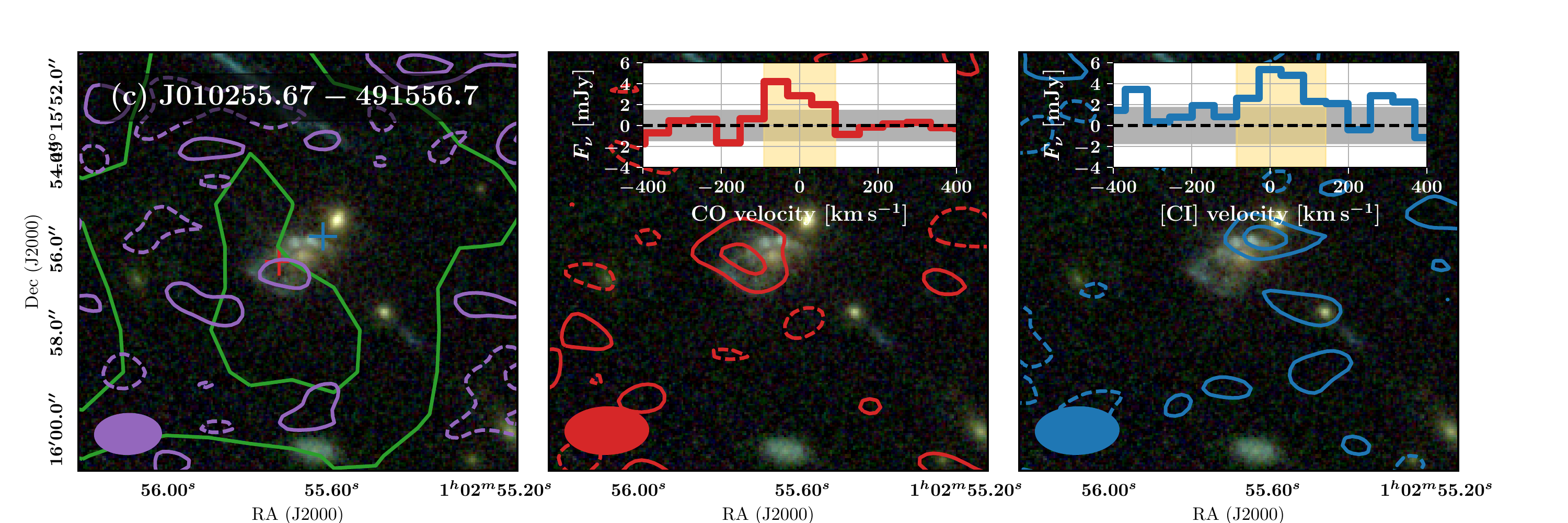}
	\caption{
		\textit{Herschel} and ALMA contours overlaid on \textit{HST}/ACS F625W + F775W + F850LP color images in 5\arcsec{} cutouts, for J0102 detections (J010252.44-491531.2 (a); J010255.50-491416.2 (b); J010255.67-491556.7 (c)).
		Left panels: \textit{Herschel}/PACS 160~\micron{} (green; multiples of $1.5~\sigma$) and ALMA dust continuum (purple; multiples of $1.5~\sigma$) contours, with ALMA synthesized beam at lower left.
		Middle panels: \CO{} integrated intensity (red; multiples of $1.5~\sigma$) contours, with synthesized beams at lower left and inset spectra showing $\pm 1~\sigma$ (gray) and velocity range used for moment map (yellow).
		Right panel: same as middle panels for the \CI{} line. 
		Red and blue crosses in left panels mark peaks of line emission in middle and right panels, respectively.
		}
\label{fig:J0102 CO detections}
\end{figure*}

\begin{figure*}
	\plotone{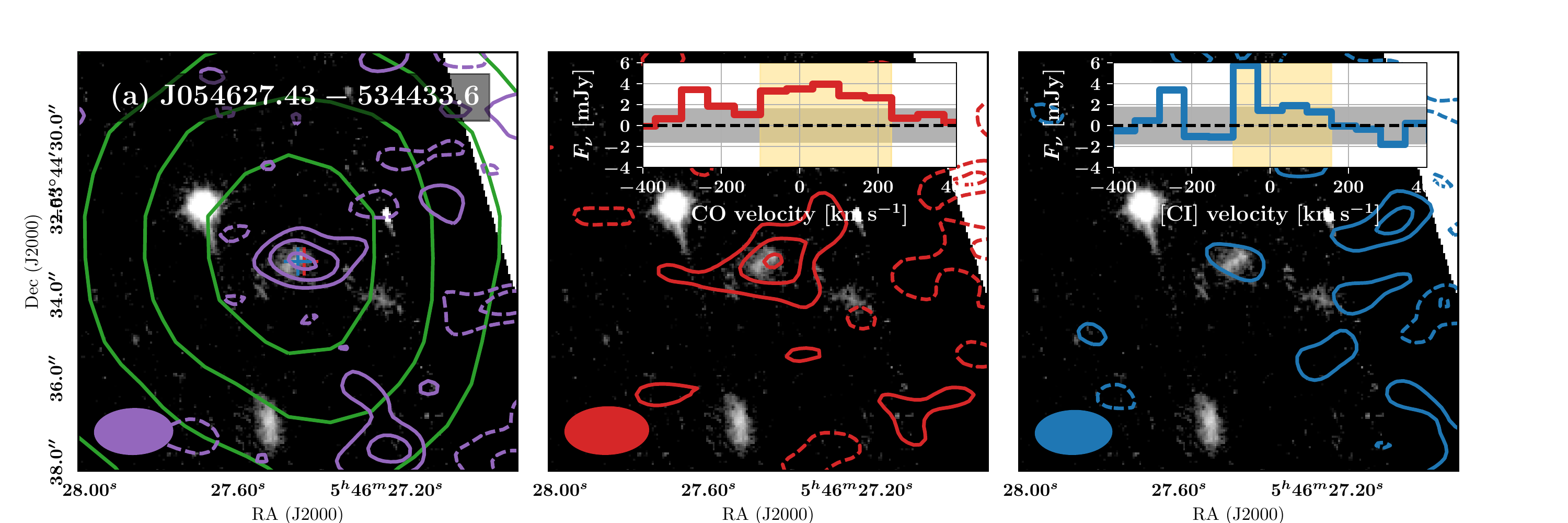}
	\plotone{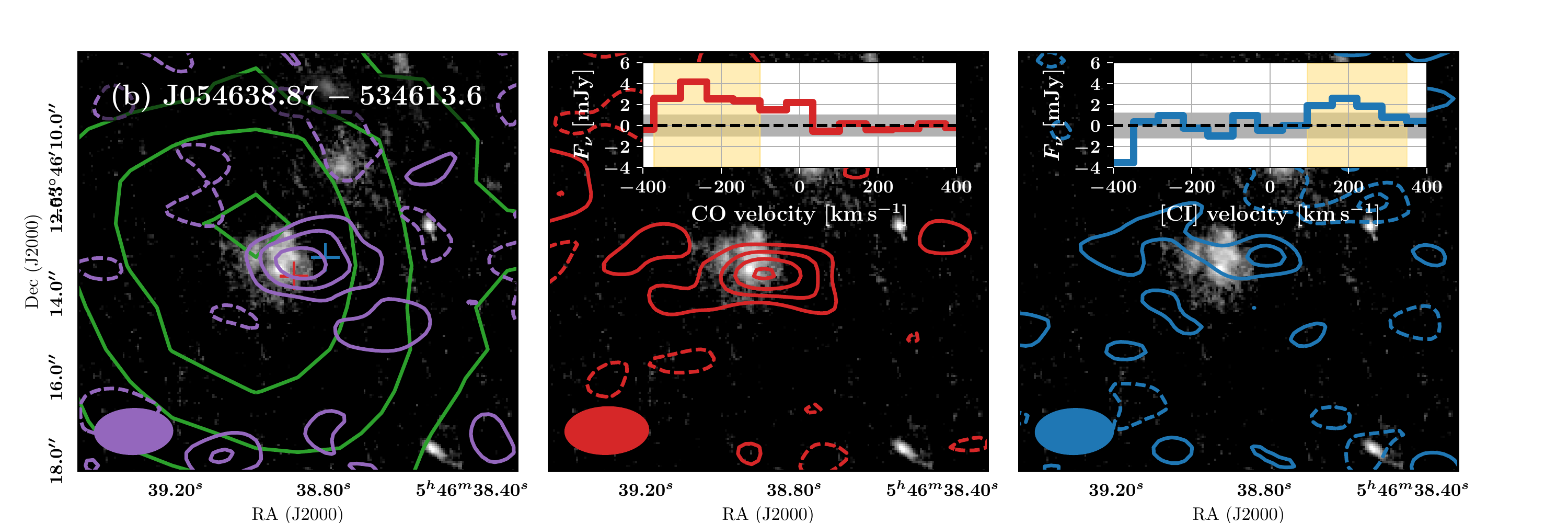}
	\plotone{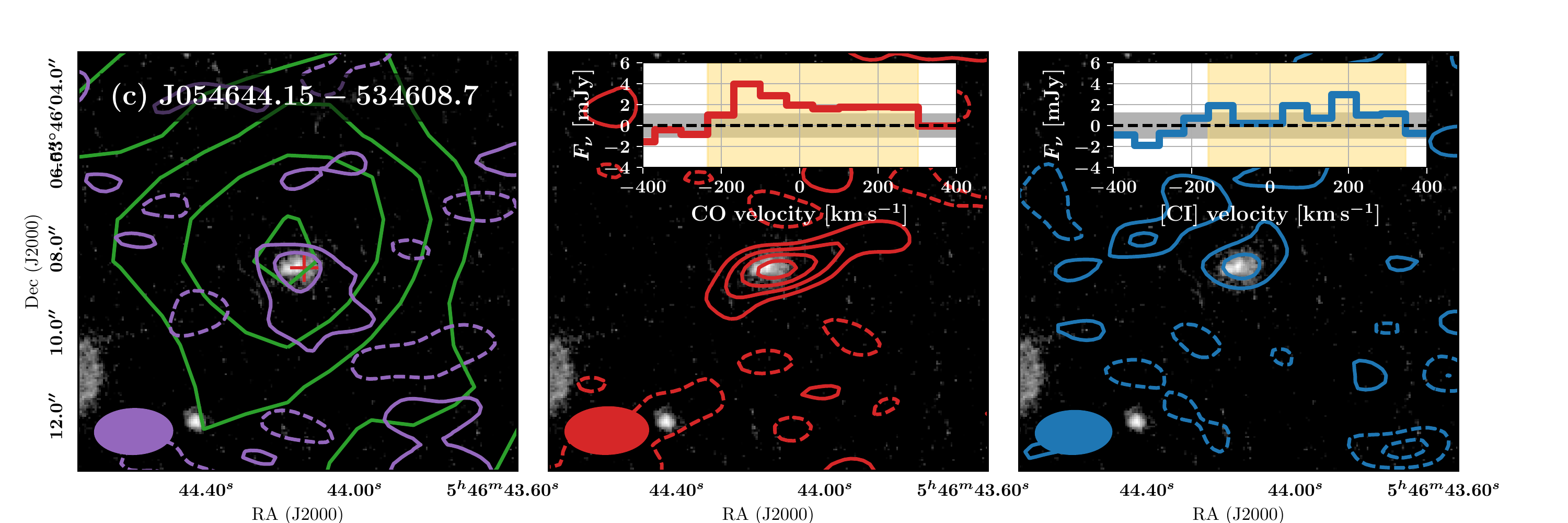}
	\caption{
		Spectral line sources in J0546 (J054627.43-534433.6 (a); J054638.87-534613.6 (b); J054644.15-534608.7 (c)).
		Notation as in Figure~\ref{fig:J0102 CO detections}. 
		Grayscale images are \textit{HST}~F814W; 160~\micron{} contours for top panel (a) are multiples of $5~\sigma$.
		}
\label{fig:J0546 CO detections}
\end{figure*}

\subparagraph{J010252.44-491531.2}
{
	
	(Figure~\ref{fig:J0102 CO detections}a) is a relatively compact galaxy at redshift $z=0.8610$.
	\CO{} is detected at about $6~\sigma$ significance, and we see a $3~\sigma$ peak to its south, suggesting that its molecular gas reservoir is large in size.
	(Our [\ion{C}{1}] spectral coverage does not extend to this redshift, so we cannot use it for comparison.)
	No radio continuum emission is detected, giving a lower limit $q_{\rm IR} > 1.7$.
}

\subparagraph{*J010252.56-491400.5}
{
	is a SoFiA-detected CO line redshifted to $z=0.8735$, with an integrated flux $I_{\rm CO}= 0.82 \pm 0.33\rm~Jy~km~s^{-1}$.
	It has an optical counterpart, a red elliptical absorption-line cluster galaxy ($z=0.8817$), which is separated by a rest-frame velocity $\Delta v \sim 2500~\rm km~s^{-1}$.
	We do not detect \textit{Herschel}, dust continuum, or [\ion{C}{1}] line emission, so this detection is likely spurious and is excluded from the list of detections.
}

\subparagraph{*J010255.27-491441.8}
{
	is a SoFiA-detected CO line at $z=0.8678$ with a spheroidal counterpart seen in \textit{HST} imaging.
	Upon visual inspection, none of the channels exceed $2~\sigma$ significance, and the emission centroids are not aligned across channels.
	Moreover, we do not detect \textit{Herschel}, dust continuum, or [\ion{C}{1}] line emission, so this detection is likely spurious and is excluded from the list of detections.
}

\subparagraph{J010255.50-491416.2}
{
	(Figure~\ref{fig:J0102 CO detections}b) is a face-on spiral galaxy detected in [\ion{O}{2}] emission by \Sifon{}.
	A nearby source 7\arcsec{} to the north has bright emission at $\lambda = 100 - 350~\micron{}$, which blends in with the galaxy.
	We measure the FIR flux density from the peak surface brightness. 
	Radio emission might also be blended with that of the nearby FIR source. 
	Based on the $\sim 2\arcsec$ separation between the radio and optical/NIR/FIR emission, the ATCA source is not associated with the cluster galaxy.
	Only a CO line is detected by SoFiA in the ALMA cubes; closer inspection reveals positive emission in the [\ion{C}{1}] line cube, but because it is not spatially coherent we only report an upper limit.
}

\subparagraph{J010255.67-491556.7}
{
	(Figure~\ref{fig:J0102 CO detections}c) is a galaxy with a redder bulge-like component surrounded by bluer spiral arm features or a ring feature with two bright knots, with a nearby red spheroidal source.
	Neither the spiral nor the red neighbor have optical redshifts, but CO and [\ion{C}{1}] lines are marginally detected in the ALMA data cubes consistent with a redshift of $z=0.8678$. 
	Only the [\ion{C}{1}] emission is detected by SoFiA.
	The CO zeroth moment map peaks about 0\farcs{}7 to the east of the spiral galaxy, and the [\ion{C}{1}] zeroth moment map peaks between the spiral and the red galaxy (about 0\farcs{}7 to the northwest of the spiral galaxy).
	No ALMA or radio continuum is detected, although there exists a faint $3~\sigma$ peak in the radio surface brightness.
}

\subparagraph{J054627.43-534433.6}
{

	(Figure~\ref{fig:J0546 CO detections}a) is a fairly compact, star-forming galaxy at redshift $z = 1.0566$.
	Due to the source's position near the edge of our mosaic, line fluxes are subject to large primary-beam corrections.
	CO is unambiguously detected by SoFiA and [\ion{C}{1}] is present at $3~\sigma$ significance. 
	This is the brightest galaxy at IR wavelengths (and the only ULIRG) in our sample, and it also has the highest ISM mass.
	We find $q_{\rm IR} = 1.9 \pm 0.4$, above the $q_{\rm IR} = 1.8$ threshold that labels it a star-forming galaxy.

}

\subparagraph{J054638.87-534613.7}
{
	(Figure~\ref{fig:J0546 CO detections}b) is a galaxy with a few near-UV-bright knots on its north and east sides.
	Both CO and [\ion{C}{1}] lines are detected by SoFiA.
	The dust continuum and CO integrated flux appear to be marginally resolved.
	The \CO{} line is at $z=1.0810$, and the [\ion{C}{1}] line is offset $\sim 400\rm~ km~ s^{-1}$ redward of the CO redshift.
	The difference in systemic velocity suggests that we may have observed a separate gas concentration, or (more likely) the [\ion{C}{1}] detection is unreliable.
	Removing the [\ion{C}{1}] detection from our sample does not change our results on the [\ion{C}{1}]-to-CO ratio (Section~\ref{sec:carbon abundance}).
	We detect an ATCA source and compute $q_{\rm IR} = 1.7 \pm 0.4$, implying the galaxy's bolometric luminosity is dominated by AGN emission.
}

\subparagraph{J054644.15-534608.7}
{
	(Figure~\ref{fig:J0546 CO detections}c) is compact and does not possess many distinguishing optical features aside from a lone star-forming clump on the east side of its nucleus.
	We have detected a CO line in the ALMA data cube consistent with J0546 membership ($z = 1.0649$) using SoFiA; this galaxy has the highest CO luminosity within our sample. 
	The CO flux extends toward the west side of the galaxy, possibly indicating the presence of cold gas in a tidal feature or faint neighbor.
	[\ion{C}{1}] emission is measured from a $3.6~\sigma$ peak in the zeroth moment map.
	The dust emission is extended to the south.
	Because no radio continuum is detected, we find that $q_{\rm IR} > 2.1$ and conclude that this galaxy's luminosity is dominated by star formation activity.

\subsection{ALMA dust-only detections} \label{sec:ALMA dust detections}

\subparagraph{*J054633.40-534454.4}
{
	is a \ion{Ca}{2} absorption-line galaxy with an elongated southern tail in \textit{HST} imaging.
	We detect a faint ALMA source separated by a distance of $0\farcs{}5$ from the \textit{HST} centroid. 
	The dust detection, if associated with the cluster galaxy, is not at odds with our upper limits on CO and [\ion{C}{1}] emission, which constrains the \textit{molecular} mass to $\lesssim 3.0 \times 10^{10}~M_\sun$, since molecular gas comprises half to two-thirds of a galaxy's ISM mass (see, e.g., \citealt{2013seg..book..491S}).
	At \textit{Herschel} wavelengths, we see unevenly-distributed positive emission totaling $S_{\rm 160 ~ \micron{}} \sim 1.5 \pm 0.4 \rm ~ mJy$, but no single pixel exceeds $2.3~\sigma$ significance; therefore, we do not claim detection of a FIR source.
	No radio emission is detected ($S_{\rm 2.1~GHz} < 24 \rm ~ \mu Jy$).
	We categorize the ALMA source as a dusty star-forming galaxy (DSFG) rather than a cluster member and rename it J054633.40-534454.5 in Table~\ref{tab:alma continuum detections} to reflect its dust continuum centroid.
}

\subparagraph{*J054636.61-534405.9}
{

	has an \textit{HST} morphology suggesting that it has several clumpy regions of star formation, but it is labeled as a \ion{Ca}{2} absorber in \Sifon{}.
	No ALMA line emission is detected.
	There is a faint ALMA continuum source displaced to the southeast of the optical light by $1\farcs{}1$, which is significantly larger than the offsets normally seen in \textit{HST}-ALMA comparisons \citep[e.g., $\sim 0\farcs{}3 - 0\farcs{}6$;][]{2017MNRAS.466..861D}. 
	In Table~\ref{tab:alma continuum detections}, it is labeled as J054636.65-534406.5 along with its submillimeter properties.

}

\subparagraph{*J054642.12-534543.9}
{
	is an [\ion{O}{2}]-emitting galaxy that is compact and relatively featureless.
	A \textit{Herschel}/PACS peak lies $1\farcs{}6$ to the northwest of the optical centroid, which itself lies 0\farcs{}7 to the northwest of a \textit{Spitzer} source.
	The \textit{Spitzer} NIR source agrees to sub-pixel accuracy with an ALMA continuum detection and a possible $3~\sigma$ radio source.
	No CO or [\ion{C}{1}] emission is detected, implying either that a large fraction of the gas reservoir is atomic ($M_{\rm ISM} = (5.5 \pm 0.8) \times 10^{10}~M_\odot$ vs. $M_{\rm mol}({\rm CO}) < 3.0 \times 10^{10}~M_\odot$ at the cluster redshift), or that the dust emission is not hosted by the cluster galaxy. 	
	We assume the latter, because the PACS source is separated by 2\farcs{}3 in projection from the ALMA source, and attribute the ALMA emission to an optically-obscured background galaxy centered on the \textit{Spitzer} source, and match the PACS source to the cluster galaxy.
	Bright SPIRE emission, unlikely to be from a cluster galaxy, is also detected at the level of $S_{\rm 250~\micron{}} = 23.8 \pm 4.0 ~ \rm mJy$, which strengthens our case for a dusty background galaxy.
	We label the ALMA and SPIRE source an DSFG rather than a cluster member and rename it J054642.09-534544.3 in Table~\ref{tab:alma continuum detections} to reflect its position centroid.
}

\subsection{Additional \textit{Herschel} detections} \label{sec:Herschel detections}

\subparagraph{*J010243.11-491408.6}
{
	is a \ion{Ca}{2} absorber with the appearance of a lenticular galaxy, and is coincident with faint \textit{Herschel}/PACS emission.
	It is bright at SPIRE wavelengths ($S_{\rm 250~\micron{}} = 14.3 \pm 3.5~\rm mJy$; $S_{\rm 350~\micron{}} = 13.3 \pm 4.0~\rm mJy$)
	It lies outside our ALMA coverage, and no significant radio emission is found.
    Given the lack of [\ion{O}{2}] emission and peak flux at SPIRE wavelengths, it is likely that a background dusty star-forming galaxy is responsible for the \textit{Herschel} emission.
}

\subparagraph{J010243.99-491744.4} 
{
	lies outside our \textit{HST} and ALMA field of view (FOV), is bright at \textit{Spitzer} wavelengths (and a LIRG), and is identified as the lone confirmed AGN in our sample on the basis of optical spectroscopy \Sifonp{}.
	Deep {\it Chandra} X-ray observations of J0102 (Hughes et al., in prep.) show an intrinsically absorbed with column density of $N_{\rm H} = 2.0 \pm 0.5 \times 10^{23}$~H atoms~cm$^{-2}$ at the galaxy's redshift of $z=0.87535$) power-law spectrum with a neutral Fe fluorescence line with fitted source-frame energy of $E_{\rm Fe} = 6.37\pm 0.08$ keV) and a 2 -- 10 keV band unabsorbed luminosity of $1.2 \times 10^{44}$~erg~s$^{-1}$, which confirms our identification of this source as an AGN.
}

\subparagraph{J010247.68-491635.7} 
{
	is an [\ion{O}{2}] emission-line galaxy lying outside our \textit{HST} and ALMA FOV.
	It is a LIRG and also an AGN due to its low $q_{\rm IR} = 1.3 \pm 0.4$.
}

\subparagraph{*J010257.79-491519.0}
{
	is a \ion{Ca}{2} absorber $1\farcs{}1$ away from an optical point source (perhaps a quasar).
	We find that the PACS flux density is $S_{\rm 160~\micron{}} = 3.7 \pm 0.2~\rm mJy$ and identify it with faint ($2-3~\sigma$) double-lobed peaks in the radio surface brightness.
	The cluster member is separated from the PACS source ($1\farcs{}8$) and ATCA centroid ($1\farcs{}4$), as well as from the weak \textit{Herschel}/SPIRE emission ($\sim 0.5~{\rm pix} = 3\arcsec$) and the NIR centroid ($1\farcs{}3$).
	We conclude that the long-wavelength emission does not come from the cluster galaxy.
}

\subparagraph{J010302.99-491458.4} 
{
	has a disturbed optical morphology suggesting that this [\ion{O}{2}]-emitting galaxy is undergoing a merger and a burst of star formation.
	It is also detected in the 250~\micron{} SPIRE band (and has weak positive emission in the 350~\micron{} band), but these appear to suffer from source confusion. 
	No ALMA or ATCA emission is detected.
}

\subparagraph{J023540.10-512255.5}
{
	is a \ion{Ca}{2} absorber detected by \textit{Herschel}/PACS.
	Based on its 160~\micron{} flux density and radio non-detection, we conclude that its bolometric luminosity is dominated by star formation ($q_{\rm IR} > 1.9$).
}

\subparagraph{J023542.4-512101.0}
{
	is a \ion{Ca}{2} absorber detected by \textit{Herschel}/PACS.
	No ATCA emission is found.
}

\subparagraph{J023547.60-512029.4}
{
	is a \ion{Ca}{2} absorber with a neighboring cluster member $2\farcs{}5$ to its northwest.
	Its \textit{Herschel}/PACS 100/160~\micron{} counterparts suffer from source blending, so we measure the PACS flux density from the brightest pixel in its surface brightness distribution.
	We do not attempt to deblend the \textit{Herschel} emission.
	No radio counterpart is found.
}

\subparagraph{J023549.20-511920.6}
{
	is a \ion{Ca}{2} absorber located $\sim 5\arcsec$ from a thin, extended source that looks like it may be a lensed background galaxy.
	Its \textit{Herschel} emission is blended, so we measure its PACS flux density from its peak surface brightness.
	It is possible that the nearby source dominates the FIR emission, although we find a slight peak in the radial profile of the \textit{Herschel}/PACS emission coincident with the cluster member.
	No radio counterpart is found.
}

\subparagraph{J023557.20-511820.8}
{
	is a \ion{Ca}{2} absorption-line galaxy with bright \textit{Herschel}/PACS emission.
	There is a $3~\sigma$ peak in the radio surface brightness, but \texttt{IMFIT} does not converge on a fit, so we report no ATCA detection.
}

\subparagraph{J043810.40-542008.1}
{
	is an [\ion{O}{2}] emitter.
	It is one of the brightest \textit{Herschel}/PACS sources in the map of J0438 and the only IRBG in either of our low-$z$ clusters.
	We also find bright SPIRE emission $S_{\rm 250~ \micron{}} = 25.6 \pm 1.6 ~ \rm mJy$ (and a debatable $S_{\rm 350~\micron{}} \sim 8.4 \pm 5.0\rm~ mJy$).
	We also find a radio counterpart and measure $q_{\rm IR} = 1.6 \pm 0.3$.
}
 
\subparagraph{J043824.40-541716.0}
{
	is a \ion{Ca}{2} absorber.
	The \textit{Herschel} emission is blended, but we are able to measure the flux density by using the matched-filter extractor at 100~\micron{} and by measuring a $4~\sigma$ peak in the lower-resolution 160~\micron{} map.
	It also has a slightly fainter neighbor to its south that is not a confirmed cluster member at a projected angular distance of $2\farcs{}4$.
	A radio source is also detected, but it is closer on the sky to the neighboring galaxy.
}

\subparagraph{J054635.39-534541.2}
{
	is an [\ion{O}{2}] emitter that appears to be an inclined star-forming disk galaxy with a partially stripped tail to the west.
	The bulk of the \textit{Herschel} emission is centered slightly to the west and offset by about $2\arcsec$. 
	We see a faint ($\sim 3~\sigma$) radio peak centered $\sim 1\farcs{}5$ to the west of the optical centroid, but are unable to reliably measure its flux density.
}

\section{Dusty star-forming galaxies} \label{sec:DSFGs}

In Table~\ref{tab:alma continuum detections}, we list all new ALMA Band 6 continuum sources detected at $4~\sigma$ significance, and their millimeter (observed frame) flux densities, as well as a brief description of the objects.
ALMA continuum counterparts to cluster galaxies, or sources separated by small projected distances to cluster galaxies, are also included in the table.
CASA \texttt{IMFIT} is used to measure flux densities and uncertainties unless otherwise stated.
Because their redshifts are not known, we cannot determine their dust masses or IR properties.
In Figure~\ref{fig:DSFGs}, we show some examples of these DSFGs along with \textit{Herschel} contours and \textit{HST} imaging.
Aguirre et al., in preparation will discuss the DSFG content of these and other cluster fields in greater detail.

\newpage
\floattable
\begin{deluxetable}{lrrrl}
\tablecaption{ALMA continuum detections} \label{tab:alma continuum detections}
\tablecolumns{5}
\tablehead{
    \colhead{Object} & 
    \colhead{RA} &        
    \colhead{Dec} &  
    \colhead{$S_{\rm 230~GHz}$} &                                         
    \colhead{Description} 
    \\
    \colhead{} &
    \colhead{[$\arcdeg$]} &
    \colhead{[$\arcdeg$]} &
    \colhead{[mJy]} &
    \colhead{}
}

\startdata
J010249.26-491438.1 & 15.705244 & -49.243920 & $1.07 \pm 0.12$ & DSFG \\
J010249.28-491506.8 & 15.705319 & -49.251876 & $3.97 \pm 0.39$ & Strongly lensed DSFG; measured manually \\
J010250.46-491541.6 & 15.710257 & -49.261566 & $0.67 \pm 0.19$ & DSFG \\
J010250.78-491409.3 & 15.711571 & -49.235906 & $1.61 \pm 0.19$ & DSFG \\
J010251.06-491538.8 & 15.712751 & -49.260790 & $3.48 \pm 0.18$ & Bright DSFG \\
J010254.58-491519.9 & 15.727436 & -49.255537 & $0.80 \pm 0.19$ & DSFG \\
J010254.89-491514.6 & 15.728714 & -49.254054 & $4.00 \pm 0.20$ & Bright DSFG; star nearby; La Flaca\tablenotemark{a} \\
J010255.66-491509.0 & 15.731921 & -49.252502 & $9.50 \pm 0.27$ & Bright DSFG; La Flaca\tablenotemark{a} \\
J010258.13-491456.2 & 15.742222 & -49.248937 & $0.82 \pm 0.18$ & DSFG \\
J054627.42-534433.6 & 86.614257 & -53.742676 & $1.21 \pm 0.18$ & Cluster galaxy (CO) \\
J054629.26-534456.9 & 86.621929 & -53.749146 & $1.17 \pm 0.19$ & DSFG \\
J054633.40-534454.5 & 86.639162 & -53.748474 & $0.38 \pm 0.09$ & DSFG; cluster galaxy (Ca II) nearby\tablenotemark{b} \\
J054633.36-534548.6 & 86.638985 & -53.763494 & $0.51 \pm 0.16$ & Strongly lensed DSFG \\
J054634.57-534552.3 & 86.644030 & -53.764525 & $2.84 \pm 0.17$ & Bright DSFG \\
J054636.65-534406.5 & 86.652722 & -53.735132 & $0.34 \pm 0.08$ & Cluster galaxy (Ca II) nearby\tablenotemark{b}\\
J054638.46-534553.6 & 86.660242 & -53.764875 & $0.89 \pm 0.13$ & DSFG \\
J054638.87-534613.8 & 86.661961 & -53.770508 & $1.08 \pm 0.15$ & Cluster galaxy (CO) \\
J054639.19-534519.2 & 86.663296 & -53.755322 & $0.95 \pm 0.16$ & DSFG; near foreground spiral \\
J054639.69-534602.6 & 86.665388 & -53.767383 & $1.02 \pm 0.12$ & DSFG \\
J054640.59-534600.5 & 86.669128 & -53.766816 & $0.64 \pm 0.05$ & DSFG \\
J054641.59-534628.5 & 86.673299 & -53.774584 & $0.89 \pm 0.11$ & No counterpart at optical or NIR wavelengths \\
J054642.09-534544.3 & 86.675374 & -53.762302 & $0.59 \pm 0.09$ & DSFG; cluster galaxy ([O II]) nearby\tablenotemark{b} \\
J054644.14-534608.9 & 86.683928 & -53.769137 & $0.81 \pm 0.11$ & Cluster galaxy (CO) \\
\enddata

\tablenotetext{a}{La Flaca is the nickname of two blended millimeter sources in the J0102 field of view detected by APEX/LABOCA \citep[see Section 4.1.1 of][]{2015ApJ...803...79L}.}
\tablenotetext{b}{See Appendix~\ref{sec:ALMA dust detections}}
\end{deluxetable}

\begin{figure*}
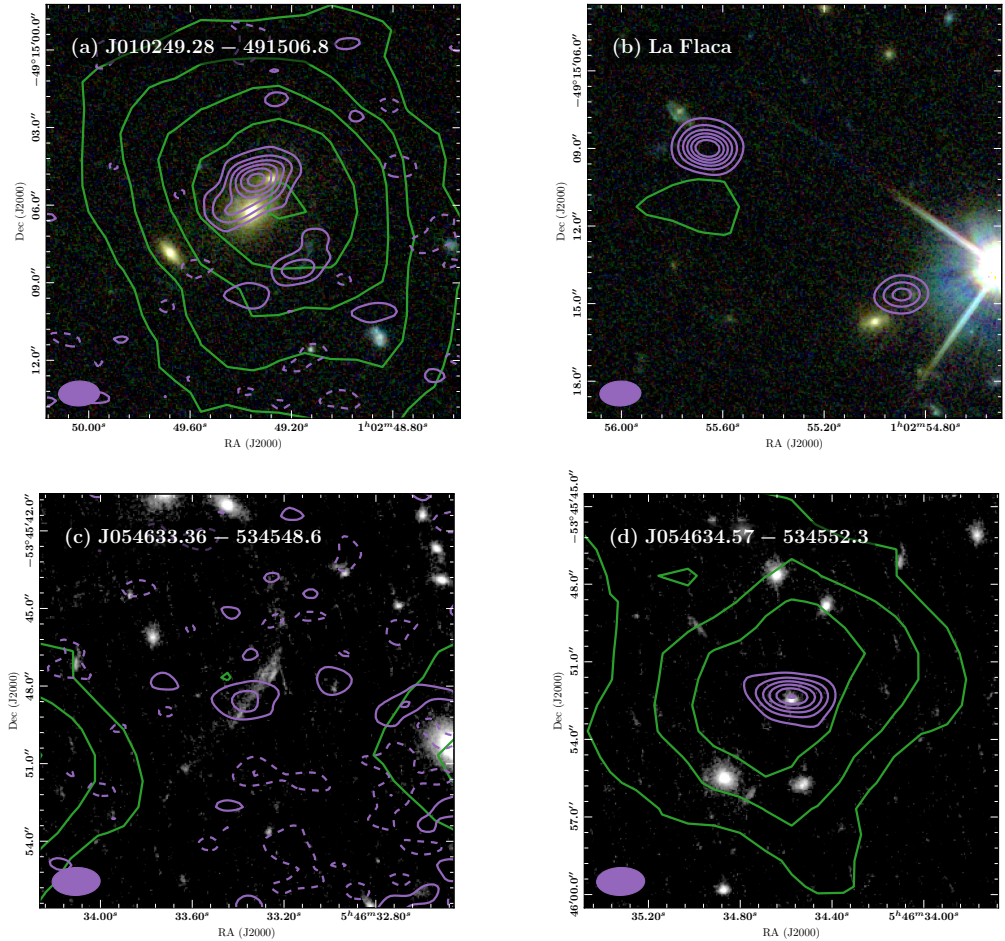

\plottwo{{J010249.28-491506.8}.pdf}{{J010255.31-491511.4_La_Flaca}.pdf}
\plottwo{{J054633.36-534548.6}.pdf}{{J054634.57-534552.3}.pdf}
\caption{
	16\arcsec{} cutouts of \textit{Herschel} and ALMA contours overlaid on \textit{HST} images.
(a) ALMA continuum contours (purple, $\pm 2~\sigma$ levels) and PACS 160~\micron{} contours (green; $\pm 2~\sigma$ levels) overlaid on \textit{HST}/ACS F625W + F775W + F850LP color imaging of a possible strongly lensed system, J010249.28-491506.8. 
Possible arcs of millimeter continuum emission are seen to the northeast and southwest, the latter of which is matched to a blue arc in \textit{HST} imaging.
The early-type galaxies at the center of the image and toward the southeast are confirmed cluster galaxies.
(b) Two bright DSFGs, J010254.89-491514.6 and J010255.66-491509.0, which were detected as a blended source \citep[and named ``La Flaca'';][]{2015ApJ...803...79L}, are shown with ALMA continuum contours (purple, $\pm 8~\sigma$ levels) and PACS 160~\micron{} contours (green; $\pm 2~\sigma$ levels) overlaid on \textit{HST} false-color imaging.
No PACS emission is detected, and whereas no short-wavelength counterpart is detected for J010254.89-491514.6 (southwest), a \textit{Spitzer} counterpart is found for J010255.66-491509.0 (northeast).
(c) A strongly lensed galaxy, J054633.36-534548.6, is shown with ALMA continuum contours (purple, $\pm 2~\sigma$ levels) and PACS 160~\micron{} contours (green; $\pm 2~\sigma$ levels) overlaid on \textit{HST}/F814W grayscale imaging.
No PACS emission is detected with the DSFG.
(d) A bright submillimeter and FIR source, J054634.57-534552.3, is shown with ALMA continuum contours (purple, $\pm 4~\sigma$ levels) and PACS 160~\micron{} contours (green; $\pm 2~\sigma$ levels) overlaid on \textit{HST}/F814W grayscale imaging.
}
\label{fig:DSFGs}
\end{figure*}

\bibliographystyle{apj}
\bibliography{bibliography} 

\begin{thebibliography}{}
\expandafter\ifx\csname natexlab\endcsname\relax\def\natexlab#1{#1}\fi

\bibitem[{{Alberts} {et~al.}(2014){Alberts}, {Pope}, {Brodwin}, {Atlee}, {Lin},
  {Dey}, {Eisenhardt}, {Gettings}, {Gonzalez}, {Jannuzi}, {Mancone},
  {Moustakas}, {Snyder}, {Stanford}, {Stern}, {Weiner}, \&
  {Zeimann}}]{2014MNRAS.437..437A}
{Alberts}, S., {Pope}, A., {Brodwin}, M., {et~al.} 2014, \mnras, 437, 437

\bibitem[{{Alberts} {et~al.}(2016){Alberts}, {Pope}, {Brodwin}, {Chung},
  {Cybulski}, {Dey}, {Eisenhardt}, {Galametz}, {Gonzalez}, {Jannuzi},
  {Stanford}, {Snyder}, {Stern}, \& {Zeimann}}]{2016ApJ...825...72A}
---. 2016, \apj, 825, 72

\bibitem[{{Astropy Collaboration} {et~al.}(2013){Astropy Collaboration},
  {Robitaille}, {Tollerud}, {Greenfield}, {Droettboom}, {Bray}, {Aldcroft},
  {Davis}, {Ginsburg}, {Price-Whelan}, {Kerzendorf}, {Conley}, {Crighton},
  {Barbary}, {Muna}, {Ferguson}, {Grollier}, {Parikh}, {Nair}, {Unther},
  {Deil}, {Woillez}, {Conseil}, {Kramer}, {Turner}, {Singer}, {Fox}, {Weaver},
  {Zabalza}, {Edwards}, {Azalee Bostroem}, {Burke}, {Casey}, {Crawford},
  {Dencheva}, {Ely}, {Jenness}, {Labrie}, {Lim}, {Pierfederici}, {Pontzen},
  {Ptak}, {Refsdal}, {Servillat}, \& {Streicher}}]{2013A&A...558A..33A}
{Astropy Collaboration}, {Robitaille}, T.~P., {Tollerud}, E.~J., {et~al.} 2013,
  \aap, 558, A33

\bibitem[{{Balogh} {et~al.}(2017){Balogh}, {Gilbank}, {Muzzin}, {Rudnick},
  {Cooper}, {Lidman}, {Biviano}, {Demarco}, {McGee}, {Nantais}, {Noble}, {Old},
  {Wilson}, {Yee}, {Bellhouse}, {Cerulo}, {Chan}, {Pintos-Castro}, {Simpson},
  {van der Burg}, {Zaritsky}, {Ziparo}, {Alonso}, {Bower}, {De Lucia},
  {Finoguenov}, {Lambas}, {Muriel}, {Parker}, {Rettura}, {Valotto}, \&
  {Wetzel}}]{2017MNRAS.470.4168B}
{Balogh}, M.~L., {Gilbank}, D.~G., {Muzzin}, A., {et~al.} 2017, \mnras, 470,
  4168

\bibitem[{{Bekki}(1999)}]{1999ApJ...510L..15B}
{Bekki}, K. 1999, \apjl, 510, L15

\bibitem[{{Bell}(2003)}]{2003ApJ...586..794B}
{Bell}, E.~F. 2003, \apj, 586, 794

\bibitem[{{Bertin} \& {Arnouts}(1996)}]{1996A&AS..117..393B}
{Bertin}, E., \& {Arnouts}, S. 1996, \aaps, 117, 393

\bibitem[{{Bleem} {et~al.}(2015){Bleem}, {Stalder}, {de Haan}, {Aird}, {Allen},
  {Applegate}, {Ashby}, {Bautz}, {Bayliss}, {Benson}, {Bocquet}, {Brodwin},
  {Carlstrom}, {Chang}, {Chiu}, {Cho}, {Clocchiatti}, {Crawford}, {Crites},
  {Desai}, {Dietrich}, {Dobbs}, {Foley}, {Forman}, {George}, {Gladders},
  {Gonzalez}, {Halverson}, {Hennig}, {Hoekstra}, {Holder}, {Holzapfel},
  {Hrubes}, {Jones}, {Keisler}, {Knox}, {Lee}, {Leitch}, {Liu}, {Lueker},
  {Luong-Van}, {Mantz}, {Marrone}, {McDonald}, {McMahon}, {Meyer}, {Mocanu},
  {Mohr}, {Murray}, {Padin}, {Pryke}, {Reichardt}, {Rest}, {Ruel}, {Ruhl},
  {Saliwanchik}, {Saro}, {Sayre}, {Schaffer}, {Schrabback}, {Shirokoff},
  {Song}, {Spieler}, {Stanford}, {Staniszewski}, {Stark}, {Story}, {Stubbs},
  {Vanderlinde}, {Vieira}, {Vikhlinin}, {Williamson}, {Zahn}, \&
  {Zenteno}}]{2015ApJS..216...27B}
{Bleem}, L.~E., {Stalder}, B., {de Haan}, T., {et~al.} 2015, \apjs, 216, 27

\bibitem[{{Bolatto} {et~al.}(2013){Bolatto}, {Wolfire}, \&
  {Leroy}}]{2013ARAA..51..207B}
{Bolatto}, A.~D., {Wolfire}, M., \& {Leroy}, A.~K. 2013, \araa, 51, 207

\bibitem[{{Boselli} \& {Gavazzi}(2006)}]{2006PASP..118..517B}
{Boselli}, A., \& {Gavazzi}, G. 2006, \pasp, 118, 517

\bibitem[{{Brodwin} {et~al.}(2013){Brodwin}, {Stanford}, {Gonzalez}, {Zeimann},
  {Snyder}, {Mancone}, {Pope}, {Eisenhardt}, {Stern}, {Alberts}, {Ashby},
  {Brown}, {Chary}, {Dey}, {Galametz}, {Gettings}, {Jannuzi}, {Miller},
  {Moustakas}, \& {Moustakas}}]{2013ApJ...779..138B}
{Brodwin}, M., {Stanford}, S.~A., {Gonzalez}, A.~H., {et~al.} 2013, \apj, 779,
  138

\bibitem[{{Bruzual} \& {Charlot}(2003)}]{2003MNRAS.344.1000B}
{Bruzual}, G., \& {Charlot}, S. 2003, \mnras, 344, 1000

\bibitem[{{Butcher} \& {Oemler}(1978)}]{1978ApJ...219...18B}
{Butcher}, H., \& {Oemler}, Jr., A. 1978, \apj, 219, 18

\bibitem[{{Caldwell} {et~al.}(1993){Caldwell}, {Rose}, {Sharples}, {Ellis}, \&
  {Bower}}]{1993AJ....106..473C}
{Caldwell}, N., {Rose}, J.~A., {Sharples}, R.~M., {Ellis}, R.~S., \& {Bower},
  R.~G. 1993, \aj, 106, 473

\bibitem[{{Carilli} \& {Walter}(2013)}]{2013ARA&A..51..105C}
{Carilli}, C.~L., \& {Walter}, F. 2013, \araa, 51, 105

\bibitem[{{Carilli} \& {Wang}(2006)}]{2006AJ....131.2763C}
{Carilli}, C.~L., \& {Wang}, R. 2006, \aj, 131, 2763

\bibitem[{{Chabrier}(2003)}]{2003PASP..115..763C}
{Chabrier}, G. 2003, \pasp, 115, 763

\bibitem[{{Chary} \& {Elbaz}(2001)}]{2001ApJ...556..562C}
{Chary}, R., \& {Elbaz}, D. 2001, \apj, 556, 562

\bibitem[{{Chengalur} {et~al.}(2001){Chengalur}, {Braun}, \&
  {Wieringa}}]{2001A&A...372..768C}
{Chengalur}, J.~N., {Braun}, R., \& {Wieringa}, M. 2001, \aap, 372, 768

\bibitem[{{Chung} {et~al.}(2009){Chung}, {van Gorkom}, {Kenney}, {Crowl}, \&
  {Vollmer}}]{2009AJ....138.1741C}
{Chung}, A., {van Gorkom}, J.~H., {Kenney}, J.~D.~P., {Crowl}, H., \&
  {Vollmer}, B. 2009, \aj, 138, 1741

\bibitem[{{Chung} {et~al.}(2011){Chung}, {Eisenhardt}, {Gonzalez}, {Stanford},
  {Brodwin}, {Stern}, \& {Jarrett}}]{2011ApJ...743...34C}
{Chung}, S.~M., {Eisenhardt}, P.~R., {Gonzalez}, A.~H., {et~al.} 2011, \apj,
  743, 34

\bibitem[{{Coia} {et~al.}(2005){Coia}, {McBreen}, {Metcalfe}, {Biviano},
  {Altieri}, {Ott}, {Fort}, {Kneib}, {Mellier}, {Miville-Desch{\^e}nes},
  {O'Halloran}, \& {Sanchez-Fernandez}}]{2005A&A...431..433C}
{Coia}, D., {McBreen}, B., {Metcalfe}, L., {et~al.} 2005, \aap, 431, 433

\bibitem[{{Comparat} {et~al.}(2015){Comparat}, {Richard}, {Kneib}, {Ilbert},
  {Gonzalez-Perez}, {Tresse}, {Zoubian}, {Arnouts}, {Brownstein}, {Baugh},
  {Delubac}, {Ealet}, {Escoffier}, {Ge}, {Jullo}, {Lacey}, {Ross}, {Schlegel},
  {Schneider}, {Steele}, {Tasca}, {Yeche}, {Lesser}, {Jiang}, {Jing}, {Fan},
  {Fan}, {Ma}, {Nie}, {Wang}, {Wu}, {Zhang}, {Zhou}, {Zhou}, \&
  {Zou}}]{2015A&A...575A..40C}
{Comparat}, J., {Richard}, J., {Kneib}, J.-P., {et~al.} 2015, \aap, 575, A40

\bibitem[{{Condon} {et~al.}(1991){Condon}, {Anderson}, \&
  {Helou}}]{1991ApJ...376...95C}
{Condon}, J.~J., {Anderson}, M.~L., \& {Helou}, G. 1991, \apj, 376, 95

\bibitem[{{Condon} {et~al.}(2002){Condon}, {Cotton}, \&
  {Broderick}}]{2002AJ....124..675C}
{Condon}, J.~J., {Cotton}, W.~D., \& {Broderick}, J.~J. 2002, \aj, 124, 675

\bibitem[{{Cowie} \& {Songaila}(1977)}]{1977Natur.266..501C}
{Cowie}, L.~L., \& {Songaila}, A. 1977, \nat, 266, 501

\bibitem[{{Daddi} {et~al.}(2007){Daddi}, {Dickinson}, {Morrison}, {Chary},
  {Cimatti}, {Elbaz}, {Frayer}, {Renzini}, {Pope}, {Alexander}, {Bauer},
  {Giavalisco}, {Huynh}, {Kurk}, \& {Mignoli}}]{2007ApJ...670..156D}
{Daddi}, E., {Dickinson}, M., {Morrison}, G., {et~al.} 2007, \apj, 670, 156

\bibitem[{{de Jong} {et~al.}(1985){de Jong}, {Klein}, {Wielebinski}, \&
  {Wunderlich}}]{1985A&A...147L...6D}
{de Jong}, T., {Klein}, U., {Wielebinski}, R., \& {Wunderlich}, E. 1985, \aap,
  147, L6

\bibitem[{{Downes} \& {Solomon}(1998)}]{1998ApJ...507..615D}
{Downes}, D., \& {Solomon}, P.~M. 1998, \apj, 507, 615

\bibitem[{{Dressler}(1980)}]{1980ApJ...236..351D}
{Dressler}, A. 1980, \apj, 236, 351

\bibitem[{{Dunlop} {et~al.}(2017){Dunlop}, {McLure}, {Biggs}, {Geach},
  {Micha{\l}owski}, {Ivison}, {Rujopakarn}, {van Kampen}, {Kirkpatrick},
  {Pope}, {Scott}, {Swinbank}, {Targett}, {Aretxaga}, {Austermann}, {Best},
  {Bruce}, {Chapin}, {Charlot}, {Cirasuolo}, {Coppin}, {Ellis}, {Finkelstein},
  {Hayward}, {Hughes}, {Ibar}, {Jagannathan}, {Khochfar}, {Koprowski},
  {Narayanan}, {Nyland}, {Papovich}, {Peacock}, {Rieke}, {Robertson},
  {Vernstrom}, {Werf}, {Wilson}, \& {Yun}}]{2017MNRAS.466..861D}
{Dunlop}, J.~S., {McLure}, R.~J., {Biggs}, A.~D., {et~al.} 2017, \mnras, 466,
  861

\bibitem[{{Dwarakanath} \& {Owen}(1999)}]{1999AJ....118..625D}
{Dwarakanath}, K.~S., \& {Owen}, F.~N. 1999, \aj, 118, 625

\bibitem[{{Eisenhardt} {et~al.}(2008){Eisenhardt}, {Brodwin}, {Gonzalez},
  {Stanford}, {Stern}, {Barmby}, {Brown}, {Dawson}, {Dey}, {Doi}, {Galametz},
  {Jannuzi}, {Kochanek}, {Meyers}, {Morokuma}, \&
  {Moustakas}}]{2008ApJ...684..905E}
{Eisenhardt}, P.~R.~M., {Brodwin}, M., {Gonzalez}, A.~H., {et~al.} 2008, \apj,
  684, 905

\bibitem[{{Elbaz} {et~al.}(2007){Elbaz}, {Daddi}, {Le Borgne}, {Dickinson},
  {Alexander}, {Chary}, {Starck}, {Brandt}, {Kitzbichler}, {MacDonald},
  {Nonino}, {Popesso}, {Stern}, \& {Vanzella}}]{2007A&A...468...33E}
{Elbaz}, D., {Daddi}, E., {Le Borgne}, D., {et~al.} 2007, \aap, 468, 33

\bibitem[{{Elbaz} {et~al.}(2010){Elbaz}, {Hwang}, {Magnelli}, {Daddi},
  {Aussel}, {Altieri}, {Amblard}, {Andreani}, {Arumugam}, {Auld}, {Babbedge},
  {Berta}, {Blain}, {Bock}, {Bongiovanni}, {Boselli}, {Buat}, {Burgarella},
  {Castro-Rodriguez}, {Cava}, {Cepa}, {Chanial}, {Chary}, {Cimatti},
  {Clements}, {Conley}, {Conversi}, {Cooray}, {Dickinson}, {Dominguez},
  {Dowell}, {Dunlop}, {Dwek}, {Eales}, {Farrah}, {F{\"o}rster Schreiber},
  {Fox}, {Franceschini}, {Gear}, {Genzel}, {Glenn}, {Griffin}, {Gruppioni},
  {Halpern}, {Hatziminaoglou}, {Ibar}, {Isaak}, {Ivison}, {Lagache}, {Le
  Borgne}, {Le Floc'h}, {Levenson}, {Lu}, {Lutz}, {Madden}, {Maffei}, {Magdis},
  {Mainetti}, {Maiolino}, {Marchetti}, {Mortier}, {Nguyen}, {Nordon},
  {O'Halloran}, {Okumura}, {Oliver}, {Omont}, {Page}, {Panuzzo},
  {Papageorgiou}, {Pearson}, {Perez Fournon}, {P{\'e}rez Garc{\'{\i}}a},
  {Poglitsch}, {Pohlen}, {Popesso}, {Pozzi}, {Rawlings}, {Rigopoulou},
  {Riguccini}, {Rizzo}, {Rodighiero}, {Roseboom}, {Rowan-Robinson},
  {Saintonge}, {Sanchez Portal}, {Santini}, {Sauvage}, {Schulz}, {Scott},
  {Seymour}, {Shao}, {Shupe}, {Smith}, {Stevens}, {Sturm}, {Symeonidis},
  {Tacconi}, {Trichas}, {Tugwell}, {Vaccari}, {Valtchanov}, {Vieira},
  {Vigroux}, {Wang}, {Ward}, {Wright}, {Xu}, \& {Zemcov}}]{2010A&A...518L..29E}
{Elbaz}, D., {Hwang}, H.~S., {Magnelli}, B., {et~al.} 2010, \aap, 518, L29

\bibitem[{{Finn} {et~al.}(2010){Finn}, {Desai}, {Rudnick}, {Poggianti}, {Bell},
  {Hinz}, {Jablonka}, {Milvang-Jensen}, {Moustakas}, {Rines}, \&
  {Zaritsky}}]{2010ApJ...720...87F}
{Finn}, R.~A., {Desai}, V., {Rudnick}, G., {et~al.} 2010, \apj, 720, 87

\bibitem[{{Foreman-Mackey} {et~al.}(2013){Foreman-Mackey}, {Hogg}, {Lang}, \&
  {Goodman}}]{2013PASP..125..306F}
{Foreman-Mackey}, D., {Hogg}, D.~W., {Lang}, D., \& {Goodman}, J. 2013, \pasp,
  125, 306

\bibitem[{{Galametz} {et~al.}(2009){Galametz}, {Madden}, {Galliano}, {Hony},
  {Schuller}, {Beelen}, {Bendo}, {Sauvage}, {Lundgren}, \&
  {Billot}}]{2009A&A...508..645G}
{Galametz}, M., {Madden}, S., {Galliano}, F., {et~al.} 2009, \aap, 508, 645

\bibitem[{{Geach} {et~al.}(2009){Geach}, {Smail}, {Coppin}, {Moran}, {Edge}, \&
  {Ellis}}]{2009MNRAS.395L..62G}
{Geach}, J.~E., {Smail}, I., {Coppin}, K., {et~al.} 2009, \mnras, 395, L62

\bibitem[{{Geach} {et~al.}(2011){Geach}, {Smail}, {Moran}, {MacArthur},
  {Lagos}, \& {Edge}}]{2011ApJ...730L..19G}
{Geach}, J.~E., {Smail}, I., {Moran}, S.~M., {et~al.} 2011, \apjl, 730, L19

\bibitem[{{Geach} {et~al.}(2006){Geach}, {Smail}, {Ellis}, {Moran}, {Smith},
  {Treu}, {Kneib}, {Edge}, \& {Kodama}}]{2006ApJ...649..661G}
{Geach}, J.~E., {Smail}, I., {Ellis}, R.~S., {et~al.} 2006, \apj, 649, 661

\bibitem[{{Genzel} {et~al.}(2015){Genzel}, {Tacconi}, {Lutz}, {Saintonge},
  {Berta}, {Magnelli}, {Combes}, {Garc{\'{\i}}a-Burillo}, {Neri}, {Bolatto},
  {Contini}, {Lilly}, {Boissier}, {Boone}, {Bouch{\'e}}, {Bournaud}, {Burkert},
  {Carollo}, {Colina}, {Cooper}, {Cox}, {Feruglio}, {F{\"o}rster Schreiber},
  {Freundlich}, {Gracia-Carpio}, {Juneau}, {Kovac}, {Lippa}, {Naab}, {Salome},
  {Renzini}, {Sternberg}, {Walter}, {Weiner}, {Weiss}, \&
  {Wuyts}}]{2015ApJ...800...20G}
{Genzel}, R., {Tacconi}, L.~J., {Lutz}, D., {et~al.} 2015, \apj, 800, 20

\bibitem[{{Gerin} \& {Phillips}(2000)}]{2000ApJ...537..644G}
{Gerin}, M., \& {Phillips}, T.~G. 2000, \apj, 537, 644

\bibitem[{{Griffin} \& {Orton}(1993)}]{1993Icar..105..537G}
{Griffin}, M.~J., \& {Orton}, G.~S. 1993, \icarus, 105, 537

\bibitem[{{Griffin} {et~al.}(2010){Griffin}, {Abergel}, {Abreu}, {Ade},
  {Andr{\'e}}, {Augueres}, {Babbedge}, {Bae}, {Baillie}, {Baluteau}, {Barlow},
  {Bendo}, {Benielli}, {Bock}, {Bonhomme}, {Brisbin}, {Brockley-Blatt},
  {Caldwell}, {Cara}, {Castro-Rodriguez}, {Cerulli}, {Chanial}, {Chen},
  {Clark}, {Clements}, {Clerc}, {Coker}, {Communal}, {Conversi}, {Cox},
  {Crumb}, {Cunningham}, {Daly}, {Davis}, {de Antoni}, {Delderfield}, {Devin},
  {di Giorgio}, {Didschuns}, {Dohlen}, {Donati}, {Dowell}, {Dowell}, {Duband},
  {Dumaye}, {Emery}, {Ferlet}, {Ferrand}, {Fontignie}, {Fox}, {Franceschini},
  {Frerking}, {Fulton}, {Garcia}, {Gastaud}, {Gear}, {Glenn}, {Goizel},
  {Griffin}, {Grundy}, {Guest}, {Guillemet}, {Hargrave}, {Harwit}, {Hastings},
  {Hatziminaoglou}, {Herman}, {Hinde}, {Hristov}, {Huang}, {Imhof}, {Isaak},
  {Israelsson}, {Ivison}, {Jennings}, {Kiernan}, {King}, {Lange}, {Latter},
  {Laurent}, {Laurent}, {Leeks}, {Lellouch}, {Levenson}, {Li}, {Li},
  {Lilienthal}, {Lim}, {Liu}, {Lu}, {Madden}, {Mainetti}, {Marliani}, {McKay},
  {Mercier}, {Molinari}, {Morris}, {Moseley}, {Mulder}, {Mur}, {Naylor},
  {Nguyen}, {O'Halloran}, {Oliver}, {Olofsson}, {Olofsson}, {Orfei}, {Page},
  {Pain}, {Panuzzo}, {Papageorgiou}, {Parks}, {Parr-Burman}, {Pearce},
  {Pearson}, {P{\'e}rez-Fournon}, {Pinsard}, {Pisano}, {Podosek}, {Pohlen},
  {Polehampton}, {Pouliquen}, {Rigopoulou}, {Rizzo}, {Roseboom}, {Roussel},
  {Rowan-Robinson}, {Rownd}, {Saraceno}, {Sauvage}, {Savage}, {Savini},
  {Sawyer}, {Scharmberg}, {Schmitt}, {Schneider}, {Schulz}, {Schwartz},
  {Shafer}, {Shupe}, {Sibthorpe}, {Sidher}, {Smith}, {Smith}, {Smith},
  {Spencer}, {Stobie}, {Sudiwala}, {Sukhatme}, {Surace}, {Stevens}, {Swinyard},
  {Trichas}, {Tourette}, {Triou}, {Tseng}, {Tucker}, {Turner}, {Vaccari},
  {Valtchanov}, {Vigroux}, {Virique}, {Voellmer}, {Walker}, {Ward}, {Waskett},
  {Weilert}, {Wesson}, {White}, {Whitehouse}, {Wilson}, {Winter}, {Woodcraft},
  {Wright}, {Xu}, {Zavagno}, {Zemcov}, {Zhang}, \&
  {Zonca}}]{2010A&A...518L...3G}
{Griffin}, M.~J., {Abergel}, A., {Abreu}, A., {et~al.} 2010, \aap, 518, L3

\bibitem[{{Guglielmo} {et~al.}(2015){Guglielmo}, {Poggianti}, {Moretti},
  {Fritz}, {Calvi}, {Vulcani}, {Fasano}, \&
  {Paccagnella}}]{2015MNRAS.450.2749G}
{Guglielmo}, V., {Poggianti}, B.~M., {Moretti}, A., {et~al.} 2015, \mnras, 450,
  2749

\bibitem[{{Gunn} \& {Gott}(1972)}]{1972ApJ...176....1G}
{Gunn}, J.~E., \& {Gott}, III, J.~R. 1972, \apj, 176, 1

\bibitem[{{Hafok} \& {Stutzki}(2003)}]{2003A&A...398..959H}
{Hafok}, H., \& {Stutzki}, J. 2003, \aap, 398, 959

\bibitem[{{Haines} {et~al.}(2009){Haines}, {Smith}, {Egami}, {Ellis}, {Moran},
  {Sanderson}, {Merluzzi}, {Busarello}, \& {Smith}}]{2009ApJ...704..126H}
{Haines}, C.~P., {Smith}, G.~P., {Egami}, E., {et~al.} 2009, \apj, 704, 126

\bibitem[{{Hayashi} {et~al.}(2010){Hayashi}, {Kodama}, {Koyama}, {Tanaka},
  {Shimasaku}, \& {Okamura}}]{2010MNRAS.402.1980H}
{Hayashi}, M., {Kodama}, T., {Koyama}, Y., {et~al.} 2010, \mnras, 402, 1980

\bibitem[{{Helou} {et~al.}(1985){Helou}, {Soifer}, \&
  {Rowan-Robinson}}]{1985ApJ...298L...7H}
{Helou}, G., {Soifer}, B.~T., \& {Rowan-Robinson}, M. 1985, \apjl, 298, L7

\bibitem[{{Hickox} {et~al.}(2009){Hickox}, {Jones}, {Forman}, {Murray},
  {Kochanek}, {Eisenstein}, {Jannuzi}, {Dey}, {Brown}, {Stern}, {Eisenhardt},
  {Gorjian}, {Brodwin}, {Narayan}, {Cool}, {Kenter}, {Caldwell}, \&
  {Anderson}}]{2009ApJ...696..891H}
{Hickox}, R.~C., {Jones}, C., {Forman}, W.~R., {et~al.} 2009, \apj, 696, 891

\bibitem[{{Hilton} {et~al.}(2010){Hilton}, {Lloyd-Davies}, {Stanford}, {Stott},
  {Collins}, {Romer}, {Hosmer}, {Hoyle}, {Kay}, {Liddle}, {Mehrtens}, {Miller},
  {Sahl{\'e}n}, \& {Viana}}]{2010ApJ...718..133H}
{Hilton}, M., {Lloyd-Davies}, E., {Stanford}, S.~A., {et~al.} 2010, \apj, 718,
  133

\bibitem[{{Hilton} {et~al.}(2013){Hilton}, {Hasselfield}, {Sif{\'o}n}, {Baker},
  {Barrientos}, {Battaglia}, {Bond}, {Crichton}, {Das}, {Devlin}, {Gralla},
  {Hajian}, {Hincks}, {Hughes}, {Infante}, {Irwin}, {Kosowsky}, {Lin},
  {Marriage}, {Marsden}, {Menanteau}, {Moodley}, {Niemack}, {Nolta}, {Page},
  {Reese}, {Sievers}, {Spergel}, \& {Wollack}}]{2013MNRAS.435.3469H}
{Hilton}, M., {Hasselfield}, M., {Sif{\'o}n}, C., {et~al.} 2013, \mnras, 435,
  3469

\bibitem[{{Hopkins} {et~al.}(2008){Hopkins}, {Hernquist}, {Cox}, \& {Kere{\v
  s}}}]{2008ApJS..175..356H}
{Hopkins}, P.~F., {Hernquist}, L., {Cox}, T.~J., \& {Kere{\v s}}, D. 2008,
  \apjs, 175, 356

\bibitem[{Hunter(2007)}]{matplotlib}
Hunter, J.~D. 2007, Computing In Science \& Engineering, 9, 90

\bibitem[{{Ivison} {et~al.}(2010){Ivison}, {Magnelli}, {Ibar}, {Andreani},
  {Elbaz}, {Altieri}, {Amblard}, {Arumugam}, {Auld}, {Aussel}, {Babbedge},
  {Berta}, {Blain}, {Bock}, {Bongiovanni}, {Boselli}, {Buat}, {Burgarella},
  {Castro-Rodr{\'{\i}}guez}, {Cava}, {Cepa}, {Chanial}, {Cimatti}, {Cirasuolo},
  {Clements}, {Conley}, {Conversi}, {Cooray}, {Daddi}, {Dominguez}, {Dowell},
  {Dwek}, {Eales}, {Farrah}, {F{\"o}rster Schreiber}, {Fox}, {Franceschini},
  {Gear}, {Genzel}, {Glenn}, {Griffin}, {Gruppioni}, {Halpern},
  {Hatziminaoglou}, {Isaak}, {Lagache}, {Levenson}, {Lu}, {Lutz}, {Madden},
  {Maffei}, {Magdis}, {Mainetti}, {Maiolino}, {Marchetti}, {Morrison},
  {Mortier}, {Nguyen}, {Nordon}, {O'Halloran}, {Oliver}, {Omont}, {Owen},
  {Page}, {Panuzzo}, {Papageorgiou}, {Pearson}, {P{\'e}rez-Fournon}, {P{\'e}rez
  Garc{\'{\i}}a}, {Poglitsch}, {Pohlen}, {Popesso}, {Pozzi}, {Rawlings},
  {Raymond}, {Rigopoulou}, {Riguccini}, {Rizzo}, {Rodighiero}, {Roseboom},
  {Rowan-Robinson}, {Saintonge}, {Sanchez Portal}, {Santini}, {Schulz},
  {Scott}, {Seymour}, {Shao}, {Shupe}, {Smith}, {Stevens}, {Sturm},
  {Symeonidis}, {Tacconi}, {Trichas}, {Tugwell}, {Vaccari}, {Valtchanov},
  {Vieira}, {Vigroux}, {Wang}, {Ward}, {Wright}, {Xu}, \&
  {Zemcov}}]{2010A&A...518L..31I}
{Ivison}, R.~J., {Magnelli}, B., {Ibar}, E., {et~al.} 2010, \aap, 518, L31

\bibitem[{{Jablonka} {et~al.}(2013){Jablonka}, {Combes}, {Rines}, {Finn}, \&
  {Welch}}]{2013A&A...557A.103J}
{Jablonka}, P., {Combes}, F., {Rines}, K., {Finn}, R., \& {Welch}, T. 2013,
  \aap, 557, A103

\bibitem[{{Jee} {et~al.}(2014){Jee}, {Hughes}, {Menanteau}, {Sif{\'o}n},
  {Mandelbaum}, {Barrientos}, {Infante}, \& {Ng}}]{2014ApJ...785...20J}
{Jee}, M.~J., {Hughes}, J.~P., {Menanteau}, F., {et~al.} 2014, \apj, 785, 20

\bibitem[{{Kauffmann} {et~al.}(2004){Kauffmann}, {White}, {Heckman},
  {M{\'e}nard}, {Brinchmann}, {Charlot}, {Tremonti}, \&
  {Brinkmann}}]{2004MNRAS.353..713K}
{Kauffmann}, G., {White}, S.~D.~M., {Heckman}, T.~M., {et~al.} 2004, \mnras,
  353, 713

\bibitem[{{Kenney} {et~al.}(2004){Kenney}, {van Gorkom}, \&
  {Vollmer}}]{2004AJ....127.3361K}
{Kenney}, J.~D.~P., {van Gorkom}, J.~H., \& {Vollmer}, B. 2004, \aj, 127, 3361

\bibitem[{{Kennicutt} \& {Evans}(2012)}]{2012ARA&A..50..531K}
{Kennicutt}, R.~C., \& {Evans}, N.~J. 2012, \araa, 50, 531

\bibitem[{{Kennicutt}(1998)}]{1998ARA&A..36..189K}
{Kennicutt}, Jr., R.~C. 1998, \araa, 36, 189

\bibitem[{{Kewley} {et~al.}(2004){Kewley}, {Geller}, \&
  {Jansen}}]{2004AJ....127.2002K}
{Kewley}, L.~J., {Geller}, M.~J., \& {Jansen}, R.~A. 2004, \aj, 127, 2002

\bibitem[{{Kirkpatrick} {et~al.}(2015){Kirkpatrick}, {Pope}, {Sajina},
  {Roebuck}, {Yan}, {Armus}, {D{\'{\i}}az-Santos}, \&
  {Stierwalt}}]{2015ApJ...814....9K}
{Kirkpatrick}, A., {Pope}, A., {Sajina}, A., {et~al.} 2015, \apj, 814, 9

\bibitem[{{Kirkpatrick} {et~al.}(2012){Kirkpatrick}, {Pope}, {Alexander},
  {Charmandaris}, {Daddi}, {Dickinson}, {Elbaz}, {Gabor}, {Hwang}, {Ivison},
  {Mullaney}, {Pannella}, {Scott}, {Altieri}, {Aussel}, {Bournaud}, {Buat},
  {Coia}, {Dannerbauer}, {Dasyra}, {Kartaltepe}, {Leiton}, {Lin}, {Magdis},
  {Magnelli}, {Morrison}, {Popesso}, \& {Valtchanov}}]{2012ApJ...759..139K}
{Kirkpatrick}, A., {Pope}, A., {Alexander}, D.~M., {et~al.} 2012, \apj, 759,
  139

\bibitem[{Kluyver {et~al.}(2016)Kluyver, Ragan-Kelley, P\'erez, Granger,
  Bussonnier, Frederic, Kelley, Hamrick, Grout, Corlay, Ivanov, Avila, Abdalla,
  Willing, \& Team}]{10.3233/978-1-61499-649-1-87}
Kluyver, T., Ragan-Kelley, B., P\'erez, F., {et~al.} 2016, Jupyter Notebooks --
  a publishing format for reproducible computational workflows (IOS Press
  Ebooks), 87--90, doi:10.3233/978-1-61499-649-1-87

\bibitem[{{Koopmann} \& {Kenney}(2004{\natexlab{a}})}]{2004ApJ...613..866K}
{Koopmann}, R.~A., \& {Kenney}, J.~D.~P. 2004{\natexlab{a}}, \apj, 613, 866

\bibitem[{{Koopmann} \& {Kenney}(2004{\natexlab{b}})}]{2004ApJ...613..851K}
---. 2004{\natexlab{b}}, \apj, 613, 851

\bibitem[{{Koyama} {et~al.}(2013){Koyama}, {Smail}, {Kurk}, {Geach}, {Sobral},
  {Kodama}, {Nakata}, {Swinbank}, {Best}, {Hayashi}, \&
  {Tadaki}}]{2013MNRAS.434..423K}
{Koyama}, Y., {Smail}, I., {Kurk}, J., {et~al.} 2013, \mnras, 434, 423

\bibitem[{{Larson} {et~al.}(1980){Larson}, {Tinsley}, \&
  {Caldwell}}]{1980ApJ...237..692L}
{Larson}, R.~B., {Tinsley}, B.~M., \& {Caldwell}, C.~N. 1980, \apj, 237, 692

\bibitem[{{Le Floc'h} {et~al.}(2005){Le Floc'h}, {Papovich}, {Dole}, {Bell},
  {Lagache}, {Rieke}, {Egami}, {P{\'e}rez-Gonz{\'a}lez}, {Alonso-Herrero},
  {Rieke}, {Blaylock}, {Engelbracht}, {Gordon}, {Hines}, {Misselt}, {Morrison},
  \& {Mould}}]{2005ApJ...632..169L}
{Le Floc'h}, E., {Papovich}, C., {Dole}, H., {et~al.} 2005, \apj, 632, 169

\bibitem[{{Lewis} {et~al.}(2002){Lewis}, {Balogh}, {De Propris}, {Couch},
  {Bower}, {Offer}, {Bland-Hawthorn}, {Baldry}, {Baugh}, {Bridges}, {Cannon},
  {Cole}, {Colless}, {Collins}, {Cross}, {Dalton}, {Driver}, {Efstathiou},
  {Ellis}, {Frenk}, {Glazebrook}, {Hawkins}, {Jackson}, {Lahav}, {Lumsden},
  {Maddox}, {Madgwick}, {Norberg}, {Peacock}, {Percival}, {Peterson},
  {Sutherland}, \& {Taylor}}]{2002MNRAS.334..673L}
{Lewis}, I., {Balogh}, M., {De Propris}, R., {et~al.} 2002, \mnras, 334, 673

\bibitem[{{Lindner} {et~al.}(2011){Lindner}, {Baker}, {Omont}, {Beelen},
  {Owen}, {Bertoldi}, {Dole}, {Fiolet}, {Harris}, {Ivison}, {Lonsdale}, {Lutz},
  \& {Polletta}}]{2011ApJ...737...83L}
{Lindner}, R.~R., {Baker}, A.~J., {Omont}, A., {et~al.} 2011, \apj, 737, 83

\bibitem[{{Lindner} {et~al.}(2015){Lindner}, {Aguirre}, {Baker}, {Bond},
  {Crichton}, {Devlin}, {Essinger-Hileman}, {Gallardo}, {Gralla}, {Hilton},
  {Hincks}, {Huffenberger}, {Hughes}, {Infante}, {Lima}, {Marriage},
  {Menanteau}, {Niemack}, {Page}, {Schmitt}, {Sehgal}, {Sievers}, {Sif{\'o}n},
  {Staggs}, {Swetz}, {Wei{\ss}}, \& {Wollack}}]{2015ApJ...803...79L}
{Lindner}, R.~R., {Aguirre}, P., {Baker}, A.~J., {et~al.} 2015, \apj, 803, 79

\bibitem[{{Madau} \& {Dickinson}(2014)}]{2014ARA&A..52..415M}
{Madau}, P., \& {Dickinson}, M. 2014, \araa, 52, 415

\bibitem[{{Magnelli} {et~al.}(2013){Magnelli}, {Popesso}, {Berta}, {Pozzi},
  {Elbaz}, {Lutz}, {Dickinson}, {Altieri}, {Andreani}, {Aussel},
  {B{\'e}thermin}, {Bongiovanni}, {Cepa}, {Charmandaris}, {Chary}, {Cimatti},
  {Daddi}, {F{\"o}rster Schreiber}, {Genzel}, {Gruppioni}, {Harwit}, {Hwang},
  {Ivison}, {Magdis}, {Maiolino}, {Murphy}, {Nordon}, {Pannella}, {P{\'e}rez
  Garc{\'{\i}}a}, {Poglitsch}, {Rosario}, {Sanchez-Portal}, {Santini}, {Scott},
  {Sturm}, {Tacconi}, \& {Valtchanov}}]{2013A&A...553A.132M}
{Magnelli}, B., {Popesso}, P., {Berta}, S., {et~al.} 2013, \aap, 553, A132

\bibitem[{{Mao} {et~al.}(2000){Mao}, {Henkel}, {Schulz}, {Zielinsky},
  {Mauersberger}, {St{\"o}rzer}, {Wilson}, \&
  {Gensheimer}}]{2000A&A...358..433M}
{Mao}, R.~Q., {Henkel}, C., {Schulz}, A., {et~al.} 2000, \aap, 358, 433

\bibitem[{{Marriage} {et~al.}(2011){Marriage}, {Acquaviva}, {Ade}, {Aguirre},
  {Amiri}, {Appel}, {Barrientos}, {Battistelli}, {Bond}, {Brown}, {Burger},
  {Chervenak}, {Das}, {Devlin}, {Dicker}, {Bertrand Doriese}, {Dunkley},
  {D{\"u}nner}, {Essinger-Hileman}, {Fisher}, {Fowler}, {Hajian}, {Halpern},
  {Hasselfield}, {Hern{\'a}ndez-Monteagudo}, {Hilton}, {Hilton}, {Hincks},
  {Hlozek}, {Huffenberger}, {Handel Hughes}, {Hughes}, {Infante}, {Irwin},
  {Baptiste Juin}, {Kaul}, {Klein}, {Kosowsky}, {Lau}, {Limon}, {Lin},
  {Lupton}, {Marsden}, {Martocci}, {Mauskopf}, {Menanteau}, {Moodley},
  {Moseley}, {Netterfield}, {Niemack}, {Nolta}, {Page}, {Parker}, {Partridge},
  {Quintana}, {Reese}, {Reid}, {Sehgal}, {Sherwin}, {Sievers}, {Spergel},
  {Staggs}, {Swetz}, {Switzer}, {Thornton}, {Trac}, {Tucker}, {Warne},
  {Wilson}, {Wollack}, \& {Zhao}}]{2011ApJ...737...61M}
{Marriage}, T.~A., {Acquaviva}, V., {Ade}, P.~A.~R., {et~al.} 2011, \apj, 737,
  61

\bibitem[{{Martini} {et~al.}(2013){Martini}, {Miller}, {Brodwin}, {Stanford},
  {Gonzalez}, {Bautz}, {Hickox}, {Stern}, {Eisenhardt}, {Galametz}, {Norman},
  {Jannuzi}, {Dey}, {Murray}, {Jones}, \& {Brown}}]{2013ApJ...768....1M}
{Martini}, P., {Miller}, E.~D., {Brodwin}, M., {et~al.} 2013, \apj, 768, 1

\bibitem[{{Mauch} \& {Sadler}(2007)}]{2007MNRAS.375..931M}
{Mauch}, T., \& {Sadler}, E.~M. 2007, \mnras, 375, 931

\bibitem[{McKinney(2010)}]{pandas}
McKinney, W. 2010, in Proceedings of the 9th Python in Science Conference, ed.
  S.~van~der Walt \& J.~Millman, 51 -- 56

\bibitem[{{McMullin} {et~al.}(2007){McMullin}, {Waters}, {Schiebel}, {Young},
  \& {Golap}}]{2007ASPC..376..127M}
{McMullin}, J.~P., {Waters}, B., {Schiebel}, D., {Young}, W., \& {Golap}, K.
  2007, in Astronomical Society of the Pacific Conference Series, Vol. 376,
  Astronomical Data Analysis Software and Systems XVI, ed. R.~A. {Shaw},
  F.~{Hill}, \& D.~J. {Bell}, 127

\bibitem[{{Menanteau} {et~al.}(2010){Menanteau}, {Gonz{\'a}lez}, {Juin},
  {Marriage}, {Reese}, {Acquaviva}, {Aguirre}, {Appel}, {Baker}, {Barrientos},
  {Battistelli}, {Bond}, {Das}, {Deshpande}, {Devlin}, {Dicker}, {Dunkley},
  {D{\"u}nner}, {Essinger-Hileman}, {Fowler}, {Hajian}, {Halpern},
  {Hasselfield}, {Hern{\'a}ndez-Monteagudo}, {Hilton}, {Hincks}, {Hlozek},
  {Huffenberger}, {Hughes}, {Infante}, {Irwin}, {Klein}, {Kosowsky}, {Lin},
  {Marsden}, {Moodley}, {Niemack}, {Nolta}, {Page}, {Parker}, {Partridge},
  {Sehgal}, {Sievers}, {Spergel}, {Staggs}, {Swetz}, {Switzer}, {Thornton},
  {Trac}, {Warne}, \& {Wollack}}]{2010ApJ...723.1523M}
{Menanteau}, F., {Gonz{\'a}lez}, J., {Juin}, J.-B., {et~al.} 2010, \apj, 723,
  1523

\bibitem[{{Menanteau} {et~al.}(2012){Menanteau}, {Hughes}, {Sif{\'o}n},
  {Hilton}, {Gonz{\'a}lez}, {Infante}, {Barrientos}, {Baker}, {Bond}, {Das},
  {Devlin}, {Dunkley}, {Hajian}, {Hincks}, {Kosowsky}, {Marsden}, {Marriage},
  {Moodley}, {Niemack}, {Nolta}, {Page}, {Reese}, {Sehgal}, {Sievers},
  {Spergel}, {Staggs}, \& {Wollack}}]{2012ApJ...748....7M}
{Menanteau}, F., {Hughes}, J.~P., {Sif{\'o}n}, C., {et~al.} 2012, \apj, 748, 7

\bibitem[{{Mihos}(2004)}]{2004cgpc.symp..277M}
{Mihos}, J.~C. 2004, Clusters of Galaxies: Probes of Cosmological Structure and
  Galaxy Evolution, 277

\bibitem[{{Moore} {et~al.}(1996){Moore}, {Katz}, {Lake}, {Dressler}, \&
  {Oemler}}]{1996Natur.379..613M}
{Moore}, B., {Katz}, N., {Lake}, G., {Dressler}, A., \& {Oemler}, A. 1996,
  \nat, 379, 613

\bibitem[{{Noble} {et~al.}(2017){Noble}, {McDonald}, {Muzzin}, {Nantais},
  {Rudnick}, {van Kampen}, {Webb}, {Wilson}, {Yee}, {Boone}, {Cooper},
  {DeGroot}, {Delahaye}, {Demarco}, {Foltz}, {Hayden}, {Lidman},
  {Manilla-Robles}, \& {Perlmutter}}]{2017ApJ...842L..21N}
{Noble}, A.~G., {McDonald}, M., {Muzzin}, A., {et~al.} 2017, \apjl, 842, L21

\bibitem[{{Noeske} {et~al.}(2007){Noeske}, {Weiner}, {Faber}, {Papovich},
  {Koo}, {Somerville}, {Bundy}, {Conselice}, {Newman}, {Schiminovich}, {Le
  Floc'h}, {Coil}, {Rieke}, {Lotz}, {Primack}, {Barmby}, {Cooper}, {Davis},
  {Ellis}, {Fazio}, {Guhathakurta}, {Huang}, {Kassin}, {Martin}, {Phillips},
  {Rich}, {Small}, {Willmer}, \& {Wilson}}]{2007ApJ...660L..43N}
{Noeske}, K.~G., {Weiner}, B.~J., {Faber}, S.~M., {et~al.} 2007, \apjl, 660,
  L43

\bibitem[{{Nulsen}(1982)}]{1982MNRAS.198.1007N}
{Nulsen}, P.~E.~J. 1982, \mnras, 198, 1007

\bibitem[{{Ott}(2010)}]{2010ASPC..434..139O}
{Ott}, S. 2010, in Astronomical Society of the Pacific Conference Series, Vol.
  434, Astronomical Data Analysis Software and Systems XIX, ed. Y.~{Mizumoto},
  K.-I. {Morita}, \& M.~{Ohishi}, 139

\bibitem[{{Papadopoulos} {et~al.}(2012){Papadopoulos}, {van der Werf},
  {Xilouris}, {Isaak}, {Gao}, \& {M{\"u}hle}}]{2012MNRAS.426.2601P}
{Papadopoulos}, P.~P., {van der Werf}, P.~P., {Xilouris}, E.~M., {et~al.} 2012,
  \mnras, 426, 2601

\bibitem[{{Peng} {et~al.}(2010){Peng}, {Lilly}, {Kova{\v c}}, {Bolzonella},
  {Pozzetti}, {Renzini}, {Zamorani}, {Ilbert}, {Knobel}, {Iovino}, {Maier},
  {Cucciati}, {Tasca}, {Carollo}, {Silverman}, {Kampczyk}, {de Ravel},
  {Sanders}, {Scoville}, {Contini}, {Mainieri}, {Scodeggio}, {Kneib}, {Le
  F{\`e}vre}, {Bardelli}, {Bongiorno}, {Caputi}, {Coppa}, {de la Torre},
  {Franzetti}, {Garilli}, {Lamareille}, {Le Borgne}, {Le Brun}, {Mignoli},
  {Perez Montero}, {Pello}, {Ricciardelli}, {Tanaka}, {Tresse}, {Vergani},
  {Welikala}, {Zucca}, {Oesch}, {Abbas}, {Barnes}, {Bordoloi}, {Bottini},
  {Cappi}, {Cassata}, {Cimatti}, {Fumana}, {Hasinger}, {Koekemoer},
  {Leauthaud}, {Maccagni}, {Marinoni}, {McCracken}, {Memeo}, {Meneux}, {Nair},
  {Porciani}, {Presotto}, \& {Scaramella}}]{2010ApJ...721..193P}
{Peng}, Y.-j., {Lilly}, S.~J., {Kova{\v c}}, K., {et~al.} 2010, \apj, 721, 193

\bibitem[{{Perley} \& {Butler}(2013)}]{2013ApJS..204...19P}
{Perley}, R.~A., \& {Butler}, B.~J. 2013, \apjs, 204, 19

\bibitem[{{Pilbratt} {et~al.}(2010){Pilbratt}, {Riedinger}, {Passvogel},
  {Crone}, {Doyle}, {Gageur}, {Heras}, {Jewell}, {Metcalfe}, {Ott}, \&
  {Schmidt}}]{2010A&A...518L...1P}
{Pilbratt}, G.~L., {Riedinger}, J.~R., {Passvogel}, T., {et~al.} 2010, \aap,
  518, L1

\bibitem[{{Poggianti} {et~al.}(2004){Poggianti}, {Bridges}, {Komiyama}, {Yagi},
  {Carter}, {Mobasher}, {Okamura}, \& {Kashikawa}}]{2004ApJ...601..197P}
{Poggianti}, B.~M., {Bridges}, T.~J., {Komiyama}, Y., {et~al.} 2004, \apj, 601,
  197

\bibitem[{{Poglitsch} {et~al.}(2010){Poglitsch}, {Waelkens}, {Geis},
  {Feuchtgruber}, {Vandenbussche}, {Rodriguez}, {Krause}, {Renotte}, {van
  Hoof}, {Saraceno}, {Cepa}, {Kerschbaum}, {Agn{\`e}se}, {Ali}, {Altieri},
  {Andreani}, {Augueres}, {Balog}, {Barl}, {Bauer}, {Belbachir}, {Benedettini},
  {Billot}, {Boulade}, {Bischof}, {Blommaert}, {Callut}, {Cara}, {Cerulli},
  {Cesarsky}, {Contursi}, {Creten}, {De Meester}, {Doublier}, {Doumayrou},
  {Duband}, {Exter}, {Genzel}, {Gillis}, {Gr{\"o}zinger}, {Henning},
  {Herreros}, {Huygen}, {Inguscio}, {Jakob}, {Jamar}, {Jean}, {de Jong},
  {Katterloher}, {Kiss}, {Klaas}, {Lemke}, {Lutz}, {Madden}, {Marquet},
  {Martignac}, {Mazy}, {Merken}, {Montfort}, {Morbidelli}, {M{\"u}ller},
  {Nielbock}, {Okumura}, {Orfei}, {Ottensamer}, {Pezzuto}, {Popesso},
  {Putzeys}, {Regibo}, {Reveret}, {Royer}, {Sauvage}, {Schreiber}, {Stegmaier},
  {Schmitt}, {Schubert}, {Sturm}, {Thiel}, {Tofani}, {Vavrek}, {Wetzstein},
  {Wieprecht}, \& {Wiezorrek}}]{2010A&A...518L...2P}
{Poglitsch}, A., {Waelkens}, C., {Geis}, N., {et~al.} 2010, \aap, 518, L2

\bibitem[{{Popesso} {et~al.}(2015){Popesso}, {Biviano}, {Finoguenov}, {Wilman},
  {Salvato}, {Magnelli}, {Gruppioni}, {Pozzi}, {Rodighiero}, {Ziparo}, {Berta},
  {Elbaz}, {Dickinson}, {Lutz}, {Altieri}, {Aussel}, {Cimatti}, {Fadda},
  {Ilbert}, {Le Floch}, {Nordon}, {Poglitsch}, {Genel}, \&
  {Xu}}]{2015A&A...579A.132P}
{Popesso}, P., {Biviano}, A., {Finoguenov}, A., {et~al.} 2015, \aap, 579, A132

\bibitem[{{Popping} {et~al.}(2015){Popping}, {Behroozi}, \&
  {Peeples}}]{2015MNRAS.449..477P}
{Popping}, G., {Behroozi}, P.~S., \& {Peeples}, M.~S. 2015, \mnras, 449, 477

\bibitem[{{Rieke} {et~al.}(2009){Rieke}, {Alonso-Herrero}, {Weiner},
  {P{\'e}rez-Gonz{\'a}lez}, {Blaylock}, {Donley}, \&
  {Marcillac}}]{2009ApJ...692..556R}
{Rieke}, G.~H., {Alonso-Herrero}, A., {Weiner}, B.~J., {et~al.} 2009, \apj,
  692, 556

\bibitem[{{Robitaille} \& {Bressert}(2012)}]{2012ascl.soft08017R}
{Robitaille}, T., \& {Bressert}, E. 2012, {APLpy: Astronomical Plotting Library
  in Python}, Astrophysics Source Code Library, ascl:1208.017

\bibitem[{{Rosario} {et~al.}(2016){Rosario}, {Mendel}, {Ellison}, {Lutz}, \&
  {Trump}}]{2016MNRAS.457.2703R}
{Rosario}, D.~J., {Mendel}, J.~T., {Ellison}, S.~L., {Lutz}, D., \& {Trump},
  J.~R. 2016, \mnras, 457, 2703

\bibitem[{{Rudnick} {et~al.}(2017){Rudnick}, {Hodge}, {Walter}, {Momcheva},
  {Tran}, {Papovich}, {da Cunha}, {Decarli}, {Saintonge}, {Willmer}, {Lotz}, \&
  {Lentati}}]{2017arXiv170906963R}
{Rudnick}, G., {Hodge}, J., {Walter}, F., {et~al.} 2017, ArXiv e-prints,
  arXiv:1709.06963

\bibitem[{{Rujopakarn} {et~al.}(2010){Rujopakarn}, {Eisenstein}, {Rieke},
  {Papovich}, {Cool}, {Moustakas}, {Jannuzi}, {Kochanek}, {Rieke}, {Dey},
  {Eisenhardt}, {Murray}, {Brown}, \& {Le Floc'h}}]{2010ApJ...718.1171R}
{Rujopakarn}, W., {Eisenstein}, D.~J., {Rieke}, G.~H., {et~al.} 2010, \apj,
  718, 1171

\bibitem[{{Salpeter}(1955)}]{1955ApJ...121..161S}
{Salpeter}, E.~E. 1955, \apj, 121, 161

\bibitem[{{Scoville} {et~al.}(2016){Scoville}, {Sheth}, {Aussel}, {Vanden
  Bout}, {Capak}, {Bongiorno}, {Casey}, {Murchikova}, {Koda},
  {{\'A}lvarez-M{\'a}rquez}, {Lee}, {Laigle}, {McCracken}, {Ilbert}, {Pope},
  {Sanders}, {Chu}, {Toft}, {Ivison}, \& {Manohar}}]{2016ApJ...820...83S}
{Scoville}, N., {Sheth}, K., {Aussel}, H., {et~al.} 2016, \apj, 820, 83

\bibitem[{{Scoville}(2013)}]{2013seg..book..491S}
{Scoville}, N.~Z. 2013, {Evolution of star formation and gas}, ed.
  J.~{Falc{\'o}n-Barroso} \& J.~H. {Knapen}, 491

\bibitem[{{Serjeant} {et~al.}(2003){Serjeant}, {Dunlop}, {Mann},
  {Rowan-Robinson}, {Hughes}, {Efstathiou}, {Blain}, {Fox}, {Ivison},
  {Jenness}, {Lawrence}, {Longair}, {Oliver}, \&
  {Peacock}}]{2003MNRAS.344..887S}
{Serjeant}, S., {Dunlop}, J.~S., {Mann}, R.~G., {et~al.} 2003, \mnras, 344, 887

\bibitem[{{Serra} {et~al.}(2015){Serra}, {Westmeier}, {Giese}, {Jurek},
  {Fl{\"o}er}, {Popping}, {Winkel}, {van der Hulst}, {Meyer}, {Koribalski},
  {Staveley-Smith}, \& {Courtois}}]{2015MNRAS.448.1922S}
{Serra}, P., {Westmeier}, T., {Giese}, N., {et~al.} 2015, \mnras, 448, 1922

\bibitem[{{Sif{\'o}n} {et~al.}(2013){Sif{\'o}n}, {Menanteau}, {Hasselfield},
  {Marriage}, {Hughes}, {Barrientos}, {Gonz{\'a}lez}, {Infante}, {Addison},
  {Baker}, {Battaglia}, {Bond}, {Crichton}, {Das}, {Devlin}, {Dunkley},
  {D{\"u}nner}, {Gralla}, {Hajian}, {Hilton}, {Hincks}, {Kosowsky}, {Marsden},
  {Moodley}, {Niemack}, {Nolta}, {Page}, {Partridge}, {Reese}, {Sehgal},
  {Sievers}, {Spergel}, {Staggs}, {Thornton}, {Trac}, \&
  {Wollack}}]{2013ApJ...772...25S}
{Sif{\'o}n}, C., {Menanteau}, F., {Hasselfield}, M., {et~al.} 2013, \apj, 772,
  25

\bibitem[{{Sif{\'o}n} {et~al.}(2016){Sif{\'o}n}, {Battaglia}, {Hasselfield},
  {Menanteau}, {Barrientos}, {Bond}, {Crichton}, {Devlin}, {D{\"u}nner},
  {Hilton}, {Hincks}, {Hlozek}, {Huffenberger}, {Hughes}, {Infante},
  {Kosowsky}, {Marsden}, {Marriage}, {Moodley}, {Niemack}, {Page}, {Spergel},
  {Staggs}, {Trac}, \& {Wollack}}]{2016MNRAS.461..248S}
{Sif{\'o}n}, C., {Battaglia}, N., {Hasselfield}, M., {et~al.} 2016, \mnras,
  461, 248

\bibitem[{{Solomon} {et~al.}(1997){Solomon}, {Downes}, {Radford}, \&
  {Barrett}}]{1997ApJ...478..144S}
{Solomon}, P.~M., {Downes}, D., {Radford}, S.~J.~E., \& {Barrett}, J.~W. 1997,
  \apj, 478, 144

\bibitem[{{Somerville} {et~al.}(2008){Somerville}, {Hopkins}, {Cox},
  {Robertson}, \& {Hernquist}}]{2008MNRAS.391..481S}
{Somerville}, R.~S., {Hopkins}, P.~F., {Cox}, T.~J., {Robertson}, B.~E., \&
  {Hernquist}, L. 2008, \mnras, 391, 481

\bibitem[{{Sunyaev} \& {Zel'dovich}(1972)}]{1972CoASP...4..173S}
{Sunyaev}, R.~A., \& {Zel'dovich}, Y.~B. 1972, Comments on Astrophysics and
  Space Physics, 4, 173

\bibitem[{{Swetz} {et~al.}(2011){Swetz}, {Ade}, {Amiri}, {Appel},
  {Battistelli}, {Burger}, {Chervenak}, {Devlin}, {Dicker}, {Doriese},
  {D{\"u}nner}, {Essinger-Hileman}, {Fisher}, {Fowler}, {Halpern},
  {Hasselfield}, {Hilton}, {Hincks}, {Irwin}, {Jarosik}, {Kaul}, {Klein},
  {Lau}, {Limon}, {Marriage}, {Marsden}, {Martocci}, {Mauskopf}, {Moseley},
  {Netterfield}, {Niemack}, {Nolta}, {Page}, {Parker}, {Staggs}, {Stryzak},
  {Switzer}, {Thornton}, {Tucker}, {Wollack}, \& {Zhao}}]{2011ApJS..194...41S}
{Swetz}, D.~S., {Ade}, P.~A.~R., {Amiri}, M., {et~al.} 2011, \apjs, 194, 41

\bibitem[{{Tacconi} {et~al.}(2013){Tacconi}, {Neri}, {Genzel}, {Combes},
  {Bolatto}, {Cooper}, {Wuyts}, {Bournaud}, {Burkert}, {Comerford}, {Cox},
  {Davis}, {F{\"o}rster Schreiber}, {Garc{\'{\i}}a-Burillo}, {Gracia-Carpio},
  {Lutz}, {Naab}, {Newman}, {Omont}, {Saintonge}, {Shapiro Griffin}, {Shapley},
  {Sternberg}, \& {Weiner}}]{2013ApJ...768...74T}
{Tacconi}, L.~J., {Neri}, R., {Genzel}, R., {et~al.} 2013, \apj, 768, 74

\bibitem[{{Topal} {et~al.}(2014){Topal}, {Bayet}, {Bureau}, {Davis}, \&
  {Walsh}}]{2014MNRAS.437.1434T}
{Topal}, S., {Bayet}, E., {Bureau}, M., {Davis}, T.~A., \& {Walsh}, W. 2014,
  \mnras, 437, 1434

\bibitem[{{Tran} {et~al.}(2010){Tran}, {Papovich}, {Saintonge}, {Brodwin},
  {Dunlop}, {Farrah}, {Finkelstein}, {Finkelstein}, {Lotz}, {McLure},
  {Momcheva}, \& {Willmer}}]{2010ApJ...719L.126T}
{Tran}, K.-V.~H., {Papovich}, C., {Saintonge}, A., {et~al.} 2010, \apjl, 719,
  L126

\bibitem[{van~der Walt {et~al.}(2011)van~der Walt, Colbert, \&
  Varoquaux}]{NumPy&SciPy}
van~der Walt, S., Colbert, S.~C., \& Varoquaux, G. 2011, Computing in Science
  \& Engineering, 13, 22

\bibitem[{{Vanderlinde} {et~al.}(2010){Vanderlinde}, {Crawford}, {de Haan},
  {Dudley}, {Shaw}, {Ade}, {Aird}, {Benson}, {Bleem}, {Brodwin}, {Carlstrom},
  {Chang}, {Crites}, {Desai}, {Dobbs}, {Foley}, {George}, {Gladders}, {Hall},
  {Halverson}, {High}, {Holder}, {Holzapfel}, {Hrubes}, {Joy}, {Keisler},
  {Knox}, {Lee}, {Leitch}, {Loehr}, {Lueker}, {Marrone}, {McMahon}, {Mehl},
  {Meyer}, {Mohr}, {Montroy}, {Ngeow}, {Padin}, {Plagge}, {Pryke}, {Reichardt},
  {Rest}, {Ruel}, {Ruhl}, {Schaffer}, {Shirokoff}, {Song}, {Spieler},
  {Stalder}, {Staniszewski}, {Stark}, {Stubbs}, {van Engelen}, {Vieira},
  {Williamson}, {Yang}, {Zahn}, \& {Zenteno}}]{2010ApJ...722.1180V}
{Vanderlinde}, K., {Crawford}, T.~M., {de Haan}, T., {et~al.} 2010, \apj, 722,
  1180

\bibitem[{{Vantyghem} {et~al.}(2017){Vantyghem}, {McNamara}, {Edge}, {Combes},
  {Russell}, {Fabian}, {Hogan}, {McDonald}, {Nulsen}, \&
  {Salom{\'e}}}]{2017arXiv170909679V}
{Vantyghem}, A.~N., {McNamara}, B.~R., {Edge}, A.~C., {et~al.} 2017, ArXiv
  e-prints, arXiv:1709.09679

\bibitem[{{Wagg} {et~al.}(2012){Wagg}, {Pope}, {Alberts}, {Armus}, {Brodwin},
  {Bussmann}, {Desai}, {Dey}, {Jannuzi}, {Le Floc'h}, {Melbourne}, \&
  {Stern}}]{2012ApJ...752...91W}
{Wagg}, J., {Pope}, A., {Alberts}, S., {et~al.} 2012, \apj, 752, 91

\bibitem[{{Walter} {et~al.}(2011){Walter}, {Wei{\ss}}, {Downes}, {Decarli}, \&
  {Henkel}}]{2011ApJ...730...18W}
{Walter}, F., {Wei{\ss}}, A., {Downes}, D., {Decarli}, R., \& {Henkel}, C.
  2011, \apj, 730, 18

\bibitem[{{Wei{\ss}} {et~al.}(2005{\natexlab{a}}){Wei{\ss}}, {Downes},
  {Henkel}, \& {Walter}}]{2005A&A...429L..25W}
{Wei{\ss}}, A., {Downes}, D., {Henkel}, C., \& {Walter}, F. 2005{\natexlab{a}},
  \aap, 429, L25

\bibitem[{{Wei{\ss}} {et~al.}(2005{\natexlab{b}}){Wei{\ss}}, {Walter}, \&
  {Scoville}}]{2005A&A...438..533W}
{Wei{\ss}}, A., {Walter}, F., \& {Scoville}, N.~Z. 2005{\natexlab{b}}, \aap,
  438, 533

\bibitem[{{Yun} {et~al.}(2001){Yun}, {Reddy}, \&
  {Condon}}]{2001ApJ...554..803Y}
{Yun}, M.~S., {Reddy}, N.~A., \& {Condon}, J.~J. 2001, \apj, 554, 803

\bibitem[{{Ziparo} {et~al.}(2014){Ziparo}, {Popesso}, {Finoguenov}, {Biviano},
  {Wuyts}, {Wilman}, {Salvato}, {Tanaka}, {Nandra}, {Lutz}, {Elbaz},
  {Dickinson}, {Altieri}, {Aussel}, {Berta}, {Cimatti}, {Fadda}, {Genzel}, {Le
  Floc'h}, {Magnelli}, {Nordon}, {Poglitsch}, {Pozzi}, {Portal}, {Tacconi},
  {Bauer}, {Brandt}, {Cappelluti}, {Cooper}, \&
  {Mulchaey}}]{2014MNRAS.437..458Z}
{Ziparo}, F., {Popesso}, P., {Finoguenov}, A., {et~al.} 2014, \mnras, 437, 458

\bibitem[{{Zitrin} {et~al.}(2013){Zitrin}, {Menanteau}, {Hughes}, {Coe},
  {Barrientos}, {Infante}, \& {Mandelbaum}}]{2013ApJ...770L..15Z}
{Zitrin}, A., {Menanteau}, F., {Hughes}, J.~P., {et~al.} 2013, \apjl, 770, L15

\end{thebibliography}

\end{document}